\documentclass[journal]{IEEEtran}

\usepackage{xcolor}
\usepackage{tikz}
\usetikzlibrary{positioning}
\usetikzlibrary{arrows}
\usetikzlibrary{circuits.ee.IEC}
\usepackage{dblfloatfix}
\usepackage{cite}
\usepackage{amssymb,amsfonts}
\usepackage{amsmath}
\usepackage{booktabs}
\usepackage[font=footnotesize]{caption}
\usepackage[font=footnotesize]{subcaption}
\usepackage[bottom]{footmisc}
\usepackage[english]{babel}
\usepackage{makecell}
\usepackage{bbold}
\usepgflibrary{shadings}
\usepackage{arydshln}
\usepackage{comment}

\let\proof\relax
\let\endproof\relax
\usepackage{amsthm}

\newtheorem{remark}{Remark}

\usepackage[capitalize,nameinlink,compress]{cleveref}
\crefname{figure}{Fig.}{Figures} % Use Figure instead of Fig.
\crefname{line}{line}{lines} % Make sure line is not capitalized
\crefname{claim}{Claim}{Claims} % Make sure line is not capitalized
\crefname{equation}{}{} % No Eq. for equations
\crefname{problem}{Problem}{Problems}
\crefname{assumption}{Assumption}{Assumptions}

\definecolor{backgroundcolor}{rgb}{0.4660, 0.6740, 0.1880}
\definecolor{backgroundcolor2}{rgb}{1, 0.4,0.5}
\definecolor{backgroundcolor_red}{rgb}{0.658,0.196,0.176}

\usepackage{soul}
\usepackage{framed}
\colorlet{shadecolor}{yellow}
\soulregister\cite7
\soulregister\cref7
\soulregister\ref7
\soulregister\pageref7
\soulregister\eqref7

\renewcommand{\baselinestretch}{1}
\begin{document}

\title{Optimal Dynamic Ancillary Services Provision Based on Local Power Grid Perception}

\author{Verena~Häberle,~\IEEEmembership{Graduate Student Member,~IEEE,}
        Xiuqiang~He,~\IEEEmembership{Member,~IEEE,}
        Linbin~Huang,~\IEEEmembership{Member,~IEEE,}\\
        Eduardo~Prieto-Araujo,~\IEEEmembership{Senior Member,~IEEE,}
        and~Florian~Dörfler,~\IEEEmembership{Senior Member,~IEEE}% <-this % stops a space

\thanks{\fontdimen2\font=0.6ex This work was supported by the European Union's Horizon 2020 and 2023 research and innovation programs (Grant Agreement Numbers 883985 and 101096197). Verena Häberle, Xiuqiang He, Linbin Huang and Florian Dörfler are with the Automatic Control Laboratory, ETH Zurich, 8092 Zurich, Switzerland. Email:\{verenhae,xiuqhe,linhuang,dorfler\}@ethz.ch. Eduardo Prieto-Araujo is a Serra Húnter Lecturer with CITCEA, Department d’Enginyeria Elèctrica, Universitat Politècnica de Catalunya, 08028 Barcelona, Spain. Email: eduardo.prieto-araujo@upc.edu}% <-this % stops a space
}

\maketitle
\begin{abstract}\fontdimen2\font=0.6ex
In this paper, we propose a systematic closed-loop approach to provide optimal dynamic ancillary services with converter-interfaced generation systems based on local power grid perception. In particular, we structurally encode dynamic ancillary services such as fast frequency and voltage regulation in the form of a \textit{parametric} transfer function matrix, which includes several parameters to define a set of different feasible response behaviors, among which we aim to find the optimal one to be realized by the converter system. Our approach is based on a so-called \textit{``perceive-and-optimize'' (P\&O)} strategy: First, we identify a grid dynamic equivalent at the interconnection terminals of the converter system. Second, we consider the closed-loop interconnection of the identified grid equivalent and the parametric transfer function matrix, which we optimize for the set of transfer function parameters, resulting in a stable and optimal closed-loop performance for ancillary services provision. In the process, we ensure that grid-code and device-level requirements are satisfied. Finally, we demonstrate the effectiveness of our approach in different numerical case studies based on a modified Kundur two-area test system.
\end{abstract}

\section{Introduction}\fontdimen2\font=0.6ex
\IEEEPARstart{T}{oday's} grid-code specifications for dynamic ancillary services provision (e.g., fast frequency and voltage regulation) with converter-based generation units are typically defined by a prescribed time-domain step-response characteristic \cite{european2016commission,oyj2021technical,Eirgrid2018,modig2019technical}. As an example, the European network code \cite{european2016commission}, which is adopted in most European national grid codes, specifies the active power provision for frequency containment reserve (FCR) in response to a frequency step change by a piece-wise linear time-domain curve (\cref{fig:fcr_intro}). Likewise, the dynamic response of reactive power for voltage regulation is defined via time-domain specifications in response to a voltage step change \cite{european2016commission}. Recent grid codes (e.g., Finland \cite{oyj2021technical}, Ireland \cite{Eirgrid2018}, Sweden/Norway/Denmark \cite{modig2019technical}) also define the activation of fast frequency reserves (FFR) or synthetic inertia response via piece-wise linear active power curves in the time domain. Such grid codes are used to involve power converters in supporting low-inertia power grids and are important measures to ensure the transition toward converter-dominated power systems.

Although the specification of the piece-wise linear time-domain curves in today's grid codes is straightforward, they only assign the lower bound of the \textit{open-loop} response characteristic for an ancillary services-providing reserve unit (see, e.g., the red curve in \cref{fig:fcr_intro}). On the one hand, different dynamic responses of a reserve unit are allowed, as long as the grid-code requirements are satisfied at or above the piece-wise linear curves. This results in a family of different feasible response behaviors, where the reserve unit injection is often pared down to the minimum by just satisfying the grid-code requirements at its boundaries. On the other hand, the current grid-code specifications are \textit{indistinguishable} for any location of converter-interfaced generation, regardless of the grid strength and dynamic characteristics at the grid-connection point. However, in future power systems with increased penetration of converter-based generation, such indistinguishable grid-code specifications in an open-loop manner will result in a sub-optimal closed-loop power system response with poor dynamic performance or even unstable behaviors \cite{geng2017small,ying2018impact}. In this regard, existing studies in \cite{poolla2019placement,ademola2018optimal,gross2017increasing} have indicated that the location of synthetic inertia and damping should be carefully chosen to provide better stability and dynamic performance. Therefore, future grid-code specifications for dynamic ancillary services provision may have to respect distinguishably the local grid dynamic characteristics in a closed-loop manner.
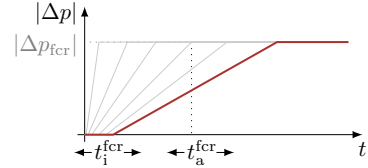
\begin{figure}[t!]
\vspace{-1mm}
    \centering
     \resizebox{0.28\textwidth}{!}{
\begin{tikzpicture}
\draw[-latex] (-2.8,0.9) -- (-2.8,2.8); 
\draw [-latex](-2.9,1) -- (1.1,1);
\node [scale=0.9]at (-3.2,2.7) {$|\Delta p|$};
\node [scale=0.9]at (1.1,0.8) {$t$};

\draw[dotted] (-1.3,2.3) -- (-1.3,1);
\node [scale=0.9] at (-1.15,0.75) {$t_\mathrm{a}^\mathrm{fcr}$};

\draw[dotted,gray](-0.1,2.3)  -- (-2.8,2.3);
\node [scale=0.9,gray] at (-3.35,2.25) {$|\Delta p_\mathrm{fcr}|$};
%\node [scale=0.7,black!60] at (-3.3,1.88) {capacity};

\node [scale=0.9] at (-2.45,0.75) {$t_\mathrm{i}^\mathrm{fcr}$};

\draw[gray!40!white](-2.8,1) -- (-2.7,1) node (v3) {} -- (-1.8,2.3) -- (0.9,2.3);

\draw[gray!40!white](-2.8,1) -- (-2.75,1) node (v5) {}  -- (-2.2,2.3) -- (0.9,2.3); 
\draw[gray!40!white] (-2.8,1) -- (-2.5,1) -- (-0.8,2.3) -- (0.9,2.3);
\draw[gray!40!white] (-2.78,1) node (v4) {}-- (-2.6,2.3) -- (0.9,2.3) node (v7) {};

\draw[gray!40!white] (-2.8,1) -- (-2.6,1) node (v6) {} -- (-1.3,2.3) -- (0.9,2.3); 
\draw [-latex](-2.25,0.75) -- (-2,0.75); 
\draw [-latex](-2.7,0.75) -- (-2.95,0.75);
\draw [-latex](-1.4,0.75) -- (-1.65,0.75); 
\draw [-latex](-0.95,0.75) -- (-0.7,0.75);
\draw[backgroundcolor_red,thick] (-2.8,1) node (v1) {}--(-2.4,1) -- (-0.1,2.3) -- (0.9,2.3) node (v2) {};

\end{tikzpicture}
}
    \vspace{-9mm}
    \caption{\fontdimen2\font=0.6ex Exemplary active power time-domain capability curve for FCR provision after a frequency step change \cite{european2016commission}. The minimum curve requirement (i.e., the lower bound of the open-loop response curve) is indicated in red.}
    \label{fig:fcr_intro}
         \vspace{-4mm}
\end{figure}
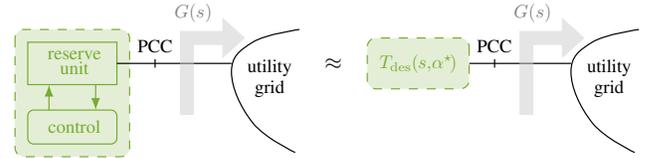
\begin{figure}[t!]
    \centering
     \resizebox{0.47\textwidth}{!}{

\begin{tikzpicture}[scale=0.5,every node/.style={scale=0.62}]

\draw [dashed,rounded corners = 3,color=backgroundcolor,fill= backgroundcolor!20] (-6.2,3.6) rectangle (-3.5,0.6);
\draw [color=backgroundcolor] (-5.9,3.3) rectangle (-3.8,2.3);
\node[backgroundcolor] at (-4.8,3) {reserve};
\node [backgroundcolor] at (-4.8,2.7) {unit};
\draw  [color=backgroundcolor, rounded corners = 2](-5.9,0.9) rectangle (-3.8,1.7);
\node [backgroundcolor] at (-4.8,1.3) {control};
\draw[-latex,color=backgroundcolor] (-4.3,2.3) -- (-4.3,1.7); 
\draw[-latex,color=backgroundcolor] (-5.4,1.7) -- (-5.4,2.3); 

\draw (-3.8,2.8) -- (-1.1,2.8); 
\draw (-2.9,2.9) -- (-2.9,2.7); 
\node at (-2.9,3.2) {PCC};
\draw plot[smooth, tension=.7] coordinates {(0.5,3.6) (-0.6,3.3)(-1.1,2.6)(-0.6,1.4)(0.4,0.7)};
\node at (-0.2,2.7) {utility};
\node at (-0.2,2.2) {grid};
\draw[fill=gray!40!white,color=gray!40!white,opacity=0.5] (-2.3,1.6) node (v2) {} -- (-2.3,3.55) -- (-1.3,3.55) -- (-1.3,3.85) -- (-0.8,3.4) -- (-1.3,2.95) -- (-1.3,3.25) -- (-2,3.25) -- (-2,1.6) -- (-2.3,1.6);
\node [color=gray] at (-2,4) {$G(s)$};

\draw (4.5,2.8) -- (6.9,2.8); 
\draw (5.1,2.9) -- (5.1,2.7); 
\node at (5.1,3.2) {PCC};
\draw plot[smooth, tension=.7] coordinates {(8.5,3.6) (7.4,3.3)(6.9,2.6)(7.4,1.4)(8.4,0.7)};
\node at (7.8,2.7) {utility};
\node at (7.8,2.2) {grid};
\draw[fill=gray!40!white,color=gray!40!white,opacity=0.5] (5.7,1.6) node (v2) {} -- (5.7,3.55) -- (6.7,3.55) -- (6.7,3.85) -- (7.2,3.4) -- (6.7,2.95) -- (6.7,3.25) -- (6,3.25) -- (6,1.6) -- (5.7,1.6);
\node [color=gray] at (6,4) {$G(s)$};
\draw [dashed, rounded corners = 3,color=backgroundcolor,fill=backgroundcolor!20] (4.5,3.4) rectangle (2.1,2.2);
\node [backgroundcolor] at (3.3,2.8) {$T_\mathrm{des}(\hspace{-0.2mm}s,\hspace{-0.5mm}\alpha^\star\hspace{-0.5mm})$};
\node[scale=1.2] at (1.3,2.8) {$\approx$};
\end{tikzpicture}
}
    \vspace{-8mm}
    \caption{\fontdimen2\font=0.6ex Sketch of a grid-connected reserve unit to provide closed-loop optimal dynamic ancillary services in the form of a desired rational transfer function matrix $T_\mathrm{des}(s,\alpha^\star)$. The identified grid dynamic equivalent is captured by $G(s)$.}
    \label{fig:grid_setup}
    \vspace{-3mm}
\end{figure}

Moreover, beyond the flaw of specifying an open-loop response characteristic and being indistinguishable for any power grid condition, the practical implementation of the piece-wise linear time-domain grid-code curves in converters is not immediate, and no systematic methods have been developed yet. In this regard, today's industrial practice is often ad-hoc and highly customized, e.g., relying on open-loop trajectory commands \cite{ruttledge2015emulated}, varying gains \cite{wu2018state}, or look-up table schemes\cite{clark2010modeling} to approximately satisfy the grid codes. The design, implementation, and tuning of such methods, however, are in general time-consuming and rigid, which impedes its applicability.

To overcome the previous 
shortcomings, we propose a \textit{closed-loop} approach to provide \textit{optimal} dynamic ancillary services based on \textit{local} power grid perception, which systematically ensures that open-loop grid code requirements and device-level limitations are reliably satisfied. To be specific, we translate the aforementioned time-domain grid-code capability curves into a rational parametric transfer function matrix $T_\mathrm{des}(s,\alpha)$ in the frequency domain, which defines a tractable desired response behavior to be realized by the converter (see \cref{fig:grid_setup}). The parameter vector $\alpha$  establishes a set of feasible response behaviors constrained by the grid-code and device-level limitations, out of which we aim to find the optimal behavior such that a stable and optimal closed-loop performance of the entire power system can be achieved. As the \textit{main contribution} of this paper, we introduce a so-called \textit{``perceive-and-optimize'' (P\&O)} strategy, which is composed of two main steps: We first use the converter-based reserve unit to identify a grid dynamic equivalent $G(s)$ at its interconnection terminals (\textit{``perceive''}). Second, we consider the closed-loop interconnection of the identified grid dynamic equivalent $G(s)$ and the converter (represented by the parametric transfer function matrix $T_\mathrm{des}(s,\alpha)$), where we optimize $\alpha$ to achieve an optimal and stable closed-loop performance of the entire power grid response, while ensuring grid-code and device-level requirements to be reliably satisfied (\textit{``optimize''}). 

Numerous studies have explored the optimization of dynamic ancillary services provision, as evidenced by \cite{poolla2019placement,ademola2018optimal}. These endeavors rely on an offline design using precise knowledge of the dynamic power system model or its approximation, which, however, is typically unavailable in practice. Moreover, these methods failed to consider potential variations in grid conditions over time. Conversely, alternative approaches, exemplified by \cite{roberts2023online}, proposed adaptive tuning methods to enhance the dynamic response behavior of grid-connected converter systems based on local power grid measurements, thereby taking a similar P\&O perspective as proposed in our work. Despite its innovation, however, the suggested concept in \cite{roberts2023online} is highly specialized and confined to providing power oscillation damping services, lacking immediate adaptability to other dynamic ancillary service products. Notably, none of the methods outlined in \cite{poolla2019placement,ademola2018optimal,roberts2023online} account for grid-code or device-level requirements. This stands in vast contrast to our method for optimal dynamic ancillary services provision, characterized by its ability to:
\begin{itemize}
\item perceive \textit{local} grid dynamics,
\item attain optimal performance in a \textit{closed loop} with the grid,
\item encode \textit{grid-code} and \textit{device-level requirements},
\item accommodate \textit{time-varying grid conditions}, and 
\item offer a \textit{systematic} and \textit{versatile} approach applicable to a wide range of dynamic ancillary services products.
\end{itemize}

The remainder of this paper is structured as follows. In \cref{sec:grid_code2tf} we demonstrate how to encode dynamic ancillary services as a parametric transfer function matrix $T_\mathrm{des}(s,\alpha)$. In \cref{sec:optimal_AS}, we present the novel P\&O strategy for optimal dynamic ancillary services provision as the main contribution of this paper, including the grid dynamic equivalent identification, and the optimization problem to compute the optimal parameter $\alpha^\star$ of $T_\mathrm{des}(s,\alpha)$. In \cref{sec:case_studies}, we provide numerical case studies to demonstrate the vastly superior performance of our approach over only fulfilling the minimum (open-loop) grid-code requirements. \cref{sec:conclusion} concludes the paper. 

\section{Encoding Dynamic Ancillary Services as Parametric Transfer Functions}\label{sec:grid_code2tf} \fontdimen2\font=0.6ex
We consider dynamic ancillary services to be encoded in the form of a rational parametric transfer function matrix $T_\mathrm{des}(s,\alpha)$ in the frequency domain (with parameter vector $\alpha=[\alpha^\mathrm{fp},\alpha^\mathrm{vq}]$), specifying a \textit{desired} decoupled frequency and voltage control behavior to be realized by a reserve unit, i.e., 
\begin{align}\label{eq:Tdes}
    \hspace{-0.2cm}
    \begin{bmatrix}
        \Delta p(s)\\ \Delta q(s)
    \end{bmatrix} = 
    \underset{=\,T_\mathrm{des}(s,\alpha)}{\underbrace{\begin{bmatrix}
        T_\mathrm{des}^\mathrm{fp}(s,\alpha^\mathrm{fp}) & 0 \\ 0 & T_\mathrm{des}^\mathrm{vq}(s,\alpha^\mathrm{vq}) 
    \end{bmatrix}}}
    \begin{bmatrix}
        \Delta f(s) \\ \Delta v(s)
    \end{bmatrix}\hspace{-0.1cm},\hspace{-0.2cm}
\end{align}
where $\Delta f$ and $\Delta v$ are the measured bus frequency and voltage magnitude deviation, and $\Delta p$ and $\Delta q$ the active and reactive power output deviations (deviating from the respective power set point). In particular, the transfer function matrix in \cref{eq:Tdes} defines a tractable response behavior which can be easily realized in standard converter control architectures as a reference model to be matched. Inspired by today's grid-code specifications for transmission networks, we stick to a classical decoupled grid-following frequency and voltage regulation in \cref{eq:Tdes}. However, our formalism directly extends to fully coupled multi-input multi-output specifications for $T_\mathrm{des}(s,\alpha)$, potentially relevant for future ancillary services or in other types of networks \cite{de2007voltage}. Moreover, also grid-forming implementations with inverse signal causality can be considered \cite{haberle2022control}.

The transfer functions in \cref{eq:Tdes} are defined as a superposition of different dynamic ancillary services products $T_\mathrm{des}^{\mathrm{fp},i}(s,\alpha^{\mathrm{fp},i})$ and $T_\mathrm{des}^{\mathrm{vq},i}(s,\alpha^{\mathrm{vq},i})$ (acting on different time-scales \cite{osmoseh2020}), i.e.,
\begin{subequations}\label{eq:superimposed_AS}
\begin{align}\label{eq:superimposed_AS_frequency}
    T_\mathrm{des}^\mathrm{fp}(s,\alpha^\mathrm{fp}) &= \textstyle\sum_i T_\mathrm{des}^{\mathrm{fp},i}(s,\alpha^{\mathrm{fp},i}),\\\label{eq:superimposed_AS_voltage}
    T_\mathrm{des}^\mathrm{vq}(s,\alpha^\mathrm{vq}) &= \textstyle\sum_i T_\mathrm{des}^{\mathrm{vq},i}(s,\alpha^{\mathrm{vq},i}).
\end{align}
\end{subequations}
where the parametric structure of $T_\mathrm{des}^{\mathrm{fp},i}(s,\alpha^{\mathrm{fp},i})$ and $T_\mathrm{des}^{\mathrm{vq},i}(s,\alpha^{\mathrm{vq},i})$ results from the different grid-code requirements of each dynamic ancillary services product $i$, as detailed below.

\subsection{Frequency Regulation}\label{sec:grid_code2tf_frequency} \fontdimen2\font=0.6ex
Nowadays, typical dynamic ancillary service products for frequency regulation are \textit{frequency containment reserve (FCR)} \cite{european2016commission}, \textit{fast frequency response (FFR)} \cite{oyj2021technical,Eirgrid2018,modig2019technical}, and \textit{other auxiliary controls such as power oscillation damping (POD)} \cite{european2016commission,de2021norma}. In this regard, we can specify $T_\mathrm{des}^\mathrm{fp}(s,\alpha^\mathrm{fp})$ in \cref{eq:superimposed_AS_frequency} as 
\begin{align}\label{eq:T_des_fp_components}
    \hspace{-3mm} T_\mathrm{des}^\mathrm{fp}(s,\alpha^\mathrm{fp}) \hspace{-0.5mm}=\hspace{-0.5mm} T_\mathrm{des}^\mathrm{fcr}(s,\alpha^\mathrm{fcr})\hspace{-0.5mm}+\hspace{-0.5mm}T_\mathrm{des}^\mathrm{ffr}(s,\alpha^\mathrm{ffr})\hspace{-0.5mm}+\hspace{-0.5mm}T_\mathrm{des}^\mathrm{aux}(s,\alpha^\mathrm{aux}),\hspace{-2mm}
\end{align}
where $T_\mathrm{des}^\mathrm{fcr}(s,\alpha^\mathrm{fcr})$ encodes the FCR provision, $T_\mathrm{des}^\mathrm{ffr}(s,\alpha^\mathrm{ffr})$ the FFR provision, and $T_\mathrm{des}^\mathrm{aux}(s,\alpha^\mathrm{aux})$ defines an auxiliary service to damp high-frequency dynamics and/or power oscillations. Of course, depending on the requirements of different grid codes and reserve units, one might also consider other types of dynamic ancillary services products. Moreover, notice that different grid codes or market formats might support either a simultaneous provision of several ancillary services products by one reserve unit (see, e.g., Section 5 in the Finish grid code \cite{oyj2021technical}), or a separate provision of services by different units. In any case, we propose a \textit{generic approach} which is compatible with either option, i.e., capable of considering one, two, or multiple types of ancillary services products in \eqref{eq:T_des_fp_components}. For the sake of simplicity (and without loss of generality), we stick to the superposition of the three products in \cref{eq:T_des_fp_components} in the following.

The structure and parameters of the transfer function terms in \cref{eq:T_des_fp_components} can be obtained from the underlying grid-code specification of each ancillary service product. For instance, in the case of the FCR and FFR provision, the associated grid-code specifications are typically defined by some prescribed piece-wise linear time-domain curves (\cref{fig:fcr_ffr_grid_code}), where the required active power response should be satisfied at or above some minimum requirements. By applying our recent method in \cite{haberle2023gridcode} (which is based on Laplace transformations followed by Padé approximations of appropriate order), we can translate such piece-wise linear time-domain curves into the aforementioned rational parametric transfer functions $T_\mathrm{des}^\mathrm{fcr}(s,\alpha^\mathrm{fcr})$ and $T_\mathrm{des}^\mathrm{ffr}(s,\alpha^\mathrm{ffr})$ in the frequency domain, respectively (the explicit expressions are stated in \cref{eq:translated_parametric_tfs1,eq:translated_parametric_tfs2} in Appendix \ref{sec:AS_constraints_appendix}). The parameter vector $\alpha$ used in these transfer functions contains the parameters of the time-domain curves, which have to satisfy certain grid-code and device-level requirements. 

As an example, for the parameters $\alpha^\mathrm{fcr} := [t_\mathrm{i}^\mathrm{fcr},t_\mathrm{a}^\mathrm{fcr}]$ of the FCR curve in \cref{fig:fcr} (depicting a more comprehensive version of \cref{fig:fcr_intro}), we require \cite{european2016commission}
\begin{subequations}\label{eq:grid_code_req_fcr}
\begin{align}\label{eq:grid_code_req_fcr1}
    0\leq t_\mathrm{i}^\mathrm{fcr} &\leq t_\mathrm{i,max}^\mathrm{fcr}\\\label{eq:grid_code_req_fcr2}
    t_\mathrm{i}^\mathrm{fcr} \leq t_\mathrm{a}^\mathrm{fcr} &\leq t_\mathrm{a,max}^\mathrm{fcr}\\\label{eq:grid_code_req_fcr3}
    |\Delta p_\mathrm{fcr}^\mathrm{max}|&\leq \left(t_\mathrm{a}^\mathrm{fcr}-t_\mathrm{i}^\mathrm{fcr}\right)\cdot r_\mathrm{max}^\mathrm{p},
\end{align}
\end{subequations}
where $t_\mathrm{i,max}^\mathrm{fcr}$ and $t_\mathrm{a,max}^\mathrm{fcr}$ are the maximum admissible FCR initial delay and activation times, $|\Delta p_\mathrm{fcr}^\mathrm{max}|=|\tfrac{1}{D_\mathrm{p}}\Delta f_\mathrm{max}|$ is the maximum FCR capacity, and $r_\mathrm{max}^\mathrm{p}$ is the maximal active power ramping rate of the reserve unit, as specified in \cref{tab:grid_code_device_level_parameters} and graphically illustrated by the minimum curve requirement (red) in \cref{fig:fcr}. Hence, as indicated by the family of light gray curves in \cref{fig:fcr}, the requirements in \eqref{eq:grid_code_req_fcr} establish a set of different feasible response behaviors for the reserve unit. For a particular feasible $\alpha^\mathrm{fcr}$ as for the bold gray curve in \cref{fig:fcr}, the unit step response of the associated translated transfer function $T_\mathrm{des}^\mathrm{fcr}(s,\alpha^\mathrm{fcr})$ is exemplarily indicated by the black dashed line. Finally, the requirements \eqref{eq:grid_code_req_fcr} can be divided into grid-code and device-level specifications as 
\begin{align}\label{eq:grid_code_req_fcr_compact}
    \alpha^\mathrm{fcr} \in \mathcal{G}^\mathrm{fcr}\cap \mathcal{D}^\mathrm{fcr},
\end{align}
where the grid-code specification set  $\mathcal{G}^\mathrm{fcr}$ is defined by the constraints in \cref{eq:grid_code_req_fcr1,eq:grid_code_req_fcr2}, 
and the  device-level limitation set $\mathcal{D}^\mathrm{fcr}$ by the constraint in \eqref{eq:grid_code_req_fcr3}, respectively. 
\renewcommand{\arraystretch}{1.2}
\begin{table}[t!]\scriptsize
    \centering
     \setlength{\tabcolsep}{1mm}
           \caption{Grid-code and device-level specification examples.}
           \vspace{-1mm}
           \begin{subtable}{1\columnwidth}
               \caption{Grid-code specifications (exemplary values adopted from \cite{european2016commission,oyj2021technical,Eirgrid2018,de2021norma}). }
               \centering
               \vspace{-2mm}
               \begin{tabular}{c||c|c}
     \toprule
          Parameter & Symbol & Value  \\ \hline
           Maximum admissible initial delay time for FCR& $t_\mathrm{i,max}^\mathrm{fcr}$ & 2\,s \\
           Maximum admissible full activation time for FCR& $t_\mathrm{a,max}^\mathrm{fcr}$& 30\,s\\\hline
        Maximum admissible full activation time for FFR&$t_\mathrm{a,max}^\mathrm{ffr}$& 2\,s\\
        Minimum support duration time for FFR &$t_\mathrm{d,min}^\mathrm{ffr}$& 8\,s + $t_\mathrm{a}^\mathrm{ffr}$\\ 
        Minimum return-to-recovery time for FFR &$t_\mathrm{r,min}^\mathrm{ffr}$&10\,s + $t_\mathrm{d}^\mathrm{ffr}$\\
        Maximum overdelivery factor during FFR &$x_\mathrm{max}^\mathrm{ffr}$&1.35 \\ \hline
        Minimum frequency for oscillation damping&$\omega_\mathrm{min}$& $2\pi\cdot0.1\tfrac{\text{rad}}{s}$\\ 
        Maximum frequency for oscillation damping &$\omega_\mathrm{max}$& $2\pi\cdot3\tfrac{\text{rad}}{s}$ \\\hline
           Maximum admissible 90\% reactive power activation time &$t_\mathrm{90,max}^\mathrm{vq}$& 5\,s\\
           Maximum admissible 100\% reactive power activation time & $t_\mathrm{100,max}^\mathrm{vq}$& 60\,s\\ 
             \bottomrule
            \end{tabular}
            \label{tab:grid_code_specifications}
    \end{subtable}
        \begin{subtable}{1\columnwidth}
    \vspace{1.6mm}
    \caption{Device-level specifications (values depend on the reserve unit).}
    \centering
         \vspace{-2mm}
                \begin{tabular}{c||c}
     \toprule
          Parameter & Symbol \\ \hline
         Maximum active power ramping rate of reserve unit &$r_\mathrm{max}^\mathrm{p}=R_\mathrm{max}^\mathrm{p}\Delta f_\mathrm{max}$ \\
        Maximum reactive power ramping rate of reserve unit &$r_\mathrm{max}^\mathrm{q}=R_\mathrm{max}^\mathrm{q}\Delta v_\mathrm{max}$  \\
         Maximum active power peak capacity of reserve unit&$m_\mathrm{max}^\mathrm{p} = M_\mathrm{max}^\mathrm{p}\Delta f_\mathrm{max}$\\
           Maximum FFR support duration time of reserve unit &$t_\mathrm{d,max}^\mathrm{ffr}$\\
            Maximum FFR return-to-recovery time of reserve unit &$t_\mathrm{r,max}^\mathrm{ffr}$\\
            %Maximum reactive power ramping rate of the reserve unit &$r_\mathrm{max}^\mathrm{q}$\\
     \bottomrule
    \end{tabular}
    \label{tab:device_level_requirements}
        \end{subtable}
    \vspace{-3mm}
     \label{tab:grid_code_device_level_parameters}
\end{table}
\renewcommand{\arraystretch}{1} \normalsize

\begin{figure}[t!]
\centering
\begin{subfigure}{0.235\textwidth}
    \centering
    \vspace{-1mm}
     \resizebox{1.05\textwidth}{!}{
\begin{tikzpicture}
\draw[-latex] (-2.8,0.9) -- (-2.8,3.1); 
\draw [-latex](-2.9,1) -- (1.1,1);
\node [scale=0.9]at (-3.2,3) {$|\Delta p|$};
\node [scale=0.9]at (1.1,0.8) {$t$};
\draw[backgroundcolor_red] (-2.8,1) node (v1) {}--(-2.4,1) -- (-0.1,2.3) -- (0.9,2.3) node (v2) {};
\draw[dotted] (-1.3,2.3) -- (-1.3,1);
\node [scale=0.9] at (-1.15,0.75) {$t_\mathrm{a}^\mathrm{fcr}$};

\draw[dotted,gray](-0.1,2.3)  -- (-2.8,2.3);
\node [scale=0.9,gray] at (-3.35,2.3) {$|\Delta p_\mathrm{fcr}|$};
%\node [scale=0.7,black!60] at (-3.3,1.88) {capacity};

\node [scale=0.9] at (-2.45,0.75) {$t_\mathrm{i}^\mathrm{fcr}$};

\draw[gray!40!white](-2.8,1) -- (-2.7,1) node (v3) {} -- (-1.8,2.3) -- (0.9,2.3);

\draw[gray!40!white](-2.8,1) -- (-2.75,1) node (v5) {}  -- (-2.2,2.3) -- (0.9,2.3); 
\draw[gray!40!white] (-2.8,1) -- (-2.5,1) -- (-0.8,2.3) -- (0.9,2.3);
\draw[gray!40!white] (-2.78,1) node (v4) {}-- (-2.6,2.3) -- (0.9,2.3) node (v7) {};

\draw[ultra thick,gray!80!white] (-2.8,1) -- (-2.6,1) node (v6) {} -- (-1.3,2.3) -- (0.9,2.3); 
\draw [-latex](-2.25,0.75) -- (-2,0.75); 
\draw [-latex](-2.7,0.75) -- (-2.95,0.75);
\draw [-latex](-1.4,0.75) -- (-1.65,0.75); 
\draw [-latex](-0.95,0.75) -- (-0.7,0.75);
\draw[thick,black,dashed] plot[smooth, tension=.7] coordinates {(-2.78,1) (-2.75,1)(-2.7,1)(-2.6,0.98) (-2.5,1.1) (-2.4,1.2) (-1.65,2.05) (-1,2.27) (0.05,2.3) (0.2,2.3) (0.5,2.3) (v7)};

\draw [ultra thick, gray!80!white](-0.8,3.2) -- (-0.5,3.2);
\node[scale=0.7] at (0,3.2){exact};
\draw [dashed,thick, black](-0.8,2.95) -- (-0.5,2.95);
\node[scale=0.7] at (0.35,2.95){$T_\mathrm{des}^\mathrm{fcr}(s,\alpha^\mathrm{fcr})$};
\draw [backgroundcolor_red](-0.8,2.7) -- (-0.5,2.7);
\node[scale=0.7] at (0.46,2.7){min. grid code};

\end{tikzpicture}
}
           \vspace{-9mm}
    \caption{Exemplary active power time-domain capability curve for FCR provision after a frequency step change \cite{european2016commission} (cf. \cref{fig:fcr_intro}): the reserve unit has to deliver a certain FCR capacity $|\Delta p_\mathrm{fcr}|$ in accordance with an initial delay time $t_\mathrm{i}^\mathrm{fcr}$ and a full activation time $t_\mathrm{a}^\mathrm{fcr}$, where the FCR capacity $|\Delta p_\mathrm{fcr}|$ is typically fixed by the allocated active power droop gain $D_\mathrm{p}$ and the amplitude of the frequency step input $\Delta f$, i.e., $|\Delta p_\mathrm{fcr}|=|\tfrac{1}{D_\mathrm{p}} \Delta f|$. The minimum grid-code curve requirements are exemplarily indicated in red.}
    \vspace{1.8mm}
    \label{fig:fcr}
\end{subfigure}
\hspace{1mm}
\begin{subfigure}{0.235\textwidth}
    \centering
     \vspace{-1mm}
     \resizebox{1.05\textwidth}{!}{
\begin{tikzpicture}
\draw[-latex] (2.3,0.9) -- (2.3,3.6); 
\draw [-latex](2.2,1) -- (6.2,1);
\node [scale=0.9]at (1.92,3.5) {$|\Delta p|$};
\node [scale=0.9]at (6.2,0.8) {$t$};
\draw[ultra thick,gray!80!white](2.3,1) node (v1) {} -- (2.6,2.9) --(3.9,2.3)-- (4.6,1) node (v5) {}--(6,1) node (v3) {};
\draw[dotted] (2.5,2.3) node (v2) {} -- (2.5,1);
\node [scale=0.9] at (4,0.75) {$t_\mathrm{d}^\mathrm{ffr}$};

\draw [dotted](3.9,2.3) -- (3.9,1); 
\draw [dotted,gray](3.9,2.3) -- (2.25,2.3);
\node  [scale=0.9,gray] at (1.775,2.2) {$|\Delta p_\mathrm{ffr}|$};

\node [scale=0.9] at (2.6,0.75) {$t_\mathrm{a}^\mathrm{ffr}$};

\draw [dotted,black](2.6,2.9) -- (2.3,2.9);

\node [scale=0.9] at (4.75,0.75) {$t_\mathrm{r}^\mathrm{ffr}$};

\node  [scale=0.7,black] at (1.55,2.9) {$|\Delta p_\mathrm{ffr}^\mathrm{peak}|$};
%\node [scale=0.7,black!60] at (1.8,2.18) {capacity};
\draw [gray!40!white](2.3,1) -- (2.7,2.3) -- (3.5,2.3) -- (4.2,1) --(6,1);

\draw [gray!40!white](2.3,1) -- (2.7,2.7) -- (3.6,2.3) -- (4.4,1) -- (6,1); 
\draw[gray!40!white](2.3,1) -- (2.5,3.2) -- (4.3,2.3) -- (5,1) -- (6,1); 
\draw [gray!40!white](2.3,1) node (v4) {}-- (2.4,3.4) -- (4.6,2.3) -- (5.5,1) -- (6,1);
\draw  [thick,black,dashed]plot[smooth, tension=.7] coordinates {(v4) (2.5,2.4) (2.55,2.95) (2.95,2.85) (3.3,2.6) (3.75,2.45) (4.05,2.1) (4.25,1.6) (4.6,1) (5,1) (5.5,1) (6,1)};
\draw[-latex] (2.8,0.75) -- (3.05,0.75); 
\draw[-latex] (2.35,0.75) -- (2.1,0.75); 
\draw [-latex] (3.75,0.75) -- (3.5,0.75); 
\draw [-latex] (4.2,0.75) -- (4.35,0.75); 
\draw[-latex]  (4.5,0.75) -- (4.35,0.75); 
\draw[-latex]  (4.95,0.75) -- (5.2,0.75);

\draw [ultra thick, gray!80!white](4.5,3.5) -- (4.8,3.5);
\node[scale=0.7] at (5.3,3.5){exact};
\draw [dashed,thick, black](4.5,3.25) -- (4.8,3.25);
\node[scale=0.7] at (5.65,3.25){$T_\mathrm{des}^\mathrm{ffr}(s,\alpha^\mathrm{ffr})$};
\draw [-latex](1.55,3.05) -- (1.55,3.3);
\draw[-latex] (1.55,2.7) -- (1.55,2.45);
\end{tikzpicture}
}
           \vspace{-9mm}
     \caption{Exemplary active power time-domain capability curve for FFR provision after a frequency step change \cite{oyj2021technical}: the reserve unit has to deliver the FFR capacity $|\Delta p_\mathrm{ffr}|$ after an activation time $t_\mathrm{a}^\mathrm{ffr}$, which has to remain activated until a particular support duration time $t_\mathrm{d}^\mathrm{ffr}$, before returning to recovery at time $t_\mathrm{r}^\mathrm{ffr}$. The overdelivery $|\Delta p_\mathrm{ffr}^\mathrm{peak}|$ is a multiple of the FFR capacity, i.e., $|\Delta p_\mathrm{ffr}^\mathrm{peak}|=x^\mathrm{ffr}|\Delta p_\mathrm{ffr}|$, where $|\Delta p_\mathrm{ffr}|$ is defined by the scaled amplitude of the frequency step input, i.e., $|\Delta p_\mathrm{ffr}|=|\tfrac{1}{K_\mathrm{p}} \Delta f|$.}
    \label{fig:ffr}
\end{subfigure}
\caption{Examples of piece-wise linear time-domain grid-code curves (simplified) and their approximation as rational parametric transfer functions.}
\label{fig:fcr_ffr_grid_code}
\vspace{-3mm}
\end{figure}

Likewise, also the parameters $\alpha^\mathrm{ffr}:=[t_\mathrm{a}^\mathrm{ffr},t_\mathrm{d}^\mathrm{ffr},t_\mathrm{r}^\mathrm{ffr},x^\mathrm{ffr}]$ of the FFR curve in \cref{fig:ffr} are subject to grid-code and device-level constraints similar to \eqref{eq:grid_code_req_fcr}, i.e., 
\begin{align}\label{eq:grid_code_req_ffr_compact}
     \alpha^\mathrm{ffr} \in \mathcal{G}^\mathrm{ffr}\cap \mathcal{D}^\mathrm{ffr},
\end{align}
where the grid-code and device-level constraint sets $\mathcal{G}^\mathrm{ffr}$ and $\mathcal{D}^\mathrm{ffr}$ encode several time and capacity bounds as listed in \cref{tab:grid_code_device_level_parameters}, which establish a feasible set of response behaviors as illustrated by the family of light gray curves in \cref{fig:ffr}. A detailed formulation of \cref{eq:grid_code_req_ffr_compact} similar to \eqref{eq:grid_code_req_fcr} can be found in the grid-code documents \cite{oyj2021technical,Eirgrid2018,modig2019technical} or more compactly in \cref{eq:grid_code_req_ffr} in Appendix \ref{sec:AS_constraints_appendix}.

\begin{figure}[t!]
\centering
\begin{subfigure}{0.235\textwidth}
    \centering
    \vspace{-1mm}
     \resizebox{1.05\textwidth}{!}{
\begin{tikzpicture}
\draw[-latex] (-2.8,0.9) -- (-2.8,3.1); 
\draw [-latex](-2.9,1) -- (0.7,1);
\node [scale=0.9] at (-3.35,3) {$|T_\mathrm{des}^\mathrm{aux}|$};
\node [scale=0.9]at (0.7,0.8) {$\omega$};

\draw  [gray!40!white]plot[smooth, tension=.7] coordinates {(-2.8,1) (-2.1,1.2) (-1.2,2.7) (-0.3,1.2) (0.4,1)};
\draw  [gray!40!white]plot[smooth, tension=.7] coordinates {(-2.8,1) (-1.7,1.2) (-1.2,2.3) (-0.7,1.2) (0.4,1)};
\draw  [gray!40!white]plot[smooth, tension=.7] coordinates {(-2.8,1) (-1.5,1.2) (-1.2,2.55) (-0.9,1.2) (0.4,1)};
\draw  [gray!80!white,ultra thick]plot[smooth, tension=.7] coordinates {(-2.8,1) (-1.9,1.2) (-1.2,2.4) (-0.5,1.2) (0.4,1)};
\draw [dotted,black](-1.2,2.4) -- (-2.8,2.4); 
\draw[dotted] (-1.6,1.8) -- (-1.6,1); 
\draw[dotted] (-0.8,1.8) -- (-0.8,1);
\draw[dotted, black] (-0.8,1.8) -- (-2.8,1.8);
\draw [gray,>=latex, <->](-2.2,2.4) -- (-2.2,1.8);
\node[scale=0.6,gray] at (-2.5,2.1) {3dB};
\node [scale=0.9] at (-1.55,0.75) {$\omega_\mathrm{l}$};
\node at (-0.7,0.75) {$\omega_\mathrm{h}$};
\node[scale=0.9] at (-3.3,2.35) {$m_\mathrm{aux}$};
\draw [-latex](-1.4,0.75) -- (-1.125,0.75); 
\draw [-latex](-1.75,0.75) -- (-2,0.75); 
\draw [-latex](-0.95,0.75) -- (-1.175,0.75); 
\draw [-latex](-0.5,0.75) -- (-0.25,0.75);
\draw [-latex](-3.3,2.2) -- (-3.3,1.95);
\draw [-latex](-3.3,2.5) -- (-3.3,2.75);
\end{tikzpicture}
}
           \vspace{-9mm}
    \caption{Exemplary magnitude Bode plot of bandpass resonator transfer function to damp high-frequency dynamics and/or power oscillations: $\omega_\mathrm{r}=\sqrt{\omega_\mathrm{l}\omega_\mathrm{h}}$ is the resonance frequency, $\Delta \omega_\mathrm{BW}=\omega_\mathrm{h}-\omega_\mathrm{l}$ the filter bandwidth, and $m_\mathrm{aux}$ the magnitude. \color{white}with the times $|\Delta q_{90}|$ and $t_{100}^\mathrm{vq}$, respectively, where the reactive power capacity levels $|\Delta q_{90}|$ and $|\Delta q_{100}|$ are fixed by the $|\Delta q_{90}|$ reactive power droop gain. allocated $|\Delta q_{90}|$ power droop gain. by the$|\Delta q_{90}|$ $\tfrac{1}{D_\mathrm{q}}$ \color{black}}
    \label{fig:pod}
    \vspace{-1mm}
\end{subfigure}
\hspace{1mm}
\begin{subfigure}{0.235\textwidth}
    \centering
    \vspace{-1mm}
 \resizebox{1.05\textwidth}{!}{
\begin{tikzpicture}
\draw[-latex] (2.3,0.9) -- (2.3,3.1); 
\draw [-latex](2.2,1) -- (6.2,1);
\node [scale=0.9]at (1.9,3) {$|\Delta q|$};
\node [scale=0.9]at (6.2,0.8) {$t$};

\draw[dotted] (2.9,2.1) node (v2) {} -- (2.9,1);

\draw [dotted](4.2,2.4) -- (4.2,1); 
\draw [dotted,gray](4.2,2.4) -- (2.3,2.4);
\node  [scale=0.9,gray] at (1.72,2.45) {$|\Delta q_{100}|$};

\node [scale=0.9] at (2.9,0.75) {$t_\mathrm{90}^\mathrm{vq}$};

\node [scale=0.9]at (4.2,0.75) {$t_\mathrm{100}^\mathrm{vq}$};

\draw[dotted,gray](2.9,2.1)  -- (2.3,2.1);

\node   [scale=0.9,gray] at (1.78,2.05) {$|\Delta q_{90}|$};
\draw [gray!40!white](2.3,1) -- (3.3,2.1) -- (4.4,2.4) --  (6,2.4); 
\draw [gray!40!white](2.3,1)  -- (3.7,2.1) -- (4.8,2.4) -- (6,2.4); 
\draw[gray!40!white](2.3,1)  -- (2.7,2.1) -- (3.5,2.4) --  (6,2.4);
\draw [gray!40!white](2.3,1)  -- (2.5,2.1) -- (2.9,2.4) --  (6,2.4);
\draw [ultra thick, gray!80!white](2.3,1) node (v1) {} --  (2.9,2.1) -- (4.2,2.4) -- (6,2.4) node (v3) {} ;
\draw [ultra thick, gray!80!white](4.4,3) -- (4.7,3);
\node[scale=0.7] at (5.2,3){exact};
\draw [dashed,thick, black](4.4,2.75) -- (4.7,2.75);
\node[scale=0.7] at (5.55,2.75){$T_\mathrm{des}^\mathrm{vq}(s,\alpha^\mathrm{vq})$}; 

\draw[-latex] (3.15,0.75) -- (3.4,0.75); 
\draw [-latex](2.65,0.75) -- (2.4,0.75); 
\draw [-latex](4.45,0.75) -- (4.7,0.75); 
\draw [-latex](3.9,0.75) -- (3.65,0.75); 
\draw [dashed, thick] plot[smooth, tension=.7] coordinates {(v1) (2.7,1.85) (3.3,2.25) (4.05,2.4) (4.7,2.4) (v3)};
\end{tikzpicture}
}
 \vspace{-9mm}
    \caption{Exemplary reactive power time-domain capability curve for voltage control after a voltage step change \cite{european2016commission}: the reserve unit has to deliver the reactive power capacity levels $|\Delta q_{90}|$ of 90\% and $|\Delta q_{100}|$ of 100\% in accordance with the times $t_{90}^\mathrm{vq}$ and $t_{100}^\mathrm{vq}$, respectively, where the reactive power capacity levels $|\Delta q_{90}|$ and $|\Delta q_{100}|$ are typically fixed by the allocated reactive power droop gain $D_\mathrm{q}$ and the amplitude of the voltage step input $\Delta v$, i.e., $|\Delta q_\mathrm{100}|=|\tfrac{1}{D_\mathrm{q}} \Delta v|$.}
    \label{fig:voltage_ctrl}
    \vspace{-1mm}
\end{subfigure}
\caption{Examples of dynamic ancillary services products (simplified) encoded as rational parametric transfer functions.}
\label{fig:pod_volt_grid_codes}
\vspace{-4mm}
\end{figure}
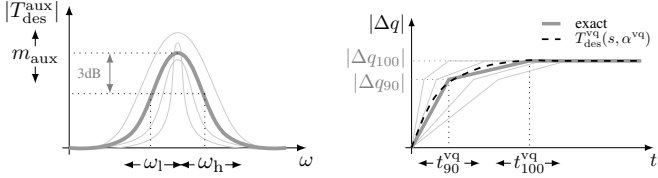

In contrast to the FCR and FFR grid-code specifications, the damping of high-frequency dynamics and/or power oscillations is usually stated less specifically in today's grid codes, despite its importance for grid-connected converters to provide damping services, similar to the power system stabilizer (PSS) in synchronous generators \cite{european2016commission}. Note that grid codes often only specify a certain range of frequencies $[\omega_\mathrm{min}, \omega_\mathrm{max}]$ in which the POD service has to be provided (e.g., Spain \cite{de2021norma}), without prescribing a particular response behavior for the reserve unit. In this regard, we define the parametric structure of $T_\mathrm{des}^\mathrm{aux}(s,\alpha^\mathrm{aux})$ in the form of a bandpass resonator transfer function as in \cref{fig:pod}, i.e.,
\begin{align}\label{eq:grid_code_req_pod_f}
    T_\mathrm{des}^\mathrm{aux}(s,\alpha^\mathrm{aux}) = m_\mathrm{aux}\cdot \tfrac{(\omega_\mathrm{h}-\omega_\mathrm{l})s}{s^2+(\omega_\mathrm{h}-\omega_\mathrm{l})s+\omega_\mathrm{l}\omega_\mathrm{h}},
\end{align}
which is parametric in $\alpha^\mathrm{aux}=[\omega_\mathrm{l},\omega_\mathrm{h},m_\mathrm{aux}]$. The auxiliary term $T_\mathrm{des}^\mathrm{aux}(s,\alpha^\mathrm{aux})$ provides additional flexibility for optimal dynamic ancillary services provision by introducing a lead-lag compensation in the frequency domain for stability enhancements and power oscillation damping (similar to PSS in synchronous generators) to supplement the (usually slower) FCR and FFR injections as indicated in \eqref{eq:T_des_fp_components}. While more general or higher-order formulations of a bandpass transfer function could be employed, we deliberately select the form in \eqref{eq:grid_code_req_pod_f} for its pragmatic structure, allowing for physical interpretation in terms of resonance frequency and filter bandwidth, thereby facilitating the formulation of parameter constraints on $\alpha^\mathrm{aux}$. Namely, $\alpha^\mathrm{aux}=[\omega_\mathrm{l},\omega_\mathrm{h},m_\mathrm{aux}]$ has to satisfy grid-code and device-level constraints as
\begin{align}\label{eq:grid_code_req_pod_compact}
    \alpha^\mathrm{aux} \in \mathcal{G}^\mathrm{aux} \cap \mathcal{D}^\mathrm{aux}.
\end{align}
In particular, the grid-code specification set $\mathcal{G}^\mathrm{aux}$ is defined by the frequency range $[\omega_\mathrm{min}, \omega_\mathrm{max}]$ to damp high-frequency dynamics and/or power oscillations, and the device-level limitation set $\mathcal{D}^\mathrm{aux}$ limits the resonance amplitude according to the reserve unit's maximum capacity (see \cref{tab:grid_code_device_level_parameters}), thereby establishing a feasible set of different resonator transfer functions as indicated by the light gray curves in \cref{fig:pod}. A detailed formulation of \eqref{eq:grid_code_req_pod_compact} can be found in \cref{eq:grid_code_req_pod} in Appendix \ref{sec:AS_constraints_appendix}.

Finally, by superimposing the previous transfer functions, we can establish the overall frequency control specification $T_\mathrm{des}^\mathrm{fp}(s,\alpha^\mathrm{fp})$ as in \cref{eq:T_des_fp_components}. In doing so, we further need to ensure that the maximum capacity and bandwidth limitations of the reserve unit are not violated during a \textit{superimposed} injection of active power. We encode these overarching constraints via an additional overall device-level constraint set $\mathcal{D}^\mathrm{fp}$ for $\mathrm{f}$-$\mathrm{p}$ control, which is specified in detail in \cref{eq:overall_fp_constraints} in Appendix \ref{sec:AS_constraints_appendix}.

\subsection{Voltage Regulation} \fontdimen2\font=0.6ex
In analogy to the frequency regulation, we in general also consider $T_\mathrm{des}^\mathrm{vq}(s,\alpha^\mathrm{vq})$ in \cref{eq:superimposed_AS_voltage} as a superposition of different voltage regulation products. The most common and often even 

\noindent only specified voltage control service in today's grid codes is the dynamic activation of reactive power proportionately to a voltage step change under particular time specifications \cite{european2016commission}. An example of the associated piece-wise linear grid-code curve is shown in \cref{fig:voltage_ctrl}, and the associated translated transfer function response (see \cref{eq:translated_parametric_tfs3} in Appendix \ref{sec:AS_constraints_appendix} for a detailed expression) is indicated via black dashed lines. Notice that the voltage control provision in \cref{fig:voltage_ctrl} might be specified differently in different grid codes. Moreover, in the same vein as for frequency regulation, one might also further populate the voltage regulation transfer function $T_\mathrm{des}^\mathrm{vq}(s,\alpha^\mathrm{vq})$ with some additional auxiliary term(s) for oscillation damping, etc.

The time-parameters $\alpha^\mathrm{vq}:=[t_{90}^\mathrm{vq},t_{100}^\mathrm{vq}]$ for the reactive power curve example in \cref{fig:voltage_ctrl} have to satisfy grid-code and device level requirements similar to \eqref{eq:grid_code_req_fcr} as
\begin{align}\label{eq:grid_code_req_vq_compact}
    \alpha^\mathrm{vq} \in \mathcal{G}^\mathrm{vq}\cap \mathcal{D}^\mathrm{vq},
\end{align}
where the grid-code and device-level constraint sets $\mathcal{G}^\mathrm{vq}$ and $\mathcal{D}^\mathrm{vq}$ encode several time constraints as listed in \cref{tab:grid_code_device_level_parameters}. A detailed formulation of \cref{eq:grid_code_req_vq_compact} similar to \eqref{eq:grid_code_req_fcr} can be found in the grid-code document\cite{european2016commission} and in \eqref{eq:vq_constraints} in Appendix \ref{sec:AS_constraints_appendix}.

Ultimately, by putting all ancillary service products together, we can establish the desired parametric transfer function matrix $T_\mathrm{des}(s,\alpha)$ with parameter vector $\alpha = [\alpha^\mathrm{fp},\alpha^\mathrm{vq}]\in\mathbb{R}^n$ in \cref{eq:Tdes}, where $\alpha^\mathrm{fp}:=[\alpha^\mathrm{fcr},\alpha^\mathrm{ffr},\alpha^\mathrm{aux}]$, to encode a set of feasible response behaviors for dynamic ancillary services provision. Out of the latter, we aim to find the optimal behavior $T_\mathrm{des}(s,\alpha^\star)$ as elaborated in the next section. It should be emphasized that our goal is to do more than cheaply satisfy the minimum open-loop grid-code requirements (e.g., by choosing $\alpha$ as the critical point where all grid-code constraints are active), with the ultimate goal being an optimal closed-loop performance regardless of the characteristics of the power grid.

 \begin{figure}[b!]
    \centering
        \vspace{-3mm}
    \resizebox {0.47\textwidth} {!} {
\begin{tikzpicture}[scale=1, every node/.style={scale=0.97}]]
\draw  [color=backgroundcolor,fill=backgroundcolor!20,rounded corners = 4,thick](2.5,-0.2) rectangle (7,-1.65);
\draw [color= gray!30!white,fill = gray!30!white,opacity=0.5] (-3.5,2.75) rectangle (2.1,-1.35);
\draw  [fill= black, rounded corners = 3](-1.7,2.5) rectangle (0.3,1.5);
\draw  [rounded corners = 3](-1.7,-0.1) rectangle (0.3,-1.1);
\node [scale = 1.2,color=white] at (-0.7,2) {${G}(s)$};
\node  [scale = 1.2] at (-0.7,-0.6) {$T_\mathrm{des}(s,\color{backgroundcolor}\alpha\color{black})$};
\draw [-latex](0.3,1.8) -- (1.8,1.8) -- (1.8,-0.6) -- (0.3,-0.6);
\draw[-latex] (-1.7,-0.6) -- (-3.2,-0.6) -- (-3.2,1.8) -- (-1.7,1.8);
\node at (-0.7,1.25) {grid equivalent};
\node at (-0.7,0.51) {ancillary services};
\node at (-0.7,0.19) {specification};
\draw[-latex] (0.3,2.2) -- (7.1,2.2); 
\draw[-latex] (-4.3,2.2) -- (-1.7,2.2);

\node at (-5.1,2.2) {input};
\node at (-5.1,1.9) {disturbance};
\node at (4.7,1.1) {performance output};
\node at (4.7,0.8) {to be minimized};
\node [color=gray] at (-0.7,3) {closed-loop dynamics $T_\mathrm{cl}(s,\alpha)$};
%\draw [fill=orange!20] (-5.9,-1.6) ellipse (1.7 and 1);
%\node at (-5.9,-1) {This method};
%\node at (-5.9,-1.4) {could also be used};
%\node at (-5.9,-1.8) {for individual};
%\node at (-5.9,-2.2) {power plants};
%\draw  [-latex]plot[smooth, tension=.7] coordinates {(2.5,-1.1) (3.4,-1.9) (4.5,-2.4)};

%\node at (3.9,-1.3) {optimization};
%\node at (4.1,-1.7) {problem};
\node [backgroundcolor] at (4.6,-0.5) {$\alpha^\star$\hspace{0.2mm}= argmin $\,\,||T_\mathrm{cl}(s,\alpha)||_2^2$};
\node [scale=0.9,backgroundcolor] at (3.9,-0.75) {$\alpha$};
\node[backgroundcolor] at (4.15,-1.05) {s.t.};
\node [backgroundcolor] at (5.75,-1.05) {grid code specs.};

\node [backgroundcolor] at (5.5,-1.4) {device limits};
\draw (4.7,2.2) -- (4.7,1.3); 
\draw [-latex](4.7,0.55) -- (4.7,-0.2);

\draw[-latex,dashed,backgroundcolor,thick] (2.5,-1.2) -- (0.1,-1.2) -- (-1.6,0.1);
\node [backgroundcolor] at (1.5,-0.95) {$\alpha^\star$};
\node at (-2.75,0.6) {$\begin{bmatrix}\Delta p\\\Delta q \end{bmatrix}$};
\node at (1.33,0.6) {$\begin{bmatrix}\Delta f\\\Delta v \end{bmatrix}$};
\end{tikzpicture}
}
    \vspace{-9mm}
    \caption{Closed-loop interconnection of the identified grid dynamic equivalent $G(s)$ with the rational parametric transfer function matrix $T_\mathrm{des}(s,\alpha)$ which is optimized for dynamic ancillary services provision.}
    \label{fig:abstract_setup}
     \vspace{-1mm}
\end{figure}
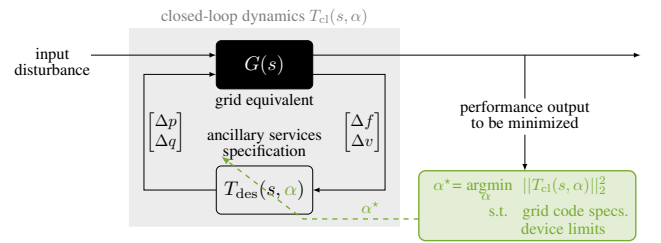

\section{\fontdimen2\font=0.5ex Optimal Dynamic Ancillary Services Provision}\label{sec:optimal_AS} \fontdimen2\font=0.6ex
We abstract the circuit topology in \cref{fig:grid_setup} and consider now a small-signal block diagram for the closed-loop interconnection of the grid dynamic equivalent ${G}(s)$ and the rational parametric transfer function matrix $T_\mathrm{des}(s,\alpha)$ for dynamic ancillary services provision as depicted in \cref{fig:abstract_setup}.

The $2\times 2$ transfer function matrix $T_\mathrm{des}(s,\alpha)$ in \cref{eq:Tdes} encodes a decoupled frequency and voltage control and relies on a grid-following signal causality, where the active and reactive power injection changes $\Delta p$ and $\Delta q$ are controlled as a function of the frequency and voltage magnitude measurements $\Delta f$ and $\Delta v$. In line with that, we approximate the power grid at the interconnection terminals of the reserve unit, i.e., the point of common coupling (PCC), by a $2\times 2$ small-signal dynamic equivalent $G(s)$, which describes the linearized power grid dynamics at the current steady-state operating point, and establishes the bus frequency and voltage magnitude deviations in response to the active and reactive power injections, i.e., 
\begin{align}\label{eq:G}
    \begin{bmatrix}
    \Delta f(s)\\ \Delta v(s) 
    \end{bmatrix} =
    \underset{=:G(s)}{\underbrace{\begin{bmatrix}
        G_\mathrm{11}(s)& G_\mathrm{12}(s)\\ G_\mathrm{21}(s)& G_\mathrm{22}(s)
    \end{bmatrix}}}
    \begin{bmatrix}
    \Delta p(s)\\ \Delta q(s) 
    \end{bmatrix}.
\end{align}
In particular, we assume that the grid operates at a fairly constant operating point, where the dynamics are nearly linear and time-invariant. In fact, during normal grid conditions, it is standard practice to build a small-signal model of the grid at a specific operating point to conduct analytical small-signal studies \cite{shair2021modeling} or design adaptive controllers \cite{shair2019mitigating}, which typically yields very accurate results. Moreover, since the dynamic model of the power system is typically not available in practice, we will use black-box identification methods to obtain the grid dynamic equivalent $G(s)$ in \eqref{eq:G} from data. The procedure is outlined in the following subsection.
 \begin{figure}[b!]
    \centering
        \vspace{-3mm}
\resizebox{0.45\textwidth}{!}{

\begin{tikzpicture}[scale=0.47,every node/.style={scale=0.47}]

\draw [rounded corners = 3](-1.9,6) rectangle (1.5,4.1);
\node at (-0.2,5.6) {Encode ancillary};
\node at (-0.2,5.2) {services as parametric};
\node at (-0.2,4.85) {transfer functions};

\node[scale=0.8] at (-0.2,4.4) {(\cref{sec:grid_code2tf})};
\draw  [color=backgroundcolor,fill=backgroundcolor!20,rounded corners = 3](-1.9,8.9) rectangle (1.5,6.4);
\node at (-0.2,8.5) {Grid dynamic};
\node at (-0.2,8.1) {equivalent};
\node at (-0.2,7.7) {identification};
\node [scale=0.8] at (-0.2,7.3) {(\cref{sec:grid_ID})};
\node [scale=0.9,color=backgroundcolor] at (-0.2,6.8) {\textbf{PERCEIVE}};
\draw [backgroundcolor,fill=backgroundcolor!20, rounded corners = 3] (2.7,8.9) rectangle (5.7,6.4);
\node at (4.2,8.5) {Closed-loop};
\node at (4.2,8.1) {power grid};
\node at (4.2,7.7) {optimization};
\node [scale=0.8] at (4.2,7.3) {(\cref{sec:optimization_formulation})};
\node [scale=0.9,color=backgroundcolor] at (4.2,6.8) {\textbf{OPTIMIZE}};
\draw  [rounded corners = 3](6.9,8.9) rectangle (11.1,6.4);
\draw  [backgroundcolor,fill=backgroundcolor!20,rounded corners=3](7.1,8.7) rectangle (10.9,7.8);
\draw  [rounded corners = 2](7.1,7.3) rectangle (10.9,6.8);
\draw[scale=1.2,color=backgroundcolor,fill=backgroundcolor] (1.1667,7) node (v1) {} -- (1.1667,6.8) -- (1.8999,6.8) -- (1.8999,6.6) -- (2.1999,6.9) -- (1.8999,7.2) -- (1.8999,7) --(1.1667,7);
\draw[scale=1.2,color=backgroundcolor,fill=backgroundcolor] (4.6333,7) node (v1) {} -- (4.6333,6.8) -- (5.3667,6.8) -- (5.3667,6.6) -- (5.6667,6.9) -- (5.3667,7.2) -- (5.3667,7) --(4.6333,7);
\node at (9,8.4) {Update converter};
\node at (9,8.1) {reference model};
\node at (9,7) {Matching control};
\node [scale=0.8] at (9,6.6) {(\cref{sec:converter_realization})};
\node [backgroundcolor]at (2.1,7.6) {$G(s)$};
\draw [dashed,-latex](1.5,5.1) -- (4.2,5.1) -- (4.2,6.4);
\draw[scale=1.2,color=backgroundcolor, fill=backgroundcolor] (7.4,7.15) node (v2) {} -- (7.6,7.15) -- (7.6,8.1) -- (-0.3,8.1) -- (-0.3,7.7) -- (-0.5,7.7) -- (-0.2,7.4) -- (0.1,7.7) -- (-0.1,7.7) -- (-0.1,7.9) -- (7.4,7.9) -- (7.4,7.1);
\draw [-latex](9,7.9) -- (9,7.2);
\node [scale=0.95] at (10,7.55) {$T_\mathrm{des}(s,\color{backgroundcolor}\alpha^\star\color{black})$};
\node [backgroundcolor]at (4.2,9.27) {repeat};
\node at (3.1,5.4) {$T_\mathrm{des}(s,\alpha)$};
\node [backgroundcolor]at (6.3,7.6) {$\alpha^\star$};
\end{tikzpicture}
}
    \vspace{-9mm}
    \caption{Flowchart of the proposed \textit{``perceive-and-optimize''} (P\&O) approach for optimal dynamic ancillary services provision based on power grid perception. The main steps are described in \cref{sec:grid_code2tf,sec:grid_ID,sec:optimization_formulation,sec:converter_realization}.}
    \label{fig:flowchart}
    \vspace{-1mm}
\end{figure}
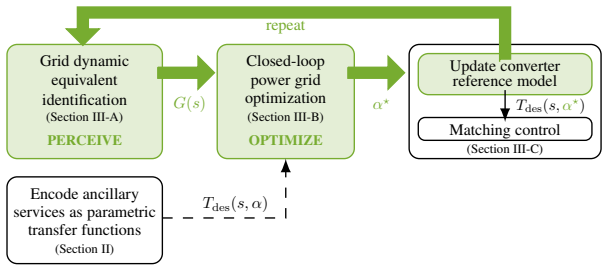

As the \textit{main contribution} of this paper, we aim to compute the optimal parameter vector $\alpha^\star$ of the desired transfer function matrix $T_\mathrm{des}(s,\alpha^\star)$ which ensures an optimal and stable closed-loop performance of the interconnection in \cref{fig:abstract_setup}, while satisfying (open-loop) grid-code and device-level requirements. Our approach is based on a so-called ``\textit{perceive-and-optimize}'' (P\&O) strategy, which is composed of two main steps (see flow-chart in \cref{fig:flowchart}) and elaborated in the following subsections, i.e.,
\begin{enumerate}
    \item \textit{``Perceive:''} We first use the converter-based reserve unit to identify a grid dynamic equivalent $G(s)$ at its interconnection terminals via black-box identification methods. 
    \item \textit{``Optimize:''} We establish a closed-loop system interconnection of the identified $G(s)$ and the parametric transfer matrix $T_\mathrm{des}(s,\alpha)$ as in \cref{fig:abstract_setup}, where we optimize for the vector of transfer function parameters $\alpha^\star$ which results in an optimal and stable closed-loop performance of the entire power grid response, while ensuring that grid-code and device-level requirements are reliably satisfied. 
\end{enumerate}
During \textit{one P\&O cycle}, we assume that grid conditions remain unchanged. Specifically, the grid conditions during the perception step are assumed to coincide with those at which the optimal $T_\mathrm{des}(s,\alpha^\star)$ is applied, ensuring that the identified grid dynamic equivalent $G(s)$ accurately reflects the prevailing grid dynamics (see \cref{sec:csI_nominal} for an example where this is not the case). In respect thereof, to address changing grid conditions, the two steps of the P\&O strategy should be repeated regularly, e.g., in a quarterly or hourly fashion when the grid operating point is changing according to the day-ahead market \cite{kirschen2018fundamentals}. Of course, also other (event-triggered or periodic) repetitions are possible. An investigation of such a multi-episodic application of the P\&O strategy will be part of future work. 

\begin{remark}
\fontdimen2\font=0.6ex Although the assumptions of having linear time-invariant grid dynamics and unchanging grid conditions during one P\&O cycle are idealized and rarely perfectly met in practice, it is important to notice that the feedback control structure of the P\&O strategy effectively linearizes and enhances the system's robustness in the face of model uncertainty.
\end{remark}

\subsection{Grid Dynamic Equivalent Identification}\label{sec:grid_ID} 
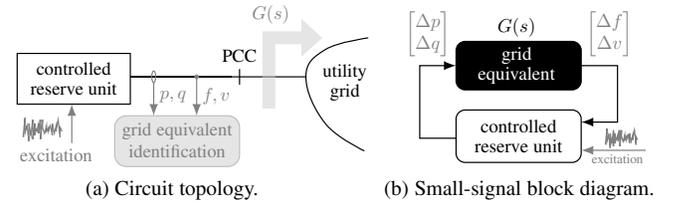
\begin{figure}[t!]
    \centering
\begin{subfigure}{0.255\textwidth}
\vspace{-1mm}
 \usetikzlibrary{circuits.ee.IEC}
\usetikzlibrary{arrows}

\resizebox {1.1\textwidth} {!} {
\tikzstyle{roundnode} =[circle, draw=blue!60, fill=blue!5, scale = 0.5]
\begin{tikzpicture}[circuit ee IEC,scale=0.47, every node/.style={scale=0.63}]

\node [color=gray] at (-0.6,5.6) {$G(s)$};

\draw plot[smooth, tension=.7] coordinates {(1.9,5) (0.8,4.7)(0.3,4)(0.8,2.8)(1.8,2.1)};

\node at (1.3,4.1) {utility};
\node at (1.3,3.6) {grid};

\draw (-1.6,4)  -- (-4.2,4) -- (0.3,4) node (v5) {};
\draw  (-4.2,4.6) rectangle (-7.1,3.3);
\node at (-5.65,4.2) {controlled};
\node at (-5.65,3.7) {reserve unit};

\draw[color=gray]  (-3.6,4) ellipse (0.05 and 0.15);
\draw [-latex,color =gray](-3.6,3.85) -- (-3.6,3);
\draw[-latex,color=gray] (-2.5,4) -- (-2.5,3);
\fill[color=gray] (-2.5,4) circle(0.6mm);
\draw [rounded corners = 4,color = gray, fill = gray!40!white,opacity=0.5] (-4.6,3) rectangle (-1.4,1.7);
\node [color = gray] at (-3,2.6) {grid equivalent};
\node [color =gray] at (-3,2.1) {identification};

\node [color=gray] at (-6.1,2) {excitation};

\draw[scale=0.5,color=gray] (-13.9,4.9) -- (-13.8,5.9) -- (-13.8,5.1) -- (-13.7,5.5) -- (-13.6,4.7) -- (-13.6,5.6) -- (-13.5,5) -- (-13.4,5.8) -- (-13.4,4.8) -- (-13.3,5.6) -- (-13.3,5) -- (-13.3,5.5) -- (-13.2,5.2) -- (-13.1,5.6) -- (-13.1,5.1) -- (-13,5.8) -- (-13,4.8) -- (-12.9,5.8) -- (-12.9,5.4) -- (-12.8,5.1) -- (-12.7,5.5) -- (-12.7,5.2) -- (-12.6,5.4) -- (-12.6,5.2) -- (-12.5,5.5) -- (-12.4,5) -- (-12.4,5.6) -- (-12.3,5.1) -- (-12.2,5.5) -- (-12.2,5.2) -- (-12.2,5.3) -- (-12.1,5.8) -- (-12,4.9) -- (-12,5.4) -- (-11.9,5.2);

\node [gray] at (-3.1,3.5) {$p,q$};
\node [gray] at (-2,3.5) {$f,v$};
\draw[fill=gray!40!white,color=gray!40!white,opacity=0.5] (-0.8,3.2) node (v2) {} -- (-0.8,5.15) -- (0.2,5.15) -- (0.2,5.45) -- (0.7,5) -- (0.2,4.55) -- (0.2,4.85) -- (-0.5,4.85) -- (-0.5,3.2) -- (-0.8,3.2);
\draw (-1.4,4.2) -- (-1.4,3.8);
\node at (-1.4,4.55) {PCC};
\draw [gray,-latex](-5.7,2.3) -- (-5.7,3.3);
\end{tikzpicture}
}
    \vspace{-9mm}
    \caption{Circuit topology.}
    \label{fig:gridID_scheme}
\end{subfigure}
\hspace{3mm}
\begin{subfigure}{0.2\textwidth}
    \centering
    \vspace{-1mm}
    \usetikzlibrary{circuits.ee.IEC}
\usetikzlibrary{arrows}

\resizebox {1\textwidth} {!} {
\tikzstyle{roundnode} =[circle, draw=blue!60, fill=blue!5, scale = 0.5]
\begin{tikzpicture}[circuit ee IEC,scale=0.47, every node/.style={scale=0.62}]

\draw [fill=black,rounded corners = 3] (3.8,4.9) rectangle (6.9,3.7);

\node at (5.3,5.25) {$G(s)$};

\draw [rounded corners = 3](6.9,3.2) rectangle (3.8,1.9);
\node at (5.35,2.8) {controlled};
\node at (5.35,2.3) {reserve unit};

\node [color=gray,scale=0.7] at (7.8,2) {excitation};

\draw[scale=0.4,color=gray] (18.75,5.85) -- (18.85,6.85) -- (18.85,6.05) -- (18.95,6.45) -- (19.05,5.65) -- (19.05,6.55) -- (19.15,5.95) -- (19.25,6.75) -- (19.25,5.75) -- (19.35,6.55) -- (19.35,5.95) -- (19.35,6.45) -- (19.45,6.15) -- (19.55,6.55) -- (19.55,6.05) -- (19.65,6.75) -- (19.65,5.75) -- (19.75,6.75) -- (19.75,6.35) -- (19.85,6.05) -- (19.95,6.45) -- (19.95,6.15) -- (20.05,6.35) -- (20.05,6.15) -- (20.15,6.45) -- (20.25,5.95) -- (20.25,6.55) -- (20.35,6.05) -- (20.45,6.45) -- (20.45,6.15) -- (20.45,6.25) -- (20.55,6.75) -- (20.65,5.85) -- (20.65,6.35) -- (20.75,6.15);

\draw [gray,-latex](8.5,2.2) -- (6.9,2.2);
\draw[-latex] (3.8,2.5) -- (2.9,2.5) -- (2.9,4.3) -- (3.8,4.3); 

\draw [-latex](6.9,4.3) -- (7.8,4.3) -- (7.8,2.9) -- (6.9,2.9);
\node [white] at (5.3,4.55) {grid};
\node [gray] at (3.1,5.1) {$\begin{bmatrix} \Delta p\\ \Delta q\end{bmatrix}$};
\node [gray] at (7.6,5.1) {$\begin{bmatrix} \Delta f\\ \Delta v\end{bmatrix}$};
\node [white]at (5.3,4.05) {equivalent};

\end{tikzpicture}
}
    \vspace{-9mm}
    \caption{Small-signal block diagram.}
    \label{fig:blockdiagram}
\end{subfigure}
\caption{Grid dynamic equivalent identification setup.}
\vspace{-3mm}
\label{fig:gridID}
\end{figure}

\fontdimen2\font=0.6ex
Since the power grid is ever-changing and usually unknown from the perspective of a generation system, the grid dynamic equivalent $G(s)$ in \cref{eq:G} has to be measured online for a real-time assessment of the grid dynamics. To do so, for the reserve unit in \cref{fig:grid_setup}, we consider a grid-equivalent identification setup as in \cref{fig:gridID_scheme}, and resort to parametric$^1$ black-box grid impedance measurement techniques, as proposed in our prior work \cite{haberle2023mimo}. More specifically, employing such a black-box identification strategy enables us to identify the complete grid dynamic equivalent solely through measurement data collected at the interconnection terminal of the reserve unit, i.e., the complex and unknown grid dynamics can be accommodated without tailoring the model to specific system characteristics.

To identify $G(s)$ in \cref{eq:G}, we inject uncorrelated wideband excitation signals with small perturbation levels (e.g., random binary sequences (RBS)) in the converter's control loop of the reserve unit, to locally excite the system during online operation as in \cref{fig:gridID_scheme} (see \cref{sec:case_studies} for details on the converter control setup). The resulting frequency and voltage magnitude responses, as well as the active and reactive power injections at the PCC, are then measured and collected as discrete-time samples to calculate an estimate of the grid dynamic equivalent $G(s)$. \cref{fig:blockdiagram} shows a block diagram of the identification problem, where the input/output perspective of the to-be-identified grid equivalent results from the electrical circuit equations.
  
Given the collected input/output data, we can apply parametric\footnote{\textit{Parametric} (i.e., model-based) system identification techniques directly identify an explicit system representation (e.g., a transfer function) and are not to be confused with \textit{parametric} (i.e., parameter-dependent) transfer functions.} system identification techniques such as prediction error (PEM) or subspace methods \cite{haberle2023mimo} to obtain an accurate estimate of the small-signal grid dynamic equivalent $G(s)$ (see \cref{sec:case_studies} for examples). Since PEM are typically simpler than subspace methods and can be computed very quickly, we consider them the preferred choice, especially for a multi-episodic application of the proposed P\&O strategy. PEM works by minimizing the prediction error between the observed output of the system and the output predicted by the model. More specifically, the goal is to find model parameters that best capture the system's dynamics, ensuring the model's predictions are as accurate as possible. One of the simplest model structures used for PEM is the so-called ARX (Auto-Regressive with eXogenous inputs) model. It represents a system's output as a linear combination of its previous outputs (auto-regressive part) and current and past inputs (exogenous part). The basic form of an ARX model is:
\begin{align}\label{eq:arx}
\begin{split}
    y(k)+a_1 &y(k-1)+\dots+a_n y(k-n)=\\
    &=b_1 u(k-1)+\dots+b_m u(k-m)+e(k).
    \end{split}
\end{align}
Here, $y(k)$ is the output at time step $k$, $u(k)$ is the input, $a_i$ and $b_j$ are the parameters to be estimated, and $e(k)$ is the prediction error. The simplicity of the ARX model lies in its straightforward linear relationship and ease of implementation, where the model parameters are estimated by solving a basic least-squares linear regression problem (for more details, see\cite{haberle2023mimo}).

Notice that in practical scenarios, measurement and process noise (e.g., external disturbances, inaccurate measurement devices) may be present during the grid dynamic equivalent identification. However, if the noise levels are not excessively high and a sufficient amount of data is collected, parametric system identification methods (especially PEM methods with a high-degree-of-freedom error model) can generally handle such noise and ensure that it only has minor impacts on the identification accuracy \cite{ljung2022system}. Larger noise levels, in turn, may necessitate special adjustments to the system identification methods\cite{ljung2022system,soderstrom2018errors}, which are out of the scope of this paper. For the practical setups in this work, however, the noise level is typically acceptable.
\begin{remark}
\fontdimen2\font=0.6ex One may alternatively choose to first identify the grid impedance in other coordinates and then transform to polar coordinates as in \eqref{eq:G} \cite{yang2019comparison,wang2017unified}.
\end{remark}
\begin{remark}
\fontdimen2\font=0.6ex If the converter system has a small capacity compared to the power grid, it can only locally identify a ``partial'' grid equivalent in its vicinity, while the rest of the grid appears as an infinite bus. If so, the converter can only affect the grid dynamics during ancillary services provision in this vicinity.
\end{remark}

\subsection{Closed-Loop Power Grid Optimization}\label{sec:optimization_formulation} \fontdimen2\font=0.6ex
After identifying the grid dynamic equivalent $G(s)$, we now compute the optimal parameter vector $\alpha^\star$ of $T_\mathrm{des}(s,\alpha^\star)$ which ensures an optimal and stable closed-loop performance of the interconnected system in \cref{fig:abstract_setup}. 

\begin{figure}[t!]
    \centering
     \resizebox {0.5\textwidth} {!} {
\begin{tikzpicture}[scale=1, every node/.style={scale=0.97}]]
\draw [color=gray!30!white,fill=gray!30!white, opacity=0.5] (-6.4,-2.6) rectangle (3.7,4.3);
\node at (-7.4,2.7) {input};
\node at (-7.4,2.4) {disturbance};
\node at (4.7,2.7) {performance};
\node at (-5.9,0.2) {$\begin{bmatrix}\Delta p\\\Delta q \end{bmatrix}$};
\node at (3.2,0.1) {$\begin{bmatrix}\Delta f\\\Delta v \end{bmatrix}$};

\node [color=gray] at (-1.3,4.65) {closed-loop dynamics $\Tilde{T}_\mathrm{cl}(s,\color{backgroundcolor}\alpha\color{black})$};
\node at (-1.3,-1) {$\dot{x}(t)=A(\color{backgroundcolor}{\alpha}\color{black})x(t)+B(\color{backgroundcolor}\alpha\color{black})\begin{bmatrix} \Delta f(t)\\ \Delta v(t)\end{bmatrix}$};
\node at (-2.4,-1.8) {$\begin{bmatrix}\Delta p(t)\\ \Delta q(t) \end{bmatrix}=C(\color{backgroundcolor}\alpha\color{black})x(t)$};
\draw  [rounded corners = 3](1.3,-0.4) rectangle (-4,-2.4);
\draw [fill=black, rounded corners = 3] (2,3.6) rectangle (-4.7,1.6);
\node at (-1.3,-0.2) {ancillary services $T_\mathrm{des}(s,\color{backgroundcolor}\alpha\color{black})$};
\node [color=white] at (-1.3,3) {$\dot{x}_\mathrm{g}(t)=A_\mathrm{g}x_\mathrm{g}(t)+B_\mathrm{g}\begin{bmatrix}\Delta p(t)\\ \Delta q(t) \end{bmatrix}+B_\mathrm{g}\begin{bmatrix} p_\mathrm{d}(t)\\q_\mathrm{d}(t)\end{bmatrix}$};
\node [color=white] at (-3.2,2.2) {$\begin{bmatrix} \Delta f(t)\\ \Delta v(t)\end{bmatrix}=C_\mathrm{g}x_\mathrm{g}(t)$};

\node at (-1.3,3.8) {grid equivalent $G(s)$};
%\node at (-1.45,1.1) {$z_\mathrm{p}(t)=\begin{bmatrix}r_\mathrm{\dot{f}}^\frac{1}{2}&0&0\\0& r_\mathrm{f}^\frac{1}{2}& 0\\ 0&0&r_\mathrm{v}^\frac{1}{2} \end{bmatrix}\begin{bmatrix} \Delta \dot{f}(t)\\ \Delta f(t) \\ \Delta v(t)\end{bmatrix}$};
\draw  [rounded corners = 3](-4.9,4.1) rectangle (2.2,0.6);
\draw[-latex] (-4,-1.4) -- (-5.4,-1.4) -- (-5.4,1.2) -- (-4.9,1.2);

\draw [-latex](2.2,1.2)--(2.7,1.2)-- (2.7,-1.4)--(1.3,-1.4); 
\draw[-latex] (-8.4,3) -- (-4.9,3);
\draw[-latex] (2.2,3) -- (5.8,3);
\node at (-7.4,3.6) {$w:=\begin{bmatrix}p_\mathrm{d}\\q_\mathrm{d} \end{bmatrix}$};

\node at (4.7,2.4) {output};
\node at (4.7,3.3) {$\Tilde{z}_\mathrm{p}$};
\node at (-3.1,1.15) {$\Tilde{z}_\mathrm{p}(t)=R^{\tfrac{1}{2}}\Tilde{E}_\mathrm{g}x_\mathrm{g}(t),$};

\node at (0.5,1.15) {$R=\text{diag}(r_{\dot{\mathrm{f}}}, r_\mathrm{f}, r_\mathrm{v})$};
\end{tikzpicture}}
    \vspace{-4mm}
    \caption{Closed-loop interconnection for $\mathcal{H}_2$ optimization. The state vector $x_\mathrm{g}$ results from the extended state-space representation in \cref{eq:extended_ss_G} of the grid dynamic equivalent, which additionally outputs the approximate integral of $\Delta f$ and $\Delta v$.}
    \label{fig:optimizaiton_problem}
        \vspace{-3mm}
\end{figure}
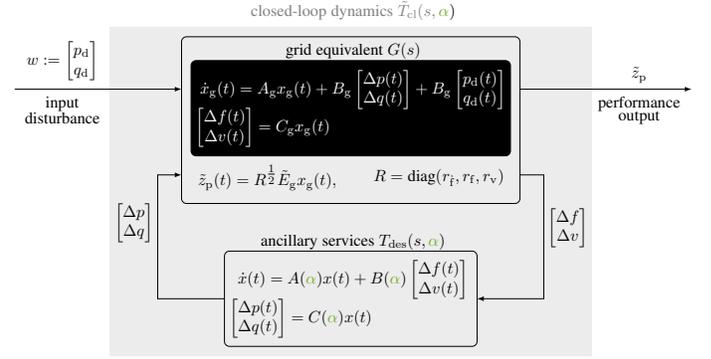

The $\alpha^\star$ computation for the closed-loop system in \cref{fig:abstract_setup} can be recast as a system norm (input-output gain) minimization problem as in \cref{fig:optimizaiton_problem}, where we select a suitable performance index to be minimized. We translate both the grid dynamic equivalent $G(s)$ and the parametric transfer function matrix $T_\mathrm{des}(s,\alpha)$ into a state-space system representation, where the control input of the former is given by the outputs of the latter (i.e., the active and reactive power deviations  $\Delta p$ and $\Delta q$). Moreover, to specify the closed-loop response behavior, we define a weighted performance output $z_\mathrm{p}$ that has to be minimized when subject to some active and reactive power disturbance $ w:=[p_\mathrm{d},q_\mathrm{d}]^\top$. Namely, we consider $z_\mathrm{p}$ to be composed of the frequency deviation $\Delta f$, the rate-of-change-of frequency (RoCoF) $\Delta \dot{f}$, and the voltage deviation $\Delta v$, i.e., 
    \begin{align}\label{eq:performance_output}
        \hspace{-1mm}z_\mathrm{p} = \left(r_{\dot{\mathrm{f}}}^{\tfrac{1}{2}} \Delta \dot{f},r^{\tfrac{1}{2}}_\mathrm{f} \Delta f, r^{\tfrac{1}{2}}_\mathrm{v} \Delta v\right)^\top,
    \end{align}
where $r_{\dot{\mathrm{f}}}, r_\mathrm{f}, r_\mathrm{v}$ are non-negative scalars trading off the relative deviations. If desired, further quantities of the closed-loop interconnection can be added to the performance output $z_\mathrm{p}$, e.g., the control efforts $\Delta p$ and $\Delta q$. Since the latter, however, are already implicitly limited by the structural constraints on the parameter $\alpha$, we refrain from including them into $z_\mathrm{p}$.

There exist different system norms (e.g., $\mathcal{H}_2$ or $\mathcal{H}_\infty$) which provide a measure of the magnitude of the closed-loop system output $z_\mathrm{p}$ in response to the disturbance input $w$. In this work, we quantify the closed-loop system performance in terms of the $\mathcal{H}_2$ norm, which results in a computationally tractable and well-understood design problem, even for complex and large systems. Moreover, the $\mathcal{H}_2$ norm is generally considered a suitable proxy for typical power system specifications \cite{gross2017increasing,poolla2019placement,sanchez2019optimal,ademola2018optimal}. In particular, given that the $\mathcal{H}_2$ norm measures the energy of the system response to impulse disturbance inputs, we can replicate the closed-loop power system response to classical step disturbances, e.g., a sudden load increase or a generation drop, by considering the integral\footnote{We \textit{approximate} the integral of $z_\mathrm{p}=T_\mathrm{cl}(s,\alpha)w$ as $\Tilde{z}_\mathrm{p}=\Tilde{T}_\mathrm{cl}(s,\alpha)w = \tfrac{1}{s+\epsilon}T_\mathrm{cl}(s,\alpha)w$, where $\epsilon$ is a small approximation factor of an ideal integrator, to ensure stability of $\Tilde{T}_\mathrm{cl}(s,\alpha)$ and the $\mathcal{H}_2$ norm to be well-defined \cite{zhou1996robust}.} of the performance output in \cref{eq:performance_output} as $\Tilde{z}_\mathrm{p}\approx\textstyle\int z_\mathrm{p}$ which, in the $\mathcal{H}_2$ norm, now reflects the system energy imbalance for \textit{step-like} disturbances. Consequently, we define the cost function as
\begin{align}\label{eq:cost_function}
        J = \textstyle\int_0^\infty \tilde{z}_\mathrm{p}^\top \tilde{z}_\mathrm{p},
    \end{align}
which corresponds to the squared $\mathcal{H}_2$ norm of the underlying closed-loop dynamical system $\Tilde{z}_\mathrm{p}=\Tilde{T}_\mathrm{cl}(s,\alpha)w$ between the disturbance input $w$ and the performance output $\Tilde{z}_\mathrm{p}$ as in \cref{fig:optimizaiton_problem}.

The closed-loop dynamical system $\Tilde{z}_\mathrm{p}=\Tilde{T}_\mathrm{cl}(s,\alpha)w$ can be obtained by closing the loop between the two state-space systems of $G(s)$ and $T_\mathrm{des}(s,\alpha)$. More specifically, we consider the state-space representation\footnote{$T_\mathrm{des}(s,\alpha)$ and $G(s)$ can be ensured to be strictly proper by selecting the transfer function structure during grid-code translation \& system identification.} of $T_\mathrm{des}(s,\alpha)$ in \cref{eq:Tdes} as
\begin{align}
   T_\mathrm{des}(s,\alpha)=C(\alpha)(sI-A(\alpha))^{-1}B(\alpha),
\end{align}
where the associated differential equations are given as
\begin{align}
\begin{split}
\dot{x}(t)&=A(\alpha)x(t)+B(\alpha)\begin{bmatrix}
    \Delta f(t)\\ \Delta v(t)
\end{bmatrix}\\
\begin{bmatrix}
    \Delta p(t)\\ \Delta q(t)
\end{bmatrix} &= C(\alpha)x(t).
\end{split}
\end{align}
To output the approximate integral $\Tilde{z}_\mathrm{p}\approx \int z_\mathrm{p}$ of the performance output in \cref{eq:performance_output}, we consider an extended transfer function of $G(s)$ in \cref{eq:G} under the disturbance input $w=[p_\mathrm{d},q_\mathrm{d}]^\top$, i.e.,
\begin{align}
    \begin{split}
        \begin{bmatrix}
         \begin{array}{c}
            \Delta f(s)\\ \Delta v(s)\\ \hline \int \Delta f(s) \\ \int \Delta v(s)
             \end{array}
        \end{bmatrix} \approx \begin{bmatrix}
        \begin{array}{c}
            G(s)\\ \hline
            \tfrac{1}{s+\epsilon}G(s)
            \end{array}
        \end{bmatrix}
        \begin{bmatrix}
            \Delta p + p_\mathrm{d}\\
            \Delta q + q_\mathrm{d}
        \end{bmatrix},
    \end{split}
\end{align}
where the associated state-space representation$^{3}$ is given as
\begin{align}\label{eq:extended_ss_G}
\begin{split}
    \dot{x}_\mathrm{g}(t) &= A_\mathrm{g}x_\mathrm{g}(t)+B_\mathrm{g}\begin{bmatrix}
         \Delta p + p_\mathrm{d}\\
            \Delta q + q_\mathrm{d}
    \end{bmatrix}\\
    \begin{bmatrix}
       \Delta f(s)\\ \Delta v(s)\\ \int \Delta f(s) \\ \int \Delta v(s)
    \end{bmatrix} &\approx \begin{bmatrix}
        C_\mathrm{g}\\ C_\mathrm{g}'
    \end{bmatrix}x_\mathrm{g}(t),
    \end{split}
\end{align}
with the grid dynamic equivalent in \cref{eq:G} as
\begin{align}
G(s) = C_\mathrm{g}(sI-A_\mathrm{g})^{-1}B_\mathrm{g}.
\end{align}

Finally, we can formulate the state-space representation of the resulting closed-loop dynamical system $\Tilde{z}_\mathrm{p}=\Tilde{T}_\mathrm{cl}(s,\alpha)w$ in  \cref{fig:optimizaiton_problem}, i.e.,
    \begin{align}\label{eq:closed_loop_ss}
    \begin{split}
       \underset{\dot{x}_\mathrm{cl}}{\underbrace{\begin{bmatrix} \dot{x}_\mathrm{g} \\ \dot{x}\end{bmatrix}}} &= \underset{=:A_\mathrm{cl}(\alpha)}{\underbrace{\begin{bmatrix} A_\mathrm{g} & B_\mathrm{g}C(\alpha) \\ B(\alpha)C_\mathrm{g} & A(\alpha)\end{bmatrix}}} \underset{=:{x}_\mathrm{cl}}{\underbrace{\begin{bmatrix} x_\mathrm{g} \\ x\end{bmatrix}}} + \underset{=:B_\mathrm{cl}(\alpha)}{\underbrace{\begin{bmatrix} B_\mathrm{g} \\ O\end{bmatrix}}}w\\
        \Tilde{z}_\mathrm{p} &= \underset{=:C_\mathrm{cl}(\alpha)}{\underbrace{\begin{bmatrix}R^{\tfrac{1}{2}}\Tilde{E}_\mathrm{g}& O \end{bmatrix}}}\underset{{x}_\mathrm{cl}}{\underbrace{\begin{bmatrix} x_\mathrm{g} \\ x\end{bmatrix}}},
        \end{split}
    \end{align}
    where $R := {\rm diag} (r_{\dot{\mathrm{f}}}, r_\mathrm{f}, r_\mathrm{v})$ and the matrix $\Tilde{E}_\mathrm{g}$ results from the extended state-space representation in \cref{eq:extended_ss_G}, which outputs $\Delta f$ and the approximate integral of $\Delta f$ and $\Delta v$, i.e.,
    \begin{align}
        \Tilde{E}_\mathrm{g} = \begin{bmatrix}
            C_\mathrm{g}[1,:]\\C_\mathrm{g}'
        \end{bmatrix}.
    \end{align}

Finally, by recalling the constraints on $\alpha$ in \cref{eq:grid_code_req_fcr_compact,eq:grid_code_req_ffr_compact,eq:grid_code_req_pod_compact,eq:grid_code_req_vq_compact}, we can state the optimization problem to be solved for $\alpha^\star$ as     
    \begin{align}\label{eq:optimization_problem_ss}
    \begin{split}
        \underset{\alpha}{\text{minimize}} \quad &J\\
        \text{subject to} \quad &\dot{x}_\mathrm{cl} = A_\mathrm{cl}(\alpha)x_\mathrm{cl}+B_\mathrm{cl}(\alpha)w\\
        &\Tilde{z}_\mathrm{p} = C_\mathrm{cl}(\alpha)x_\mathrm{cl}\\
        & \alpha \in \mathcal{G} \cap \mathcal{D}, 
        \end{split}
    \end{align}
    where $\mathcal{G}:=\mathcal{G}^\mathrm{fcr}\times \mathcal{G}^\mathrm{ffr}\times\mathcal{G}^\mathrm{aux}\times \mathcal{G}^\mathrm{vq}$ and $\mathcal{D}:=[[\mathcal{D}^\mathrm{fcr}\times \mathcal{D}^\mathrm{ffr}\times\mathcal{D}^\mathrm{aux}]\cap \mathcal{D}^\mathrm{fp}]\times\mathcal{D}^\mathrm{vq}$ (with $\times$ denoting the Cartesian product of sets).
Due to the parametric nature of $A_\mathrm{cl}(\alpha),B_\mathrm{cl}(\alpha)$ and $C_\mathrm{cl}(\alpha)$, the problem in \cref{eq:optimization_problem_ss} is generally non-convex. However, since the objective function is smooth, an explicit gradient of $J$ can be derived and directly used to solve the optimization problem via scalable first-order methods (e.g., projected gradient descent), similar as in \cite{gross2017increasing,poolla2019placement,sanchez2019optimal,ademola2018optimal}. A detailed computational approach to derive an expression for the gradient of $J$ is presented in Appendix B.

\subsection{Realization in Converter Control Architecture}\label{sec:converter_realization}
The obtained desired transfer function matrix $T_\mathrm{des}(s,\alpha^\star)$ defines a frequency and voltage control behavior in the frequency domain which can be realized in a converter-based generation system. More specifically, $T_\mathrm{des}(s,\alpha^\star)$ can be incorporated as a \textit{reference model} into conventional cascaded converter control architectures which are designed for reference tracking, and thus enable a simple matching control implementation, e.g., via classical PI controllers (see \cref{sec:converter_system} for a detailed implementation). Moreover, such an implementation can handle changing reference models $T_\mathrm{des}(s,\alpha^\star)$ for different grid conditions with sufficient accuracy, such that a re-tuning of the PI gains is only needed in exceptional cases. Nevertheless, if one desires more advanced and robust matching control implementations, one could alternatively resort to multivariable linear parameter varying (LPV) $\mathcal{H}_\infty$ methods \cite{haberle2021control,haberle2022control}, which can robustly include the time-varying parameter vector $\alpha$, and hence directly account for time-varying grid conditions without re-computing the matching controller. However, the high-dimensionality and non-linear dependency of the parameters in $T_\mathrm{des}(s,\alpha^\star)$ might lack practical realizability of LPV methods.  

Beyond that, instead of realizing $T_\mathrm{des}(s,\alpha^\star)$ with one single reserve unit, one might alternatively consider a heterogeneous collection of multiple energy sources (i.e., a so-called dynamic virtual power plant (DVPP)) to provide $T_\mathrm{des}(s,\alpha^\star)$ in aggregation \cite{haberle2021control,haberle2022control}. This offers the advantage of providing dynamic ancillary services via $T_\mathrm{des}(s,\alpha^\star)$ across higher energy and power levels and a wider range of time-scales, even during time-varying conditions of individual energy sources.

\subsection{Multi-Converter Scenarios}\label{sec:multi_converter}
So far, the proposed P\&O strategy has been presented for a single reserve unit, which is interconnected with the grid. However, an immediate question is how to apply the P\&O strategy in the case of multiple reserve units. Given the novelty and underexplored nature of the P\&O strategy, a straightforward approach is to study a scenario where each reserve unit simply applies the P\&O strategy \textit{sequentially} (and repeatedly) one after another, while the remaining units keep using their previous $T_{\mathrm{des},i}(s,\alpha^\star)$. Indeed, as demonstrated later in our case studies in \cref{sec:case_studies}, such a sequential application of the P\&O strategy ensures closed-loop stability of each converter system with the remaining grid at every P\&O cycle, meanwhile being independent of the number of other converter systems and their locations in the interconnected grid from the local perspective of each reserve unit.

Alternatively, a more intricate but conceivable approach would be the simultaneous application of the P\&O strategy to multiple converter systems, which, however, presents various unexplored challenges, such as suboptimality, gaming behavior, stability, and undesired mutual interference of the converter systems. Therefore, we advocate for further investigation into these complexities in future research endeavors.

\section{Case Studies}\label{sec:case_studies} 
\begin{figure}[b!]
    \centering
        \vspace{-3mm}
     \usetikzlibrary{arrows}
\resizebox {1\columnwidth} {!} {
\begin{tikzpicture}[scale=0.39, every node/.style={scale=0.62}]

	\draw(-14.55,9.3) -- (-8.6,9.3);
	\draw [ultra thick](-11.1,10.1) -- (-11.1,8.5);
	\draw [ultra thick](-16.4,10.1) -- (-16.4,8.5);
		\draw [ultra thick](-8.6,10.1) -- (-8.6,8.5);
	\draw [ultra thick](-3.6,10.1) -- (-3.6,8.5);
		\draw [ultra thick](-1.1,12.6) -- (-1.1,11);
	\draw [ultra thick](1.6,7.9) -- (1.6,6.3);
			\draw [ultra thick](-13.8,10.1) -- (-13.8,8.5);
	\draw [ultra thick](1.6,10.1) -- (1.6,8.5);
		\draw [ultra thick](-16.4,12.6) -- (-16.4,11);
		\draw [ultra thick](-1.1,10.1) -- (-1.1,8.5);

	\draw [ultra thick](4.2,10.1) -- (4.2,8.5);
		\draw [ultra thick](-6.1,10.1) -- (-6.1,8.5);

	\node at (-16.4,8) {1};
	\node at (-8.6,10.6) {7};
	\node at (-11.1,10.6) {6};
	\node at (1.6,10.6) {11};
	\node at (4.2,10.6) {3};
	\node at (-3.6,10.6) {9};
	\node at (-6.1,10.6) {8};

	\node at (-13.8,5.8) {2};

	%\draw [-latex] (-4.5,4) rectangle (-2.5,2.5);
	\draw[fill=black!20](-17.9,9.3) node (v3) {} circle (7 mm); 
	\draw(-2.2,11.8)  circle (3.5 mm); 
	\draw(-1.8,11.8)  circle (3.5 mm); 
		\draw(-15.3,9.3)  circle (3.5 mm); 
	\draw(-14.9,9.3)  circle (3.5 mm); 
	\draw  plot[smooth, tension=.7] coordinates {(-18.4,9.3) (-18.187,9.8) (v3) (-17.66,8.8) (-17.4,9.3)};
	
	%\node at (-3.5,3.25) {DVPP};

	\draw(-12.6,7.1)  circle (3.5 mm); 
	\draw(-13,7.1)  circle (3.5 mm); 
		\draw [fill=black!20](2.8,7.1) node (v8) {} circle (7 mm); 
	\draw  plot[smooth, tension=.7] coordinates {(2.21,7.1) (2.5,7.6) (v8) (3.1,6.6) (3.37,7.1)};
		\draw (4.7,9.3) -- (3.45,9.3);
	\draw(0.8,7.1)  circle (3.5 mm); 
	\draw(0.4,7.1)  circle (3.5 mm); 
	\draw (-17.2,9.3) -- (-15.65,9.3);
\draw [fill=black!20](5.4,9.3) node (v8) {} circle (7 mm); 
	\draw  plot[smooth, tension=.7] coordinates {(4.81,9.3) (5.1,9.8) (v8) (5.7,8.8) (5.97,9.3)};
	
	\draw (-13.35,7.1) -- (-14.5,7.1);
	
	\draw(3.1,9.3)  circle (3.5 mm); 
	\draw(2.7,9.3)  circle (3.5 mm);  
	
	\draw[fill=black!20](-15.2,7.1) node {} circle (7 mm); 
	%(-12.7,16.5)
	\draw  plot[smooth, tension=.7] coordinates {(-15.7,7.1) (-15.487,7.6)(-15.2,7.1)(-14.94,6.6) (-14.7,7.1)};
	\node at (-17.9,8.2) {SG 1};

	\node at (2.8,6) {SG 4};

	\node at (-15.2,6) {SG 2};

\draw[ultra thick] (-13.8,7.9) -- (-13.8,6.3);

\draw (-12.25,7.1) -- (-11.9,7.1) -- (-11.6,8.8) -- (-11.1,8.8);

\node at (-13.8,10.6) {5};
\draw (-8.6,9.7) -- (-3.6,9.7) node (v5) {}; 
\draw (-8.6,8.9) -- (-3.6,8.9);
\draw[-latex] (-8.6,8.7) node (v1) {} -- (-9.1,8.7) -- (-9.1,7.4);
\draw  (-8.6,8.7) -- (-8.1,8.7) -- (-8.1,8.2); 
\draw (-8.35,8.2) -- (-7.85,8.2); 
\draw (-8.35,8) -- (-7.85,8);  
\draw (-8.1,8) -- (-8.1,7.4);  
\draw (-8.2,7.4) -- (-8,7.4); 
\draw [-latex](-3.6,8.7) node (v2) {} -- (-4.1,8.7) -- (-4.1,7.4); 
\draw (-3.6,8.7) -- (-3.1,8.7) -- (-3.1,8.2); 
\draw (-3.1,8) -- (-3.1,7.4); 
\draw (-3.2,7.4) -- (-3,7.4);  
\draw (-3.35,8.2) -- (-2.85,8.2); 
\draw (-3.35,8) -- (-2.85,8);
\draw (-3.6,9.3) -- (2.35,9.3);
\node at (-1.1,8.1) {10};
\draw (-1.1,8.8) -- (-0.6,8.8) -- (-0.3,7.1) -- (0.05,7.1);
\draw (1.15,7.1) -- (2.1,7.1);
\node at (5.4,8.2) {SG 3};
\node at (1.6,5.8) {4};
\node at (-1.1,10.6) {13};
\draw (-13.8,9.8) -- (-14.3,9.8) -- (-14.4,11.8) -- (-14.85,11.8);
\draw (-15.9,11.8) -- (-16.7,11.8);
\draw[color=backgroundcolor,fill=backgroundcolor!20]  (-19.7,12.4) rectangle (-17.7,11.2); 

\node [backgroundcolor] at (-18.7,12) {reserve};
\node [backgroundcolor] at (-18.7,11.6) {unit 1};

\draw (-3.6,9.7)-- (-3.1,9.7) -- (-3,11.8) -- (-2.55,11.8);
	\draw(-15.6,11.8)  circle (3.5 mm); 
	\draw(-15.2,11.8)  circle (3.5 mm);

\node at (-16.4,10.6) {12};
\draw (-1.5,11.8) -- (-0.8,11.8);
\draw  [magenta,fill=magenta!20](0.3,12.4) rectangle (2.3,11.2);

\node [magenta] at (1.3,12) {reserve};
\node [magenta] at (1.3,11.6) {unit 2};

\draw [backgroundcolor](-17.7,11.8) -- (-17.4,11.8) -- (-16.8,12.4);
\node [backgroundcolor]at (-17.2,12.6) {$S_1$};
\draw[magenta] (0.3,11.8) -- (0,11.8) -- (-0.7,12.4);
\node[magenta] at (-0.2,12.6) {$S_2$};
\end{tikzpicture}

}
    \vspace{-9mm}
    \caption{One-line diagram of the three-phase two-area test system with two converter-based reserve units for optimal dynamic ancillary services provision.}
    \label{fig:2area_system}
\end{figure}
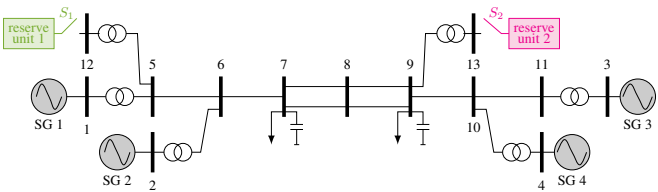

\fontdimen2\font=0.6ex
To demonstrate the effectiveness of the proposed P\&O approach, we use Simscape Electrical in MATLAB/Simulink to perform a detailed electromagnetic transients (EMT) simulation based on a \textit{nonlinear} model of the three-phase two-area test system in \cref{fig:2area_system}. In a first case study, we investigate how one single reserve unit, after applying the P\&O strategy, can significantly improve the overall closed-loop power grid performance during nominal grid conditions, while satisfying grid-code and device-level requirements. For the same setup, we additionally run Monte Carlo simulations to investigate how an inaccurate grid dynamic equivalent model during changing grid conditions is affecting the effectiveness of the P\&O strategy. In a second case study, we consider oscillatory grid conditions caused by weakly-damped inter-area modes, and reveal how a sequential application of the P\&O strategy by multiple reserve units allows us to iteratively optimize the overall frequency and voltage response behavior of the power grid while attenuating the inter-area oscillations. 

\subsection{Two-Area Test System Model} \fontdimen2\font=0.6ex
The two-area system in \cref{fig:2area_system} consists of two areas connected by a weak tie. It contains four thermal-based synchronous generators (SGs) and two additional converter-based reserve units connected to buses 12 and 13. The transmission lines are modeled via nominal $\pi$ sections (i.e., with RLC dynamics), and the step-up transformers via three-phase linear models. We adopt an 8th-order model for the synchronous machines equipped with an IEEE type 1 voltage regulator (AVR) combined with an exciter, and a power system stabilizer (PSS). The governors are modeled as a proportional speed-droop control with first-order delay, and the turbines as reheat steam turbines \cite{kundur2007power}. The power grid parameters are chosen similarly as in \cite{kundur2007power}, and the main parameters of the SGs are summarized in \cref{tab:sg_parameters}.

\subsection{Converter-Interfaced Generation System}\label{sec:converter_system} 
\begin{figure}[t!]
    \centering
    \usetikzlibrary{circuits.ee.IEC}
\resizebox {0.5\textwidth}{!}{
\begin{tikzpicture}[circuit ee IEC, scale =1, every node/.style={scale = 0.8}]
\draw [rounded corners = 3,color=backgroundcolor!,fill=backgroundcolor!20] (0.3,-1.2) rectangle (5,-2.8);
\draw  (-1.4,1.8) rectangle (-0.4,0.8);
\node at (-1.65,1.3) {$v_\mathrm{dc}$};
\draw (-1.4,0.8) -- (-0.4,1.8);
\node [scale=1.1] at (-1.1,1.5) {$=$};
\node [scale=1.1] at (-0.7,1.1) {$\approx$};
\node [scale=0.9] at (-0.9,2.15) {power};
\node [scale=0.9] at (-0.9,1.95) {converter};
\draw (-1.4,1.7) -- (-2.2,1.7) node (v1) {}; 
\draw (-1.4,0.9) -- (-2.2,0.9) node (v2) {}; 
\draw (-2.2,1.7) node (v3) {} -- (-2.2,1.35); 
\draw (-2.2,0.9) node (v4) {} -- (-2.2,1.25); 
\draw (-2.35,1.35) -- (-2.05,1.35); 
\draw (-2.35,1.25) -- (-2.05,1.25); 
\draw [-latex](-1.9,1.6) -- (-1.9,1); 
\draw(-2.2,1.7) -- (-3.3,1.7) node (v5) {}; 
\draw (-2.2,0.9) -- (-3.3,0.9) node (v6) {}; 
\draw  (-3.3,1.3) ellipse (0.2 and 0.2); 
\draw (-3.3,1.5) -- (-3.3,1.7) ; 
\draw (-3.3,1.1) -- (-3.3,0.9); 
\draw[-latex] (-3.3,1.15) -- (-3.3,1.45); 
\draw [-latex](-3.9,1.3)-- (-3.5,1.3);
\draw  (-3.9,1.6) rectangle (-4.9,1);
\node at (-3.65,1.55) {$i_\mathrm{dc}$};
\node at (-4.4,1.3) {$\frac{1}{\tau_\mathrm{dc}s+1}$}; 
\draw [-latex](-5.4,1.3) --(-4.9,1.3);
\draw [dashed,rounded corners = 2] (-3,0.8) rectangle (-5.2,1.8);
\node [scale=0.9] at (-4.1,1.97) {primary source};
\draw  (-6,1.55) rectangle (-5.4,1.05); 
\draw (-5.9,1.15) -- (-5.8,1.15) -- (-5.6,1.45) -- (-5.5,1.45); 
\draw[-latex] (0.5,-1.7) -- (-0.1,-1.7); 

\draw [rounded corners = 2] (0.5,-1.5) rectangle (1.1,-1.9);
\node [scale=0.9] at (0.8,-1.7) {PI};

\node at (3.6,-1.7) {$T_\mathrm{des}^\mathrm{fp}(s,\color{backgroundcolor}\alpha^\star\color{black})\Delta f_\mathrm{pll}-\Delta p$};

\draw[-latex] (-0.9,0.4) -- (-0.9,0.8);

\draw [rounded corners = 2] (0.5,-2.1) rectangle (1.1,-2.5);
\draw [rounded corners = 2] (-5.4,0.4) rectangle (-6,0); 
\draw [-latex](0.5,-2.3) -- (-0.1,-2.3); 

\node[scale=0.9] at (0.8,-2.3) {PI};
\node [scale=0.9] at (-5.7,0.2) {PI};

\node at (-5.7,-0.6) {$v_\mathrm{dc}^\star\hspace{-1.1mm}-\hspace{-0.8mm}v_\mathrm{dc}$};

\node at (3.5,-2.3) {$\Delta q-T_\mathrm{des}^\mathrm{vq}(s,\color{backgroundcolor}\alpha^\star\color{black})\Delta v$};

\draw (-0.4,1.5) to [inductor={yscale=1.2,xscale=0.8}] (1,1.5);
\draw  (1,1.6) rectangle (1.6,1.4); 
\draw (1.6,1.5) -- (2,1.5) node (v7) {}; 

\node at (-2.65,1.3) {$C_\mathrm{dc}$};

\fill(-2.2,1.7)circle(0.2mm);
\fill(-2.2,0.9)circle(0.2mm);
\node at (0.3,1.8) {$L_\mathrm{f}$};
\node at (1.3,1.8) {$R_\mathrm{f}$};

\draw  (4,1.5) ellipse (0.035 and 0.1);
\fill (3.2,1.5) node (v8) {}circle(0.3mm);
\draw [-latex](3.2,1.5) -- (3.2,1); 
\draw [-latex](4,1.4) -- (4,1);
\draw  [rounded corners = 3](2.7,1) rectangle (4.5,0.3);
\node at (3.6,0.8) {PLL \& power};

\node at (3.6,0.5) {computation};
\draw [color=gray,rounded corners = 3, fill=gray!40!white,opacity=0.5] (2.5,-0.1) rectangle (4.7,-0.8);
\node [gray] at (3.6,-0.3) {grid equivalent};
\node [gray] at (3.6,-0.6) {identification};
\draw [-latex](3.2,0.3) -- (3.2,-0.1); 
\draw [-latex](4,0.3) -- (4,-0.1);
\node at (2.9,0.1) {$p,q$};
\node at (4.5,0.1) {$f_\mathrm{pll}, v$};
\node at (2.9,1.3) {$v_\mathrm{abc}$};
\node at (4.3,1.3) {$i_\mathrm{abc}$};
\draw [rounded corners = 3] (-1.7,0.4) rectangle (-0.1,0);
\node at (-0.9,0.2) {modulation};
\draw  (-1.25,-0.5) rectangle (-0.55,-1); 
\draw (-1.25,-1) -- (-0.55,-0.5);
\node [scale=0.8] at (-1.02,-0.62) {abc};
\node [scale=0.8] at (-0.75,-0.85) {dq};
\draw [-latex](2.7,0.65) -- (2.2,0.65);
\node at (2.3,0.9) {$\theta_\mathrm{pll}$};
\draw[-latex] (-0.2,-0.75) -- (-0.55,-0.75);
\node at (0.1,-0.75) {$\theta_\mathrm{pll}$};  
\draw [-latex](-0.9,-0.5) -- (-0.9,0);
\draw (1.6,-1.7) circle(1mm);
\draw (1.6,-2.3) circle(1mm);
\draw [-latex](-0.7,-1.5)--(-0.7,-1); 
\draw [-latex](-1.1,-1.5) -- (-1.1,-1);

\draw [-latex,gray](1.6,-1) -- (1.6,-1.6); 
\draw [-latex,gray](1.6,-3) -- (1.6,-2.4);

\draw[scale=0.25,color=gray] (5.3,-3.5) -- (5.4,-2.1) -- (5.4,-3.3) -- (5.5,-2.9) -- (5.6,-3.7) -- (5.6,-2.8) -- (5.7,-3.4) -- (5.8,-2.6) -- (5.8,-3.6) -- (5.9,-2.8) -- (5.9,-3.4) -- (5.9,-2.9) -- (6,-3.2) -- (6.1,-2.8) -- (6.1,-3.3) -- (6.2,-2.6) -- (6.2,-3.6) -- (6.3,-2.6) -- (6.3,-3) -- (6.4,-3.3) -- (6.5,-2.9) -- (6.5,-3.2) -- (6.6,-3) -- (6.6,-3.2) -- (6.7,-2.9) -- (6.8,-3.4) -- (6.8,-2.8) -- (6.9,-3.3) -- (7,-2.9) -- (7,-3.2) -- (7,-3.1) -- (7.1,-2.6) -- (7.2,-3.5) -- (7.2,-3) -- (7.3,-3.2);
\draw[scale=0.25,color=gray] (5.4,-13.5) -- (5.5,-12.5) -- (5.5,-13.3) -- (5.6,-12.9) -- (5.7,-13.7) -- (5.7,-12.8) -- (5.8,-13.4) -- (5.9,-12.6) -- (5.9,-13.6) -- (6,-12.8) -- (6,-13.4) -- (6,-12.9) -- (6.1,-13.2) -- (6.2,-12.8) -- (6.2,-13.3) -- (6.3,-12.6) -- (6.3,-13.6) -- (6.4,-12.6) -- (6.4,-13) -- (6.5,-13.3) -- (6.6,-12.9) -- (6.6,-13.2) -- (6.7,-13) -- (6.7,-13.2) -- (6.8,-12.9) -- (6.9,-13.4) -- (6.9,-12.8) -- (7,-13.3) -- (7.1,-12.9) -- (7.1,-13.2) -- (7.1,-13.1) -- (7.2,-12.6) -- (7.3,-13.5) -- (7.3,-13) -- (7.4,-13.2);
\node [scale=0.8,gray] at (1.6,0) {wideband};
\node [scale=0.8,gray] at (1.6,-0.2) {excitation};
\node [scale=0.8,gray] at (1.6,-0.4) {signal injection};

\draw [rounded corners = 3] (-1.7,-1.5) rectangle (-0.1,-2.5);
\node at (-0.9,-1.7) {inner loop:};
\node at (-0.9,-2) {current};
\draw[-latex] (-2.2,-2) -- (-1.7,-2);
\node at (-2.5,-1.75) {$\theta_\mathrm{pll}$};

\draw[-latex](-5.7,0.4) -- (-5.7,1.05);
\node at (-5.9,0.7) {$i_\mathrm{dc}^\star$};

\node at (0.1,-1.5) {$i_\mathrm{d}^\star$};

\node at (0.1,-2.5) {$i_\mathrm{q}^\star$};

\draw [-latex](1.5,-1.7) -- (1.1,-1.7); 
\draw  [-latex](1.5,-2.3) -- (1.1,-2.3); 
\draw [-latex] (-5.7,-0.4) -- (-5.7,0);
%\draw  [dashed, rounded corners=3,backgroundcolor](-6.15,-0.55) rectangle (-5.15,-2.45);
%\node [backgroundcolor] at (-5.1,-0.5) {$^*$};

\draw[scale=0.65,fill=gray!40!white,color=gray!40!white,opacity=0.5] (7.4308,1.2231) node (v2) {} -- (7.4308,3.1731) -- (8.2308,3.1731) -- (8.2308,3.4731) -- (8.7308,3.0231) -- (8.2308,2.5731) -- (8.2308,2.8731) -- (7.7308,2.8731) -- (7.7308,1.2231) -- (7.4308,1.2231);

\node [gray] at (4.9,2.3) {$G(s)$};
\draw [dashed](5.2,1.5) node (v9) {} -- (6,1.5);
\node at (5.7,1.25) {grid};
\draw (4.5,1.6) -- (4.5,1.4);
\node [scale=0.9] at (4.45,1.75) {PCC};

\node at (-0.9,-2.3) {control};
\draw[backgroundcolor,-latex] (5.3,-2) -- (5,-2);
\node [backgroundcolor] at (5.7,-2) {$\alpha^\star(t)$};

\draw  (2.18,1.5) ellipse (0.17 and 0.17); 
\draw  (2.38,1.5) ellipse (0.17 and 0.17);
\draw (2.55,1.5) -- (5.2,1.5);
\node at (-2.5,-2) {$i_\mathrm{abc}$};
\node at (-2.5,-2.25) {$v_\mathrm{abc}$};
\draw[-latex] (2.2,-1.7) -- (1.7,-1.7);
\draw[-latex] (2.2,-2.3) -- (1.7,-2.3);

\node [scale=0.8,backgroundcolor] at (3.5,-1.375) {outer loop: matching control};
\node [scale=0.8] at (1.9,-1.85) {$+$};
\node [scale=0.8,gray] at (1.75,-1.4) {$+$};
\node [scale=0.8] at (1.9,-2.15) {$+$};
\node [scale=0.8,gray] at (1.75,-2.6) {$-$};
\end{tikzpicture}
}
    \vspace{-9mm}
    \caption{One-line diagram of three-phase power converter interface of one reserve unit including grid equivalent identification and matching control.}
    \label{fig:converter_model}
    \vspace{-2mm}
\end{figure}
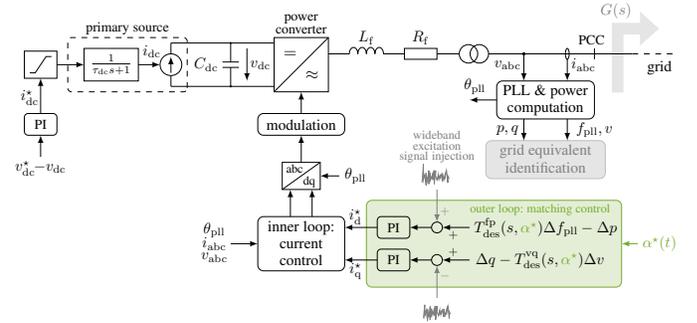

\fontdimen2\font=0.6ex
The proposed grid-side converter model is shown in \cref{fig:converter_model} with the associated parameters in \cref{tab:conv_parameters}. It represents an aggregation of multiple commercial converter modules and is based on a state-of-the-art converter control scheme \cite{yazdani2010voltage}. We can easily incorporate the required grid dynamic equivalent identification setup and the transfer-function matching control in such a setting. While \cref{fig:converter_model} shows only one exemplary converter control architecture, our method is compatible with any control architecture that accepts active and reactive power references.

Similar to \cite{tayyebi2020frequency}, we assume the dc current $i_\mathrm{dc}$ to be supplied by a controllable dc current source, e.g., schematically representing the machine-side converter of a direct-drive wind turbine, a PV system, etc. In particular, we consider a generic coarse-grain model of the primary source and model its response time by a first-order delay with a time constant $\tau_\mathrm{dc}$ \cite{tayyebi2020frequency}, e.g., representing the dynamics of the resource as well as communication and/or actuation delays. Further, we limit the primary source input by a saturation element, which, e.g., corresponds to the power/current limits of a machine-side converter or an energy storage system, or PV/wind power generation limits. For wind or PV generation, we assume that they are operated under deloaded conditions with respect to their maximum power point, allowing them to put an active power reserve aside for participating in frequency regulation. While we stick to such an abstract representation of the primary source, one could also consider more detailed models tailored to the application at hand.
 \renewcommand{\arraystretch}{1.2}
\begin{table}[t!]\scriptsize
    \centering
     \setlength{\tabcolsep}{1mm}
           \caption{Main parameters of the SGs in the nominal two-area system.}
            \vspace{-2mm}
               \begin{tabular}{c|c||c|c}
     \toprule
          Parameter & Value & Parameter & Value \\ \hline
          SG power rating & 900\,MVA&Exciter gain & 200 \\
          SG voltage rating & 20\,kV &Exciter time const. & 0.001\,s\\ \hline
          Governor droop gain & 0.05 &PSS gain & 25\\
          Governor time constant & 0.2\,s &PSS wash-out time const.& 1\,s \\
          Steam chest time constant & 0.5\,s &Lead-lag time const. \# 1& 50e-3\,s, 20e-3\,s\\
          Turbine power fraction& 0.3 & Lead-lag time const. \# 2& 3\,s, 5.4\,s \\\cline{3-4}
         Reheat time constant & 8\,s&Equivalent inertia & \makecell{4\,s (SG 1\&2)\\ 4.35\,s (SG 3\&4)} \\     
     \bottomrule
    \end{tabular}
        \vspace{-3mm}
     \label{tab:sg_parameters}
\end{table}
\renewcommand{\arraystretch}{1} \normalsize
 \renewcommand{\arraystretch}{1.2}
\begin{table}[t!]\scriptsize
    \centering
     \setlength{\tabcolsep}{1mm}
           \caption{Converter parameters of reserve units 1 and 2.}
            \vspace{-2mm}
               \begin{tabular}{c||c|c}
     \toprule
          Parameter & Symbol & Value  \\ \hline
          Voltage, power \& frequency base & $V_\mathrm{b}$, $S_\mathrm{b}$, $f_\mathrm{b}$ & 1\,kV, 500\,MVA, 50\,Hz\\
            DC-link capacitor& $C_\mathrm{dc}$ & 0.24\,p.u.\\
            $RL$-filter components & $L_\mathrm{f}$, $R_\mathrm{f}$ & 0.1\,p.u., 0.01\,p.u. \\
            DC-source time constants & $\tau_\mathrm{dc,1}$, $\tau_\mathrm{dc,2}$  & 0.1\,s, 0.04\,s\\ \hline
            Norm. max. active power ramp. rate & $R_\mathrm{max,1}^\mathrm{p}$, $R_\mathrm{max,2}^\mathrm{p}$ & 111\,s$^{-1}$, 150\,s$^{-1}$\\ 
            Norm. max. reactive power ramp. rate & $R_\mathrm{max,1}^\mathrm{q}$, $R_\mathrm{max,2}^\mathrm{q}$ & 150\,s$^{-1}$, 150\,s$^{-1}$\\
            Norm. max. active power peak capac. & $M_\mathrm{max,1}^\mathrm{p}$, $M_\mathrm{max,2}^\mathrm{p}$ & 70, 100\\
           Max. FFR support duration &  $t_\mathrm{d,max,1}^\mathrm{ffr}$, $t_\mathrm{d,max,2}^\mathrm{ffr}$ & 20\,s, 20\,s\\
           Max. return-to-recovery time & $t_\mathrm{r,max,1}^\mathrm{ffr}$, $t_\mathrm{r,max,2}^\mathrm{ffr}$& 20\,s, 20\,s\\
     \bottomrule
    \end{tabular}
    \tiny
    \begin{tabular}{l}
         \makecell{*Indices 1 and 2 refer to reserve units 1 and 2, respectively. No index implies identical values for reserve units 1 and 2.}
         \end{tabular}
        \vspace{-4mm}
     \label{tab:conv_parameters}
\end{table}
\renewcommand{\arraystretch}{1} \normalsize

The ac-side control of the grid-side converter is used to control the output current magnitude. It is implemented in a $\mathrm{dq}$-coordinate frame generated via a phase-locked loop (PLL), which tracks the system frequency to keep the converter synchronized with the grid voltage \cite{yazdani2010voltage}.

To identify the grid dynamic equivalent $G(s)$, we inject uncorrelated wideband excitation signals with small perturbation levels in the converter's control loop to excite the power grid during online operation as indicated in \cref{fig:converter_model}. The resulting frequency and voltage magnitude responses, as well as the active and reactive power injections at the PCC are then measured and collected in the form of discrete-time samples to calculate an estimate of the grid dynamic equivalent $G(s)$. 

Finally, the transfer function matching control is included in the outer control loop of the converter via simple PI controllers to track the desired dynamic response behavior for frequency and voltage regulation as $T_\mathrm{des}^\mathrm{fp}(s,\alpha^\star)$ and $T_\mathrm{des}^\mathrm{vq}(s,\alpha^\star)$, respectively. 

\subsection{Benchmark Ancillary Services Specification} 
\begin{figure}[t!]
    \centering
     \scalebox{0.46}{\includegraphics[]{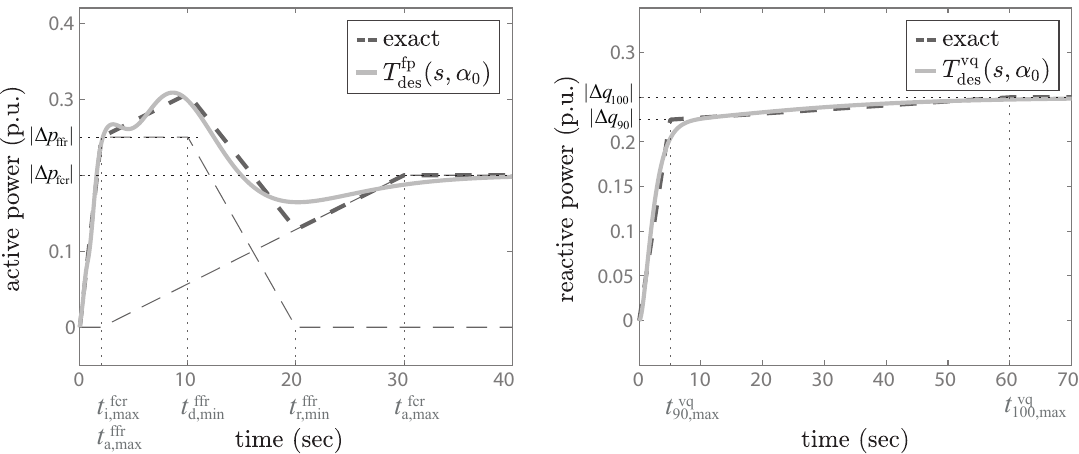}}
     \vspace{-1mm}
    \caption{Active and reactive power response after a negative frequency and voltage step change for the cheap $T_\mathrm{des}(s,\alpha_0)$ which satisfies minimal grid-code requirements (cf. grid-code examples in \cref{fig:fcr_ffr_grid_code,fig:pod_volt_grid_codes}).}
        \vspace{-3mm}
    \label{fig:min_grid_code_response}
\end{figure}

\fontdimen2\font=0.6ex
To demonstrate the efficiency of the proposed P\&O strategy in the following case studies, we define a \textit{cheap} desired transfer function matrix $T_\mathrm{des}(s,\alpha_0)$ as a benchmark ancillary services specification to compare with. More specifically, we select $\alpha_0$ as the critical point within $\mathcal{G}$ where all grid-code constraints in \cref{eq:grid_code_req_fcr_compact,eq:grid_code_req_ffr_compact,eq:grid_code_req_pod_compact,eq:grid_code_req_vq_compact} are active, i.e., 
\begin{align}
\nonumber
     \text{FCR:}\,\,\,& t_\mathrm{i}^\mathrm{fcr} = t_\mathrm{i,max}^\mathrm{fcr}, \,\, t_\mathrm{a}^\mathrm{fcr} = t_\mathrm{a,max}^\mathrm{fcr}\\\nonumber
     \text{FFR:}\,\,\,& t_\mathrm{a}^\mathrm{ffr} = t_\mathrm{a,max}^\mathrm{ffr}, \,\, t_\mathrm{d}^\mathrm{ffr} = t_\mathrm{d,min}^\mathrm{ffr},\,\, t_\mathrm{r}^\mathrm{ffr} = t_\mathrm{r,min}^\mathrm{ffr},\,\,
    x^\mathrm{ffr}=1\\\label{eq:min_grid_code_specifications}
     \text{AUX:}\,\,\,& m_\mathrm{aux} = 0\\\nonumber
     \text{VQ:}\,\,\,& t_\mathrm{90}^\mathrm{vq} = t_\mathrm{90,max}^\mathrm{vq}, \,\, t_\mathrm{100}^\mathrm{vq} = t_\mathrm{100,max}^\mathrm{vq},
\end{align}
thereby encoding the \textit{minimum} open-loop grid-code requirements as in \cref{tab:grid_code_specifications}. In particular, such a basic choice of $\alpha_0$ results in a cheap, but feasible dynamic ancillary services provision, where the effort of the reserve unit is pared down to the minimum. The active and reactive power responses of the cheap control $T_\mathrm{des}(s,\alpha_0)$ (and the underlying piece-wise linear time-domain curves) after a frequency and voltage step change are depicted in \cref{fig:min_grid_code_response}. The associated active and reactive power capacity levels are fixed by the allocated active and reactive power droop gains $D_\mathrm{p}=-0.05$, $K_\mathrm{p}=-0.04$ and $D_\mathrm{q}=-0.04$ (cf. \cref{fig:fcr_ffr_grid_code,fig:voltage_ctrl}). Notice that we will use the same droop gains also for the optimal $T_\mathrm{des}(s,\alpha^\star)$ computation in the following case studies, since they are typically directly given by the system operator and might not allow for flexibility.
 
\subsection{Case Study I: Optimal Ancillary Services Provision During Nominal Grid Conditions} \label{sec:csI_nominal}\fontdimen2\font=0.6ex
We commence our validation by examining nominal grid conditions, where only reserve unit 1 is interconnected with the two-area system (S1 closed in \cref{fig:2area_system}), while reserve unit 2 remains disconnected (S2 open). As a starting point, we consider the cheap transfer function matrix $T_\mathrm{des,1}(s,\alpha_0)$ to be realized in the outer loop of the converter system of reserve unit 1. Based on this, we aim to improve the overall closed-loop power grid performance by applying the P\&O strategy to compute an optimal $T_\mathrm{des,1}(s,\alpha_\mathrm{nom}^\star)$ for reserve unit 1.

\subsubsection{Grid Dynamic Equivalent Identification} 
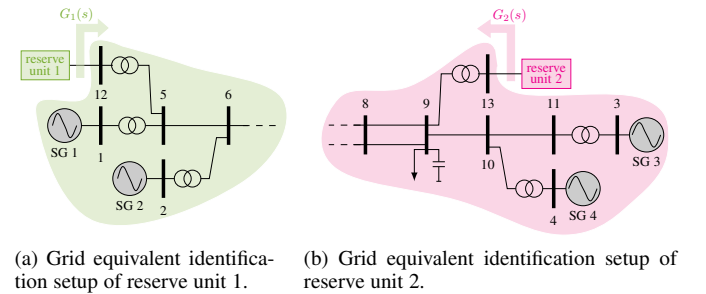
\begin{figure}[b!]
    \centering
\begin{subfigure}{0.39\columnwidth}
\vspace{-5mm}
     \usetikzlibrary{arrows}
\resizebox {1.1\columnwidth} {!} {
\begin{tikzpicture}[scale=0.39, every node/.style={scale=0.62}]

\draw [white,fill=backgroundcolor!40!white,opacity = 0.5] plot[smooth, tension=.7] coordinates {(-15.4,2.3) (-16.7,2.4) (-16.9,0.2) (-19,-0.4) (-16.5,-5) (-10.5,-3.4)(-8.7,0) (-12.7,1.5)(-15.4,2.3)};
	\draw(-14.45,-1) -- (-10.2,-1);
	\draw [ultra thick](-11,-0.2) -- (-11,-1.8);
	\draw [ultra thick](-16.3,-0.2) -- (-16.3,-1.8);

			\draw [ultra thick](-13.7,-0.2) -- (-13.7,-1.8);

		\draw [ultra thick](-16.3,2.3) -- (-16.3,0.7);

	\node at (-16.3,-2.3) {1};

	\node at (-11,0.3) {6};

	\node at (-13.7,-4.5) {2};

	%\draw [-latex] (-4.5,4) rectangle (-2.5,2.5);
	\draw[fill=black!20](-17.8,-1) node (v3) {} circle (7 mm);

		\draw(-15.2,-1)  circle (3.5 mm); 
	\draw(-14.8,-1)  circle (3.5 mm); 
	\draw  plot[smooth, tension=.7] coordinates {(-18.3,-1) (-18.087,-0.5) (v3) (-17.56,-1.5) (-17.3,-1)};
	
	%\node at (-3.5,3.25) {DVPP};

	\draw(-12.5,-3.2)  circle (3.5 mm); 
	\draw(-12.9,-3.2)  circle (3.5 mm);

	\draw (-17.1,-1) -- (-15.55,-1);

	\draw (-13.25,-3.2) -- (-14.4,-3.2);

	\draw[fill=black!20](-15.1,-3.2) node {} circle (7 mm); 
	%(-12.7,16.5)
	\draw  plot[smooth, tension=.7] coordinates {(-15.6,-3.2) (-15.387,-2.7)(-15.1,-3.2)(-14.84,-3.7) (-14.6,-3.2)};
	\node at (-17.8,-2.1) {SG 1};

	\node at (-15.1,-4.3) {SG 2};

\draw[ultra thick] (-13.7,-2.4) -- (-13.7,-4);

\draw (-12.15,-3.2) -- (-11.8,-3.2) -- (-11.5,-1.5) -- (-11,-1.5);

\node at (-13.7,0.3) {5};

\draw (-13.7,-0.5) -- (-14.2,-0.5) -- (-14.3,1.5) -- (-14.75,1.5);
\draw (-15.8,1.5) -- (-17.6,1.5);
\draw[color=backgroundcolor,fill=backgroundcolor!20]  (-19.6,2.1) rectangle (-17.6,0.9); 

\node [backgroundcolor] at (-18.6,1.7) {reserve};
\node [backgroundcolor] at (-18.6,1.3) {unit 1};

\node [backgroundcolor] at (-17.3,3.6) {$G_1(s)$};
\draw[fill=backgroundcolor!40!white,color=backgroundcolor!40!white,opacity=0.5] (-17.3,1) node (v4) {} -- (-17,1) -- (-17,2.9) -- (-16.4,2.9) -- (-16.4,2.7) -- (-16,3.05) -- (-16.4,3.4) -- (-16.4,3.2) -- (-17.3,3.2) --  (-17.3,1);

	\draw(-15.5,1.5)  circle (3.5 mm); 
	\draw(-15.1,1.5)  circle (3.5 mm);

\node at (-16.3,0.3) {12};

\draw [dashed](-10.4,-1) -- (-8.9,-1);
\end{tikzpicture}
}
    \vspace{-9mm}
    \caption{Grid equivalent identification setup of reserve unit 1.}
    \label{fig:2area_system_unit1_ID}
    \vspace{-1mm}
\end{subfigure}
\hspace{2mm}
\begin{subfigure}{0.56\columnwidth}
\vspace{-5mm}
     \usetikzlibrary{arrows}
\resizebox {1\columnwidth} {!} {
\begin{tikzpicture}[scale=0.39, every node/.style={scale=0.62}]
		\draw [white,fill=magenta!40!white,opacity = 0.5] plot[smooth, tension=.7] coordinates {(-1.6,33.9) (-2.4,34.1) (-5.1,32.4) (-8,31.9) (-8.2,29.1) (-3.1,27.7) (1.3,26.3) (4.3,27.9) (5.5,30) (4.4,31.9)(-0.6,32.1) (-1.5,33.2)};

\node [magenta] at (-1,35.3) {$G_2(s)$};
\draw[fill=magenta!40!white,color=magenta!40!white,opacity=0.5] (-1.1,32.6) -- (-1.1,34.6) -- (-1.8,34.6) -- (-1.8,34.4) -- (-2.3,34.75) -- (-1.8,35.1) -- (-1.8,34.9) -- (-0.8,34.9) --  (-0.8,32.6);

	\draw [ultra thick](-4.4,31.2) -- (-4.4,29.6);
		\draw [ultra thick](-1.9,33.7) -- (-1.9,32.1);
	\draw [ultra thick](0.8,29) -- (0.8,27.4);

	\draw [ultra thick](0.8,31.2) -- (0.8,29.6);

		\draw [ultra thick](-1.9,31.2) -- (-1.9,29.6);

	\draw [ultra thick](3.4,31.2) -- (3.4,29.6);
		\draw [ultra thick](-6.9,31.2) -- (-6.9,29.6);

	\node at (0.8,31.7) {11};
	\node at (3.4,31.7) {3};
	\node at (-4.4,31.7) {9};
	\node at (-6.9,31.7) {8};

	%\draw [-latex] (-4.5,4) rectangle (-2.5,2.5);

	\draw(-3,32.9)  circle (3.5 mm); 
	\draw(-2.6,32.9)  circle (3.5 mm);

	%\node at (-3.5,3.25) {DVPP};

		\draw [fill=black!20](2,28.2) node (v8) {} circle (7 mm); 
	\draw  plot[smooth, tension=.7] coordinates {(1.41,28.2) (1.7,28.7) (v8) (2.3,27.7) (2.57,28.2)};
		\draw (3.9,30.4) -- (2.65,30.4);
	\draw(0,28.2)  circle (3.5 mm); 
	\draw(-0.4,28.2)  circle (3.5 mm); 

\draw [fill=black!20](4.6,30.4) node (v8) {} circle (7 mm); 
	\draw  plot[smooth, tension=.7] coordinates {(4.01,30.4) (4.3,30.9) (v8) (4.9,29.9) (5.17,30.4)};

	\draw(2.3,30.4)  circle (3.5 mm); 
	\draw(1.9,30.4)  circle (3.5 mm);

	%(-12.7,16.5)

	\node at (2,27.1) {SG 4};

\draw (-7.4,30.8) -- (-4.4,30.8) node (v5) {}; 
\draw (-7.4,30) -- (-4.4,30);

\draw [-latex](-4.4,29.8) node (v2) {} -- (-4.9,29.8) -- (-4.9,28.5); 
\draw (-4.4,29.8) -- (-3.9,29.8) -- (-3.9,29.3); 
\draw (-3.9,29.1) -- (-3.9,28.5); 
\draw (-4,28.5) -- (-3.8,28.5);  
\draw (-4.15,29.3) -- (-3.65,29.3); 
\draw (-4.15,29.1) -- (-3.65,29.1);
\draw (-4.4,30.4) -- (1.55,30.4);
\node at (-1.9,29.2) {10};
\draw (-1.9,29.9) -- (-1.4,29.9) -- (-1.1,28.2) -- (-0.75,28.2);
\draw (0.35,28.2) -- (1.3,28.2);
\node at (4.6,29.3) {SG 3};
\node at (0.8,26.9) {4};
\node at (-1.9,31.7) {13};

\draw (-4.4,30.8)-- (-3.9,30.8) -- (-3.8,32.9) -- (-3.35,32.9);

\draw (-2.3,32.9) -- (-0.5,32.9);
\draw  [magenta,fill=magenta!20](-0.5,33.5) rectangle (1.5,32.3);

\node [magenta] at (0.5,33.1) {reserve};
\node [magenta] at (0.5,32.7) {unit 2};

\draw[dashed] (-7.1,30.8) -- (-8.4,30.8);
\draw [dashed](-7.1,30) -- (-8.4,30);
\end{tikzpicture}

}
    \vspace{-9mm}
    \caption{Grid equivalent identification setup of reserve unit 2.}
    \label{fig:2area_system_unit2_ID}
    \vspace{-1mm}
\end{subfigure} 
\caption{Grid dynamic equivalent identification of the two-area system.}
\label{eq:2_area_grid_ID_setups}
\end{figure}

We use the converter interface of reserve unit 1 to identify the grid dynamic equivalent $G_1(s)$ as in \cref{fig:2area_system_unit1_ID}. During the online grid identification experiment, we consider constant stationary grid conditions. To excite the power grid, we inject two uncorrelated superimposed RBS signals, each with an amplitude of 0.03 p.u. and a sampling rate of 1~kHz, in the power loop of the converter control architecture in \cref{fig:converter_model} for 40 seconds. When doing so, we ensure that the perturbation level is rather small to not deteriorate the ongoing grid operation.

\begin{figure}[t!]
    \centering
    \vspace{-1mm}
     \scalebox{0.45}{\includegraphics[]{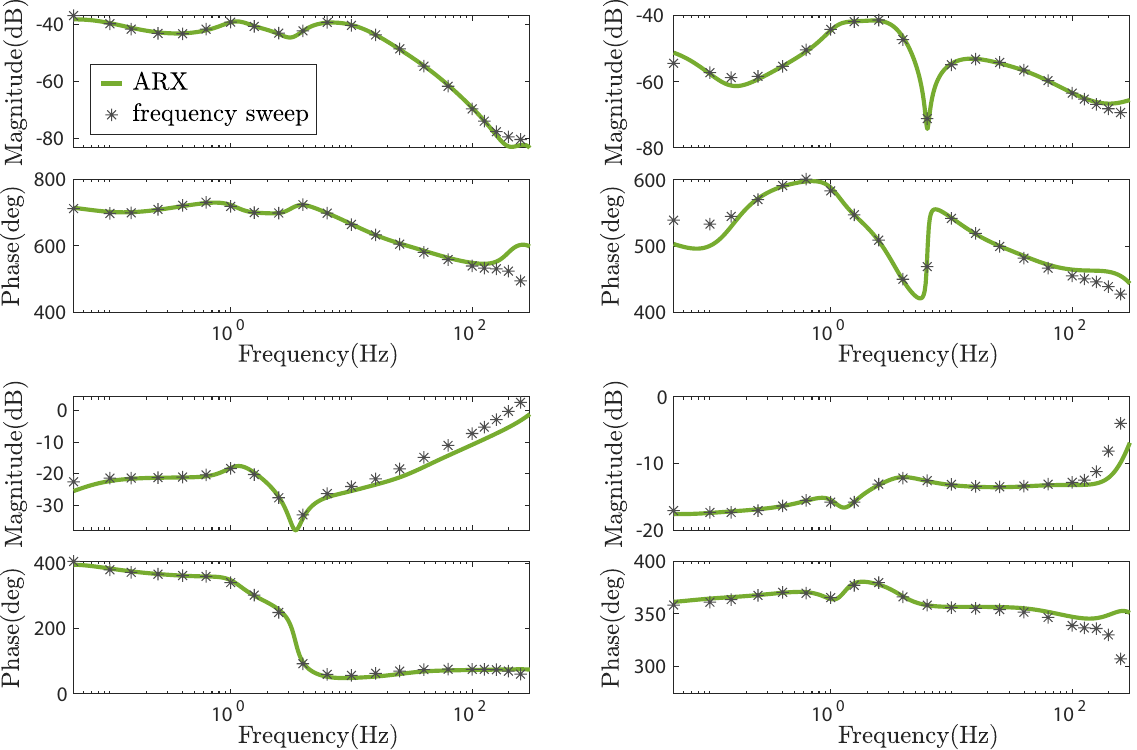}}
     \vspace{-1mm}
    \caption{Bode diagrams of the identified $2\times2$ grid dynamic equivalent $G_1(s)$ for the nominal two-area system in \cref{fig:2area_system_unit1_ID}.}
    \label{fig:gridID_nominal}
        \vspace{-3mm}
\end{figure}

The small-signal frequency and voltage magnitude responses, as well as the active and reactive power injections at the PCC of reserve unit 1, are then measured and collected (in the presence of measurement noise) at a sampling rate of 1 kHz. Since the recorded data set is based on discrete-time samples, the parametric grid impedance model is (initially) identified in the discrete domain. Namely, following a similar procedure as in \cite{haberle2023mimo}, we compute an ARX model by applying parametric system identification techniques. We additionally include other processing steps (e.g., prefiltering of the data) to ensure the accuracy of the system identification in the frequency range of interest. A proper model structure and order selection is done iteratively by testing a certain model structure and order and checking the fitting performance with some validation data. Once an appropriate ARX transfer function model has been identified, we apply model reduction techniques and convert it from discrete into continuous domain to finally obtain $G_1(s)$ of order 17. The resulting Bode diagram of $G_1(s)$ is illustrated by the solid green line in \cref{fig:gridID_nominal}. We can see that an accurate fitting of the reference grid model (obtained via \textit{noise-free} sinusoidal sweep methods and indicated by the star symbols) is achieved.

\subsubsection{Optimal $T_\mathrm{des}$ Computation}
\begin{figure}[t!]
    \centering
     \scalebox{0.46}{\includegraphics[]{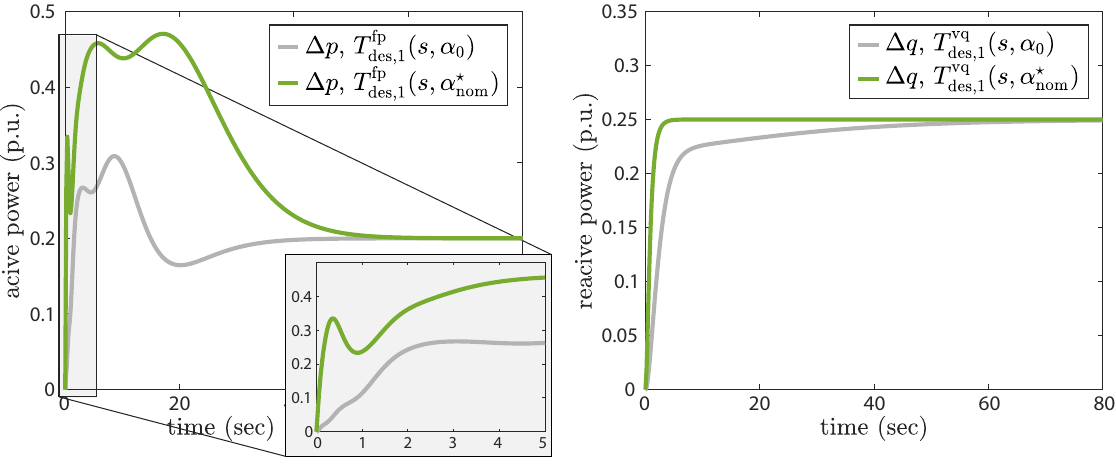}}
     \vspace{-1mm}
    \caption{Open-loop active and reactive power step responses after a negative frequency and voltage step change for the optimal $T_\mathrm{des,1}(s,\alpha_\mathrm{nom}^\star)$ and the cheap $T_\mathrm{des,1}(s,\alpha_0)$, respectively.}
        \vspace{-3mm}
    \label{fig:ol_step_nominal}
\end{figure}

\begin{figure}[t!]
    \centering
     \scalebox{0.46}{\includegraphics[]{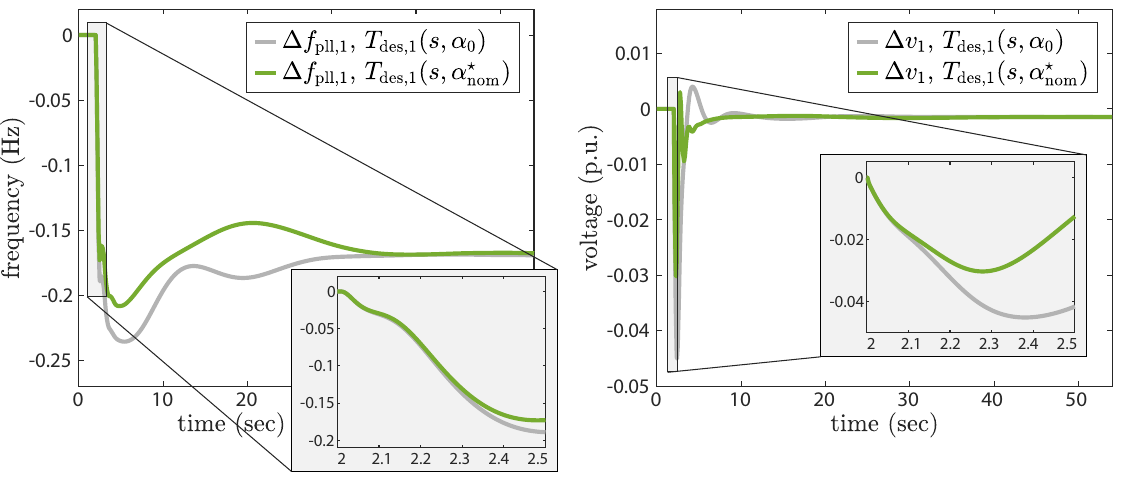}}
     \vspace{-1mm}
    \caption{Closed-loop system response behavior of the nominal two-area system after a load increase at bus 7 for the optimal $T_\mathrm{des,1}(s,\alpha_\mathrm{nom}^\star)$ and the cheap $T_\mathrm{des,1}(s,\alpha_0)$, respectively.}
    \label{fig:cl_step_nominal}
    \vspace{-3mm}
\end{figure}

Once the dynamic grid equivalent $G_1(s)$ is identified, we proceed to solve the closed-loop power grid optimization problem as detailed in \cref{sec:optimization_formulation}, to compute a locally optimal solution for $\alpha_\mathrm{nom}^\star$, while taking both grid-code and local device-level limitations of reserve unit 1 (see \cref{tab:conv_parameters}) into consideration. It is crucial to note that $G_1(s)$ is a full $2\times2$ matrix, therefore capturing the inherent coupling of active and reactive power with both frequency and voltage during closed-loop optimization. The resulting optimal parameter vector $\alpha_\mathrm{nom}^\star$ is presented in \cref{tab:Tdes_parameters}. A comparative analysis of the open-loop step response behaviors of the optimal $T_\mathrm{des,1}(s,\alpha_\mathrm{nom}^\star)$ and the cheap  $T_\mathrm{des,1}(s,\alpha_0)$ in \cref{fig:ol_step_nominal} allows us to conclusively assess the reliable satisfaction of minimum grid-code requirements of $T_\mathrm{des,1}(s,\alpha_\mathrm{nom}^\star)$. Finally, by realizing the obtained $T_\mathrm{des,1}(s,\alpha_\mathrm{nom}^\star)$ in the converter control loop of reserve unit 1, we observe a significant enhancement of the closed-loop system response behavior. Namely, despite being bandwidth and capacity-limited, and with only half the rating of one SG, the optimal dynamic ancillary services provision by reserve unit 1 demonstrates a substantial improvement of the system response during a load increase at bus 7 (\cref{fig:cl_step_nominal}). Specifically, we achieve a \textbf{12.6\%} improvement in RoCoF, an \textbf{11.6\%} improvement in frequency nadir, and a \textbf{32.9\%} reduction in voltage peak compared to the initial cheap $T_\mathrm{des,1}(s,\alpha_0)$ implementation.

\subsubsection{Changing Grid Conditions}
Given the exceptional performance of the optimal $T_{\mathrm{des,1}}(s,\alpha_{\mathrm{nom}}^\star)$ under ``laboratory conditions'', we now aim to investigate the effectiveness of the proposed P\&O strategy when dealing with inaccurate grid dynamic equivalent models. Possible sources of model inaccuracy include scenarios where the grid is not operating in a small-signal regime, changes in grid conditions after identifying the grid dynamics during one P\&O cycle, or significant identification errors caused by high noise levels during system identification. In the following, we exemplarily investigate changing grid conditions after the grid identification during one P\&O cycle as one possible source of model inaccuracy.

\begin{figure}[b!]
    \centering
        \vspace{-3mm}
     \scalebox{0.465}{\includegraphics[]{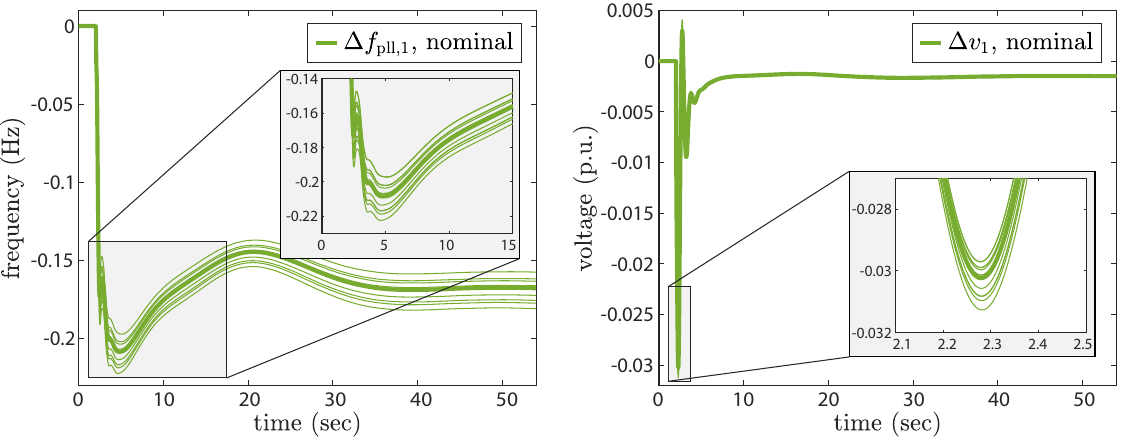}}
     \vspace{-5mm}
    \caption{Closed-loop response behavior of the two-area system following a load increase at bus 7, analyzed for the optimal $T_\mathrm{des,1}(s, \alpha_\mathrm{nom}^\star)$ in a nominal scenario (fat curve) and in Monte Carlo simulations with randomly varied active power generation set points of the SGs within a $\pm 10 \%$ range (thin curves).}
    \label{fig:monte_carlo_results}
    \vspace{-1mm}
\end{figure}

To further this investigation, we apply the previously obtained optimal $T_{\mathrm{des,1}}(s,\alpha_{\mathrm{nom}}^\star)$ to various grid conditions that differ from those initially perceived in \cref{fig:gridID_nominal}. More specifically, we perform Monte Carlo simulations by applying random active power generation set points to the four SGs in the 2-area system, varying within a range of $\pm 10\%$ from the nominal scenario, and study the system's response behavior during the same load increase at bus 7 as before. The simulation results, illustrated in \cref{fig:monte_carlo_results}, reveal that even with an inaccurate grid dynamic equivalent model, the P\&O strategy enhances the overall system response with close to optimal performance. This demonstrates the robustness of the P\&O strategy against minor inaccuracies in dynamic grid equivalent models. Nevertheless, the suboptimal response at the same time also highlights the necessity for accurate grid models, which can be achieved by reducing identification and model errors due to noise, or by adopting a multi-episodic application of the P\&O strategy where grid dynamics are continuously updated online in response to significantly changing grid conditions.

\subsection{Case Study II: Optimal Ancillary Services Provision During Oscillatory Grid Conditions with Multiple Reserve Units} \fontdimen2\font=0.6ex
We now consider oscillatory grid conditions caused by weakly-damped inter-area modes, arising from long transmission lines, fast exciters, and ill-tuned PSS gains (we decreased the PSS gains by a factor of five). Both reserve units are connected to the 2-area system (S1 and S2 closed), initially providing ancillary services as specified by the cheap transfer function matrices $T_\mathrm{des,1}(s,\alpha_0)$ and $T_\mathrm{des,2}(s,\alpha_0)$, respectively. Being limited in energy, reserve unit 2, does not provide the FCR service, i.e., we eliminate the $T_\mathrm{des,2}^\mathrm{fcr}$ term in \cref{eq:T_des_fp_components}.

Related to our discussion in \cref{sec:multi_converter}, we study a multi-converter scenario of the proposed method for optimal dynamic ancillary services provision, where each reserve unit applies the P\&O strategy sequentially one after another, while the remaining units stay connected with their previous $T_{\mathrm{des},i}(s,\alpha^\star)$.

\subsubsection{First P\&O Cycle} 
\begin{figure}[t!]
    \centering
    \vspace{-2mm}
     \scalebox{0.45}{\includegraphics[]{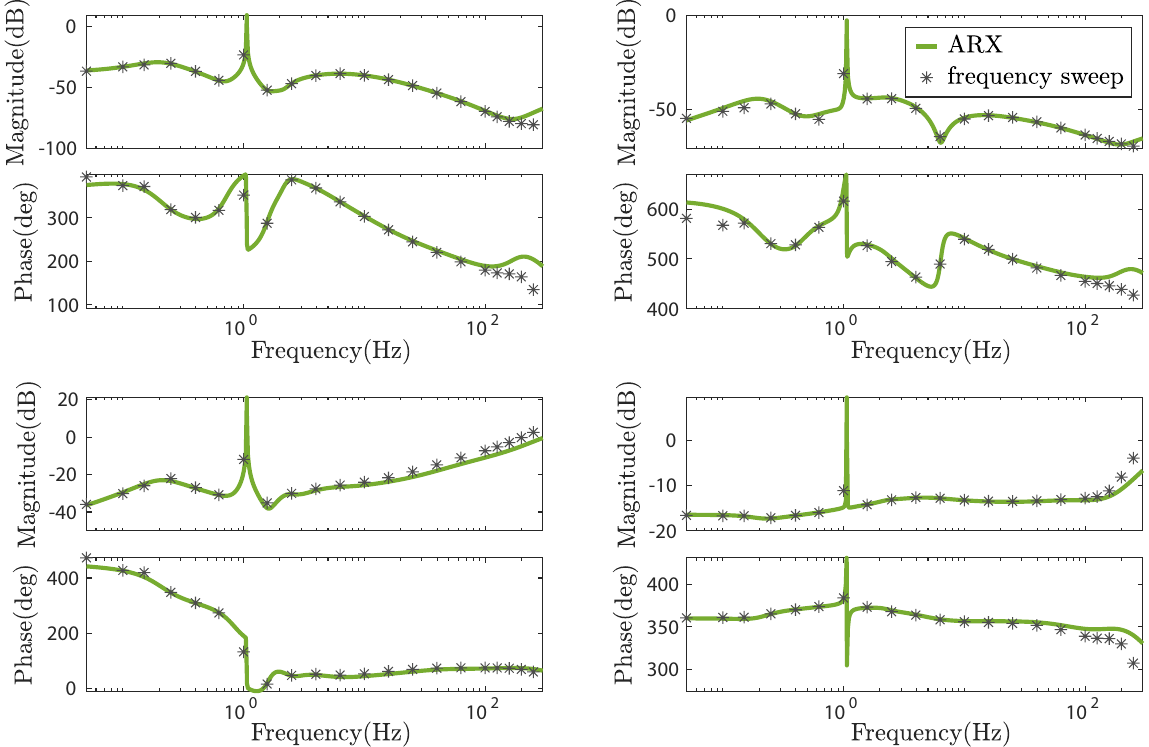}}
     \vspace{-1mm}
    \caption{Bode diagrams of the identified $2\times2$ grid dynamic equivalent $G_1(s)$ for the oscillatory two-area system in \cref{fig:2area_system_unit1_ID} during the first P\&O cycle.}
        \vspace{-3mm}
    \label{fig:gridID_marginal_u1}
\end{figure}

During grid identification, we obtain $G_1(s)$ of order 20. The resulting Bode diagram is illustrated by the solid green line in \cref{fig:gridID_marginal_u1}, with a significant resonance peak at approximately 1 Hz, indicating the oscillation frequency of the weakly-damped inter-area modes of the two-area system. 

\begin{figure}[t!]
    \centering
     \scalebox{0.46}{\includegraphics[]{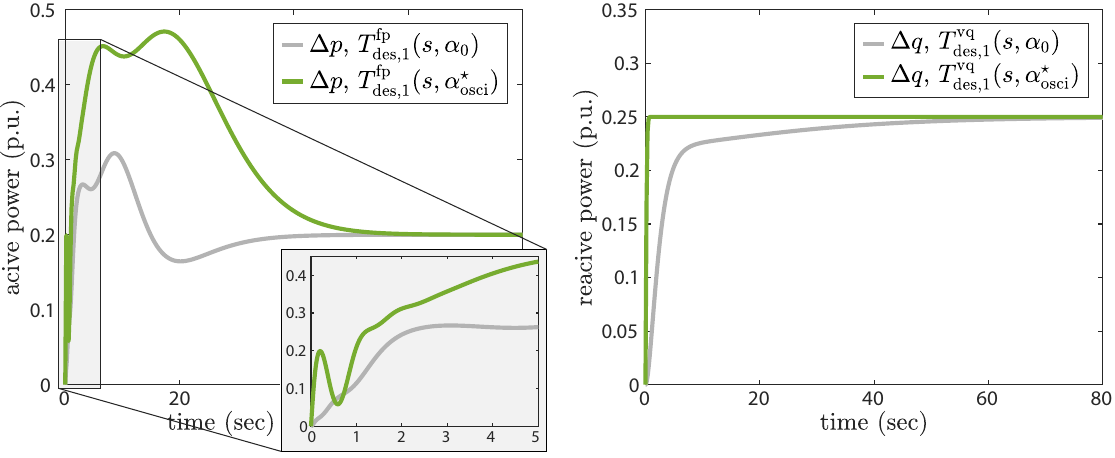}}
     \vspace{-1mm}
    \caption{Open-loop active and reactive power step responses after a negative frequency and voltage step change for the optimal $T_\mathrm{des,1}(s,\alpha_\mathrm{osci}^\star)$ and the cheap $T_\mathrm{des,1}(s,\alpha_0)$, respectively.}
    \vspace{-3mm}
    \label{fig:ol_step_marginal_u1}
\end{figure}

\begin{figure}[t!]
    \centering
     \scalebox{0.46}{\includegraphics[]{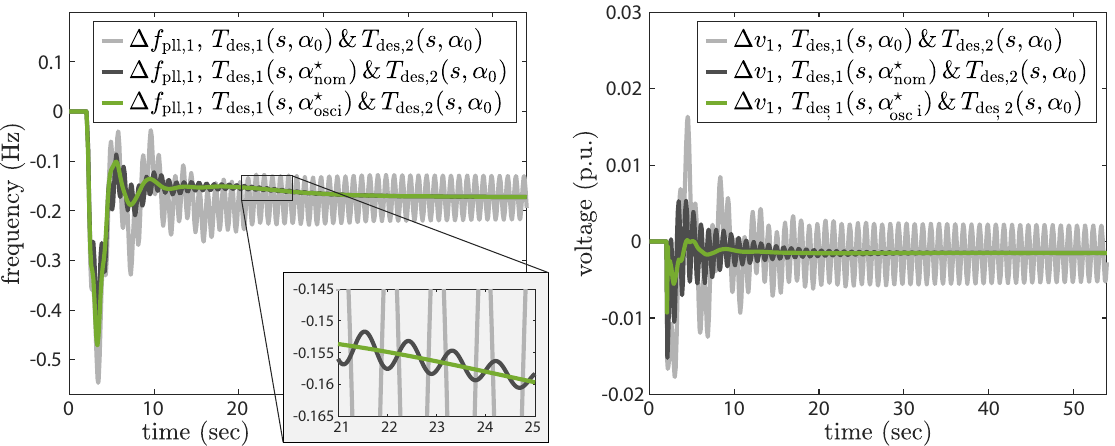}}
     \vspace{-1mm}
    \caption{Closed-loop system response behavior of the oscillatory two-area system after a load increase at bus 7. We compare the optimal $T_\mathrm{des,1}(s,\alpha_\mathrm{osci}^\star)$, the nominal $T_\mathrm{des,1}(s,\alpha_\mathrm{nom}^\star)$, and the cheap $T_\mathrm{des,1}(s,\alpha_0)$, of reserve unit 1, respectively. Reserve unit 2 is always realizing the cheap $T_\mathrm{des,2}(s,\alpha_0)$.}
        \vspace{-3mm}
    \label{fig:cl_step_marginal_u1}
\end{figure}

For identical cost-function weights and gradient descent algorithm settings as in case study I, we compute an optimal $\alpha_\mathrm{osci}^\star$ as listed in \cref{tab:Tdes_parameters}. The associated open-loop step response of $T_\mathrm{des,1}(s,\alpha_\mathrm{osci}^\star)$ is depicted in \cref{fig:ol_step_marginal_u1}. By comparing $\alpha_\mathrm{osci}^\star$ with the optimal $\alpha_\mathrm{nom}^\star$ from the previous case study I during nominal grid conditions, it becomes apparent how different grid conditions generally result in a different optimal dynamic ancillary services provision (for the same optimization settings).
\renewcommand{\arraystretch}{1.2}
\begin{table}[t!]\scriptsize
    \centering
     \setlength{\tabcolsep}{1mm}
           \caption{Optimal transfer function parameters $\alpha^\star$ for case studies I and II.}
            \vspace{-2mm}
               \begin{tabular}{l||c|c|c|c|c|c|c|c|c|c|c}
                   \toprule
                & $t_\mathrm{i}^\mathrm{fcr}$ &   $t_\mathrm{a}^\mathrm{fcr}$ & $t_\mathrm{a}^\mathrm{ffr}$ & $t_\mathrm{d}^\mathrm{ffr}$ & $t_\mathrm{r}^\mathrm{ffr}$ & 
                $x^\mathrm{ffr}$ &
                $\omega_\mathrm{l}$ & $\omega_\mathrm{h}$ & $m_\mathrm{aux}$ & $t_\mathrm{90}^\mathrm{vq}$ &$t_\mathrm{100}^\mathrm{vq}$\\ \hline
$\alpha_0$ &2&30&2&10&20&1&-&-&0&5&60 \\
               $\alpha_\mathrm{nom}^\star$ &0.01&3.73&1.32&21.32&41.32&1&2.4&7&-43.15&1.56&2.87 \\
               $\alpha_\mathrm{osci}^\star$ \#1 &0.01&5.64&1.65&21.65&41.65&1&5.36&9.96&-40.16&0.29&0.3 \\
               $\alpha_\mathrm{osci}^\star$ \#2 &-&-&0.84&20.84&40.84&1.35&0.95&5.55&-64&0.48&1.27\\
     \bottomrule
    \end{tabular}
        \vspace{-4mm}
     \label{tab:Tdes_parameters}
\end{table}
\renewcommand{\arraystretch}{1} \normalsize
With $T_\mathrm{des,1}(s,\alpha_\mathrm{osci}^\star)$, we can improve the oscillatory closed-loop system response of the two-area system quite significantly, i.e., once again, the performance enhancement of $\alpha_\mathrm{osci}^\star$ over $\alpha_0$ is astonishing. As illustrated in \cref{fig:cl_step_marginal_u1}, the inter-area oscillations at 1 Hz are significantly attenuated during a load increase at bus 7. Moreover, compared to the system response with the cheap $T_\mathrm{des,1}(s,\alpha_0)$, we can achieve a \textbf{10.5\%} improvement in RoCoF, a \textbf{13.8\%} improvement in frequency nadir, and a voltage peak reduction of \textbf{47.5\%}. Finally, the optimal $T_\mathrm{des,1}(s,\alpha_\mathrm{nom}^\star)$ for nominal grid conditions in case study I does not result in a satisfying response behavior when deployed in the oscillatory grid, which, in alignment with our observations during the Monte Carlo Simulations in \cref{sec:csI_nominal}, justifies the necessity of perceiving the grid characteristics in a online manner, especially during significantly changing grid conditions.

\subsubsection{Second P\&O Cycle}
\begin{figure}[t!]
    \centering
     \scalebox{0.46}{\includegraphics[]{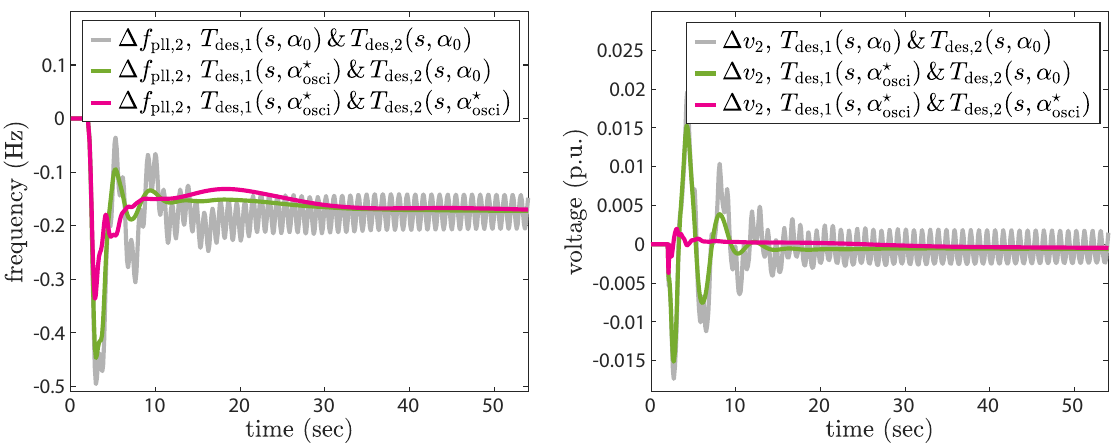}}
     \vspace{-1mm}
    \caption{Closed-loop system response behavior of the oscillatory two-area system after a load increase at bus 7. We compare the initial system configuration based on the cheap $T_\mathrm{des,1}(s,\alpha_0)$ and $T_\mathrm{des,2}(s,\alpha_0)$ of both units, with the optimal $T_\mathrm{des,1}(s,\alpha_\mathrm{osci}^\star)$ of unit 1 after the first P\&O cycle and the optimal $T_\mathrm{des,2}(s,\alpha_\mathrm{osci}^\star)$ of unit 2 after the second P\&O cycle. }
    \label{fig:cl_step_marginal_u2}
    \vspace{-3mm}
\end{figure}

Keeping $T_\mathrm{des,1}(s,\alpha_\mathrm{osci}^\star)$ for reserve unit 1, we now apply the P\& O strategy to reserve unit 2. After a grid identification as in \cref{fig:2area_system_unit2_ID}, we obtain $G_2(s)$ of order 24 and compute the optimal $T_\mathrm{des,2}(s,\alpha_\mathrm{osci}^\star)$ with $\alpha_\mathrm{osci}^\star$ as in \cref{tab:Tdes_parameters}. Compared to the first P\&O cycle, we can clearly observe a further improvement of the overall closed-loop grid response behavior during a load increase at bus 7 (\cref{fig:cl_step_marginal_u2}), i.e., a RoCoF improvement of \textbf{16.1\%}, a frequency nadir improvement of \textbf{24.7\%}, and a voltage peak improvement of \textbf{75.9\%}. 

\section{Conclusion}\label{sec:conclusion}\fontdimen2\font=0.6ex
\subsection{Summary}
We have presented a systematic approach to provide optimal dynamic ancillary services with converter-interfaced generation systems based on local power grid perception. To do so, we used the \textit{``perceive-and-optimize''} (P\&O) strategy: We first identified a grid dynamic equivalent at the interconnection terminals of the converter, and then established a closed-loop system interconnection of the identified grid equivalent and a desired transfer function matrix, where we optimize for the latter to provide optimal dynamic ancillary services. In the process, we ensure that grid-code and device-level requirements are satisfied. Our numerical experiments verify the superiority of our approach over cheap ancillary services provision based on minimum (open-loop) grid-code requirements, especially during changing grid-conditions.

\subsection{Outlook}
The provision of our proposed ``high-performing dynamic ancillary services'' computed via the P\&O strategy is at a first glance mainly of interest to system operators, as it allows for improved performance of the overall power grid behavior. In this regard, an immediate use case of our approach could be the implementation in STATCOMs or HVDC stations, as they are typically owned by the system operator. However, also from the viewpoint of external generating units not owned by the system operator, the P\&O strategy offers several \textit{local benefits}, such as improved small-signal stability, equipment protection, increased power quality, damping of local oscillations, etc. Nevertheless, despite these advantages, the current common practice typically considers it to be more convenient and profitable to just take cheap and minimal efforts to satisfy (open-loop) grid-code requirements, albeit being less performant and even prone to failure under particular grid conditions (cf. simulation results of case study II). Since this will no longer be acceptable for a reliable future power system operation, it is crucial to make the provision of high-performing dynamic ancillary services more attractive to external generating units, and thus realizable in practice. In particular, in addition to the aforementioned local benefits, we envision that the generating units could be encouraged with the following arguments (to be addressed as future work), i.e.,
\begin{enumerate}
        \item \textbf{Future dynamic ancillary service markets.} The provision of high-performing dynamic ancillary services should be incentivized by future dynamic ancillary services markets. Achieving this goal necessitates ensuring accountability and quantifiability of the dynamic ancillary services products, to enable a sale option on future ancillary services markets. In particular, given that our optimal dynamic ancillary service product is encoded in the form of a transfer function matrix $T_\mathrm{des}(s,\alpha)$, we consider this to be a measurable and thus well-suited format for being sold on future dynamic ancillary services markets. Indeed, already now, the first attempts how to assimilate novel high-performing ancillary service products into future markets have been proposed \cite{fernandez2020fast,meng2019fast,feng2023joint}.
        \item \textbf{Tender-based contract requirements.} Aside from incentives via ancillary services market profits, there should also be tender-based contract requirements for grid connection in future power systems, which ask for optimal and advanced dynamic ancillary services provision. Indeed, as an example, such tender-based requirements for high-performing dynamic ancillary services provision beyond minimal efforts are already common practice in the UK or Ireland \cite{fernandez2020fast}.
\end{enumerate}

\appendices

\section{Ancillary Services as Transfer Functions}\fontdimen2\font=0.6ex
\subsection{Translated Parametric Transfer Functions}\label{sec:parametric_TFs_appendix}
The rational parametric transfer functions $T_\mathrm{des}^\mathrm{fcr}(s,\alpha^\mathrm{fcr})$, $T_\mathrm{des}^\mathrm{ffr}(s,\alpha^\mathrm{ffr})$ and $T_\mathrm{des}^\mathrm{vq}(s,\alpha^\mathrm{vq})$ in \cref{sec:grid_code2tf} are obtained by translating the associated piece-wise linear time-domain grid-code curves in \cref{fig:fcr_ffr_grid_code,fig:voltage_ctrl} into the frequency domain. This can be achieved by applying our recent method in \cite{haberle2023gridcode}, which is based on piece-wise affine time-domain parametrizations and Laplace transformations, followed by Padé approximations of appropriate order $n$. In particular, the resulting parametric transfer function expressions as a function of the parameters $\alpha$ turn out to be quite intricate, i.e.,
\newpage
\begin{subequations}\label{eq:translated_parametric_tfs}
    \begin{align}\label{eq:translated_parametric_tfs1}
        T_\mathrm{des}^\mathrm{fcr}(s,\alpha^\mathrm{fcr}) &= \tfrac{1/D_\mathrm{p}}{s(t_\mathrm{a}^\mathrm{fcr}-t_\mathrm{i}^\mathrm{fcr})} \left( \tfrac{(1-\tfrac{t_\mathrm{i}^\mathrm{fcr}}{2n}s)^n}{(1+\tfrac{t_\mathrm{i}^\mathrm{fcr}}{2n}s)^n} -\tfrac{(1-\tfrac{t_\mathrm{a}^\mathrm{fcr}}{2n}s)^n}{(1+\tfrac{t_\mathrm{a}^\mathrm{fcr}}{2n}s)^n}\right)\\\label{eq:translated_parametric_tfs2}
         T_\mathrm{des}^\mathrm{ffr}(s,\alpha^\mathrm{ffr}) &= \tfrac{1/K_\mathrm{p}}{s t_\mathrm{a}^\mathrm{ffr}} \left(  1-\tfrac{(1-\tfrac{t_\mathrm{p}^\mathrm{ffr}}{2n}s)^n}{(1+\tfrac{t_\mathrm{p}^\mathrm{ffr}}{2n}s)^n}\right )\\\nonumber
         &\quad+\tfrac{1/K_\mathrm{p}(1-x^\mathrm{ffr})}{s(t_\mathrm{d}^\mathrm{ffr}-t_\mathrm{p}^\mathrm{ffr})} \left( \tfrac{(1-\tfrac{t_\mathrm{p}^\mathrm{ffr}}{2n}s)^n}{(1+\tfrac{t_\mathrm{p}^\mathrm{ffr}}{2n}s)^n} -\tfrac{(1-\tfrac{t_\mathrm{d}^\mathrm{ffr}}{2n}s)^n}{(1+\tfrac{t_\mathrm{d}^\mathrm{ffr}}{2n}s)^n}\right)\\\nonumber
         &\quad-\tfrac{1/K_\mathrm{p}}{s(t_\mathrm{r}^\mathrm{ffr}-t_\mathrm{d}^\mathrm{ffr})} \left( 
\tfrac{(1-\tfrac{t_\mathrm{d}^\mathrm{ffr}}{2n}s)^n}{(1+\tfrac{t_\mathrm{d}^\mathrm{ffr}}{2n}s)^n} -\tfrac{(1-\tfrac{t_\mathrm{r}^\mathrm{ffr}}{2n}s)^n}{(1+\tfrac{t_\mathrm{r}^\mathrm{ffr}}{2n}s)^n}
\right)\\\label{eq:translated_parametric_tfs3}
T_\mathrm{des}^\mathrm{vq}(s,\alpha^\mathrm{vq}) &= \tfrac{0.9/D_\mathrm{q}}{t_{90}s} \left(  1-\tfrac{(1-\tfrac{t_{90}}{2n}s)^n}{(1+\tfrac{t_{90}}{2n}s)^n}\right)\\\nonumber
&\quad +\tfrac{0.1/D_\mathrm{q}}{(t_{100}-t_{90})s} \left( \tfrac{(1-\tfrac{t_{90}}{2n}s)^n}{(1+\tfrac{t_{90}}{2n}s)^n} - \tfrac{(1-\tfrac{t_{100}}{2n}s)^n}{(1+\tfrac{t_{100}}{2n}s)^n}\right),
\end{align}
\end{subequations}
where $t_\mathrm{p}^\mathrm{ffr}=x^\mathrm{ffr}t_\mathrm{a}^\mathrm{ffr}$.
\subsection{Ancillary Services Constraints}\label{sec:AS_constraints_appendix}
\subsubsection*{FFR Provision} The constraint set which defines the grid-code and device-level requirements in \eqref{eq:grid_code_req_ffr_compact} of the active power capability curve for FFR provision in \cref{fig:ffr} is given as
\begin{subequations}\label{eq:grid_code_req_ffr}
\begin{align}
\label{eq:grid_code_req_ffr1}
0 \leq t_\mathrm{a}^\mathrm{ffr} &\leq t_\mathrm{a,max}^\mathrm{ffr}\\\label{eq:grid_code_req_ffr2}
t_\mathrm{d,min}^\mathrm{ffr}&\leq t_\mathrm{d}^\mathrm{ffr}\\\label{eq:grid_code_req_ffr3}
t_\mathrm{r,min}^\mathrm{ffr}&\leq t_\mathrm{r}^\mathrm{ffr}\\\label{eq:grid_code_req_ffr4}
1\leq x^\mathrm{ffr}&\leq x^\mathrm{ffr}_\mathrm{max}\\\label{eq:grid_code_req_ffr5}
|\Delta p_\mathrm{ffr}^\mathrm{max}|&\leq t_\mathrm{a}^\mathrm{ffr}\cdot r_\mathrm{max}^\mathrm{p}\\\label{eq:grid_code_req_ffr6}
t_\mathrm{d}^\mathrm{ffr}&\leq t_\mathrm{d,max}^\mathrm{ffr}\\\label{eq:grid_code_req_ffr7}
t_\mathrm{r}^\mathrm{ffr}&\leq t_\mathrm{r,max}^\mathrm{ffr}\\\label{eq:grid_code_req_ffr8}
x^\mathrm{ffr}&\leq \tfrac{m_\mathrm{max}^\mathrm{p}}{|\Delta p_\mathrm{ffr}^\mathrm{max}|},
\end{align}
\end{subequations}
where $t_\mathrm{a,max}^\mathrm{ffr}$ is the maximum admissible full activation time for the FFR provision. The FFR support duration time $t_\mathrm{d}^\mathrm{ffr}$ is lower and upper bounded by the minimum and maximum support duration $t_\mathrm{d,min}^\mathrm{ffr}$ and $t_\mathrm{d,max}^\mathrm{ffr}$, respectively\footnote{In our work, we assume $t_\mathrm{d,min}^\mathrm{ffr}>>t_\mathrm{a,max}^\mathrm{ffr}+t_\mathrm{a}^\mathrm{ffr}$, which is typically the case in today's grid codes.}. Likewise, the minimum and maximum FFR return-to-recovery times are given by $t_\mathrm{r,min}^\mathrm{ffr}$ and $t_\mathrm{r,max}^\mathrm{ffr}$, respectively\footnote{In our work, we assume $t_\mathrm{r,min}^\mathrm{ffr}\geq t_\mathrm{a,max}^\mathrm{ffr}+t_\mathrm{d}^\mathrm{ffr}$, which is typically the case in today's grid codes.}. The FFR overdelivery factor $x^\mathrm{ffr}$ must not exceed $\tfrac{m_\mathrm{max}^\mathrm{p}}{|\Delta p_\mathrm{ffr}^\mathrm{max}|}$, where $m_\mathrm{max}^\mathrm{p}$ is the reserve unit's maximum active power peak capacity, and $|\Delta p_\mathrm{ffr}^\mathrm{max}|=|\tfrac{1}{K_\mathrm{p}}\Delta f_\mathrm{max}|$ is the maximum FFR capacity. Additionally, $x^\mathrm{ffr}$ has to be smaller than the maximum tolerable overdelivery factor $x_\mathrm{max}^\mathrm{ffr}\in [1,2]$ specified in the grid code. As for the FCR constraints in \cref{eq:grid_code_req_fcr} in the revised manuscript, $r_\mathrm{max}^\mathrm{p}$ is the maximal active power ramping rate of the reserve unit. Finally, the grid-code and device-level specification sets as in \cref{eq:grid_code_req_ffr_compact} are defined as $\mathcal{G}^\mathrm{ffr}= \{ t_\mathrm{a}^\mathrm{ffr},t_\mathrm{d}^\mathrm{ffr},t_\mathrm{r}^\mathrm{ffr},x^\mathrm{ffr}\,|\, \eqref{eq:grid_code_req_ffr1}-\eqref{eq:grid_code_req_ffr4}\}$ and $\mathcal{D}^\mathrm{ffr}= \{t_\mathrm{a}^\mathrm{ffr},t_\mathrm{d}^\mathrm{ffr},t_\mathrm{r}^\mathrm{ffr},x^\mathrm{ffr}\,|\, \eqref{eq:grid_code_req_ffr5}-\eqref{eq:grid_code_req_ffr8} \}$, respectively, and all constraint parameters are further specified in \cref{tab:grid_code_device_level_parameters}.

\subsubsection*{Auxiliary Control Provision} The constraint set which defines the grid-code and device-level requirements in \cref{eq:grid_code_req_pod_compact} of the bandpass resonator transfer function in \cref{fig:pod} for auxiliary control provision (e.g., POD) is given as
\begin{subequations}\label{eq:grid_code_req_pod}
    \begin{align}\label{eq:grid_code_req_pod1}
        \omega_\mathrm{min}&\leq \omega_\mathrm{l}\\\label{eq:grid_code_req_pod2}
        \omega_\mathrm{h}&\leq \omega_\mathrm{max}\\\label{eq:grid_code_req_pod3}
        \omega_\mathrm{l}&\leq \omega_\mathrm{h}\\\label{eq:grid_code_req_pod4}
        |\Delta p_\mathrm{aux}^\mathrm{peak, max}(\alpha^\mathrm{aux})|&\leq m_\mathrm{max}^\mathrm{p}\\\label{eq:grid_code_req_pod5}
        0\leq t_\mathrm{aux}^\mathrm{s}(\alpha^\mathrm{aux})&\leq t_\mathrm{a,max}^\mathrm{ffr}\\\label{eq:grid_code_req_pod6}
        0\leq t_\mathrm{aux}^\mathrm{s}(\alpha^\mathrm{aux})&\leq t_\mathrm{i,max}^\mathrm{fcr},
    \end{align}
\end{subequations}
where $\omega_\mathrm{min}$ and $\omega_\mathrm{max}$ are the lower and upper bounds of the bandpass frequency range. Again, $m_\mathrm{max}^\mathrm{p}$ is the reserve unit's maximum active power peak capacity, and $|\Delta p_\mathrm{aux}^\mathrm{peak,max}(\alpha^\mathrm{aux})|$ is the parametric time-domain expression of the maximum peak power injection after a step disturbance, which can be obtained from the time-domain expression of $T_\mathrm{des}^\mathrm{aux}(s,\alpha^\mathrm{aux})$ as
\begin{align}\label{eq:p_aux_peak_max}
    |\Delta p_\mathrm{aux}^\mathrm{peak,max}(\alpha^\mathrm{aux})| = \left | \tfrac{2m_\mathrm{aux}}{\sqrt{\omega_\mathrm{d}^2+1}}e^{\tfrac{-\mathrm{arctan}(\omega_\mathrm{d})}{\omega_\mathrm{d}}}\cdot \Delta f_\mathrm{max} \right |,
    \end{align}
where $\omega_\mathrm{d} = \sqrt{4\tfrac{\omega_\mathrm{l}\omega_\mathrm{h}}{(\omega_\mathrm{h}-\omega_\mathrm{l})^2}-1} \in \mathbb{R}$. Moreover, the constraints in \eqref{eq:grid_code_req_pod5} and \eqref{eq:grid_code_req_pod6} ensure that the time-domain pulse width $t_\mathrm{aux}^\mathrm{s}(\alpha^\mathrm{aux})=\tfrac{9.2}{\omega_\mathrm{h}-\omega_\mathrm{l}}$ of the auxiliary active power injection is not interfering with the minimum grid-code requirements of the FCR and FFR injection.

Finally, the grid-code and device-level specification sets of the auxiliary control provision as in \cref{eq:grid_code_req_pod_compact} are defined as $\mathcal{G}^\mathrm{aux}= \{ \omega_\mathrm{l},\omega_\mathrm{h},m_\mathrm{aux}\,|\, \eqref{eq:grid_code_req_pod1}-\eqref{eq:grid_code_req_pod3},\eqref{eq:grid_code_req_pod5}, \eqref{eq:grid_code_req_pod6}\}$ and $\mathcal{D}^\mathrm{aux}= \{\omega_\mathrm{l},\omega_\mathrm{h},m_\mathrm{aux}\,|\, \eqref{eq:grid_code_req_pod4}\}$, respectively, and all constraint parameters are further specified in \cref{tab:grid_code_device_level_parameters}.

\subsubsection*{Superimposed Frequency Control} As indicated in the last paragraph in Section \cref{sec:grid_code2tf_frequency}, by superimposing the FCR, FFR and auxiliary control transfer functions, we can establish the overall frequency control specification $T_\mathrm{des}^\mathrm{fp}(s,\alpha^\mathrm{fp})$ as in \eqref{eq:T_des_fp_components}. While doing so, we further need to ensure that the maximum capacity and bandwidth limitations of the reserve unit are not violated during such a superimposed injection of active power. This requires additional overall device-level constraints for the $\mathrm{f}$-$\mathrm{p}$ control, which we approximate as
\begin{subequations}\label{eq:overall_fp_constraints}
\begin{align}\label{eq:max_cap_lim}
    \hspace{-2mm}|\Delta p_\mathrm{fcr}^\mathrm{max}| + |x^\mathrm{ffr}\Delta p_\mathrm{ffr}^\mathrm{max}| + |\Delta p_\mathrm{aux}^\mathrm{peak,max}(\alpha^\mathrm{aux})| &\leq m_\mathrm{max}^\mathrm{p}\\ \label{eq:max_bandw_lim}
\tfrac{|\Delta p_\mathrm{fcr}^\mathrm{max}|}{t_\mathrm{a}^\mathrm{fcr}-t_\mathrm{i}^\mathrm{fcr}}+\tfrac{|\Delta p_\mathrm{ffr}^\mathrm{max}|}{t_\mathrm{a}^\mathrm{ffr}}+\tfrac{|\Delta p_\mathrm{aux}^\mathrm{peak,max}(\alpha^\mathrm{aux})|}{t_\mathrm{aux}^\mathrm{peak}(\alpha^\mathrm{aux})} &\leq r_\mathrm{max}^\mathrm{p},
\end{align} 
\end{subequations}
where $|\Delta p_\mathrm{fcr}^\mathrm{max}|=|\tfrac{1}{D_\mathrm{p}}\Delta f_\mathrm{max}|$ is the maximum FCR capacity, $|\Delta p_\mathrm{ffr}^\mathrm{max}|= |\tfrac{1}{K_\mathrm{p}}\Delta f_\mathrm{max}|$ the maximum FFR capacity, $t_\mathrm{aux}^\mathrm{peak}(\alpha^\mathrm{aux}) = \tfrac{\mathrm{arctan}(\omega_\mathrm{d})}{\zeta \omega_\mathrm{d}}$ the peak power injection time of $T_\mathrm{des}^\mathrm{aux}(s,\alpha^\mathrm{aux})$ after a step disturbance with $\zeta = \tfrac{\omega_\mathrm{h}-\omega_\mathrm{l}}{2}$, and $|\Delta p_\mathrm{aux}^\mathrm{peak,max}(\alpha^\mathrm{aux})|$ the maximum peak power injection of $T_\mathrm{des}^\mathrm{aux}(s,\alpha^\mathrm{aux})$ after a step disturbance as in \eqref{eq:p_aux_peak_max}. In particular, as indicated in the last paragraph in Section \cref{sec:grid_code2tf_frequency}, we encode these overall $\mathrm{f}$-$\mathrm{p}$ device-level constraints in \eqref{eq:overall_fp_constraints} via an additional constraint set $\mathcal{D}^\mathrm{fp}=\{\alpha^\mathrm{fcr}, \alpha^\mathrm{ffr}, \alpha^\mathrm{aux}\,|\, \eqref{eq:overall_fp_constraints}\}$.

\subsubsection*{Voltage Regulation} The constraint set which defines the grid-code and device-level requirements in \cref{eq:grid_code_req_vq_compact} of the reactive power capability curve in \cref{fig:voltage_ctrl} is given as
\begin{subequations}\label{eq:vq_constraints}
 \begin{align}\label{eq:vq_constraints1}
    0\leq t_{90}^\mathrm{vq} &\leq t_\mathrm{90,max}^\mathrm{vq}\\\label{eq:vq_constraints2}
    t_{90}^\mathrm{vq} \leq t_{100}^\mathrm{vq} &\leq t_\mathrm{100,max}^\mathrm{vq}\\\label{eq:vq_constraints3}
    |\Delta q_{90}^\mathrm{max}| &\leq t_{90}^\mathrm{vq} \cdot r_\mathrm{max}^\mathrm{q}\\\label{eq:vq_constraints4}
    0.1\cdot |\Delta q_{100}^\mathrm{max}|&\leq \left(t_{100}^\mathrm{vq}-t_{90}^\mathrm{vq}\right) \cdot r_\mathrm{max}^\mathrm{q}
    \end{align}
\end{subequations}
where $t_\mathrm{90,max}^\mathrm{vq}$ and $t_\mathrm{90,max}^\mathrm{vq}$ are the maximum admissible activation times for the 90\% and 100\% reactive power capacity provision, respectively, $|\Delta q_{90}^\mathrm{max}|= 0.9\cdot |\Delta q_{100}^\mathrm{max}|$ and $|\Delta q_{100}^\mathrm{max}|$ are the maximum 90\% and 100\% reactive power capacity levels, and $r_\mathrm{max}^\mathrm{q}$ is the maximum reactive power ramping rate of the reserve unit. Finally, the grid-code and device-level specification sets as in \cref{eq:grid_code_req_vq_compact} are defined as $\mathcal{G}^\mathrm{vq}= \{ t_{90}^\mathrm{vq},t_{100}^\mathrm{vq}\,|\, \eqref{eq:vq_constraints1}-\eqref{eq:vq_constraints2}\}$ and $\mathcal{D}^\mathrm{vq}= \{t_{90}^\mathrm{vq},t_{100}^\mathrm{vq}\,|\, \eqref{eq:vq_constraints3}-\eqref{eq:vq_constraints4}\}$, respectively, and all constraint parameters are further specified in \cref{tab:grid_code_device_level_parameters}.

\section{$\mathcal{H}_2$ Optimization Problem}\label{sec:optimization}\fontdimen2\font=0.6ex 
The $\mathcal{H}_2$-norm between the disturbance input $w$ and the performance output $\Tilde{z}_\mathrm{p}$ of the system in \cref{eq:closed_loop_ss} is given by \cite{zhou1996robust}
    \begin{align}\label{eq:H2_norm}
    \begin{split}
        J &= \|\Tilde{T}_\mathrm{cl}(s,\alpha)\|^2_2 \\
        &=\text{trace}(C_\mathrm{cl}(\alpha) PC_\mathrm{cl}(\alpha)^\top)\\
        &= \text{trace}(B_\mathrm{cl}(\alpha)^\top QB_\mathrm{cl}(\alpha)),\hspace{-1mm}
        \end{split}
    \end{align}
where $P$ and $Q$ are the observability and controllability Gramain obtained as the positive definite solutions of the Lyapunov equation and its dual
\begin{subequations}\label{eq:lyapunov}
\begin{align}
    A_\mathrm{cl}(\alpha)P+PA_\mathrm{cl}(\alpha)^\top + B_\mathrm{cl}(\alpha)B_\mathrm{cl}(\alpha)^\top &= \mathbb{0}
    \\A_\mathrm{cl}(\alpha)^\top Q +QA_\mathrm{cl}(\alpha)+C_\mathrm{cl}(\alpha)^\top C_\mathrm{cl}(\alpha) &= \mathbb{0},
\end{align}
\end{subequations}
parameterized in $\alpha$ for the given closed-loop system matrices $A_\mathrm{cl}(\alpha)$, $B_\mathrm{cl}(\alpha)$ and $C_\mathrm{cl}(\alpha)$. Based on the latter, the optimization problem in \cref{eq:optimization_problem_ss} to compute the locally optimal $\alpha^\star$ with respect to the $\mathcal{H}_2$ norm $||\Tilde{T}_\mathrm{cl}(s,\alpha)||_2^2$ can be recasted as 
\begin{align}\label{eq:H2_optimization}
\begin{split}
    \underset{\alpha, P, Q}{\text{minimize}}\quad &\text{trace}(C_\mathrm{cl}(\alpha)PC_\mathrm{cl}(\alpha)^\top)\\
    \text{subject to}\quad & \alpha \in \mathcal{G}\cap \mathcal{D}, \, \eqref{eq:lyapunov}, \, P \succ 0,\, Q \succ 0.
    \end{split}
\end{align}
The constraints \eqref{eq:lyapunov} make the problem \cref{eq:H2_optimization} non-convex and difficult to solve. However, since the objective function is smooth, an explicit gradient $\nabla_\alpha J= \nabla_\alpha \text{trace}(C_\mathrm{cl}(\alpha)PC_\mathrm{cl}(\alpha)^\top)$ can be derived and directly used to solve \cref{eq:H2_optimization} for the (locally) optimal parameter vector $\alpha^\star$ via scalable first-order methods such as projected gradient descent, i.e.,
\begin{align}
    \alpha^{k+1} = \mathrm{proj}_{\mathcal{G}\cap\mathcal{D}} \left[ \alpha^k-\gamma^k\mathrm{grad}J(\alpha^k)\right],
\end{align}
where $\gamma^k\in\mathbb{R}$ is the step-size parameter, and $\mathrm{proj}_{\mathcal{G}\cap\mathcal{D}}[\cdot]$ the projection operator which is defined as
\begin{align}
\begin{split}
     \mathrm{proj}_{\mathcal{G}\cap\mathcal{D}} \left[ \xi \right] = \underset{y\in\mathbb{R}^n}{\text{argmin}}\,\, & \tfrac{1}{2}|| y-\xi ||_2^2\\
     \text{s.t.}\,\, & y \in \mathcal{G}\cap\mathcal{D}.
\end{split}
\end{align}
The term $\mathrm{grad}J(\alpha^k)$ specifies a gradient direction which allows for several implementation options or variants thereof, e.g.,
\begin{itemize}
    \item $\mathrm{grad}J(\alpha^k)= \nabla_\alpha J(\alpha^k)$ for classical gradient descent methods \cite{bertsekasnonlinear}, or
    \item $\mathrm{grad}J(\alpha^k)= \mathrm{sgn}(\nabla_\alpha J(\alpha^k))$ for sign gradient descent methods \cite{goodfellow2014explaining,moulay2019properties,riedmiller1993direct},
    \item etc.
\end{itemize}
In particular, sign gradient descent methods represent a widely adopted metaheuristic approach known for their typically faster convergence properties and increased likelihood of finding the global optimum compared to classical gradient descent methods. Moreover, they demonstrate reduced sensitivity to step size and initial conditions (further details can be found in \cite{goodfellow2014explaining,moulay2019properties,riedmiller1993direct}). In any case, for all the above implementation options, by using steps similar to the ones in \cite{ademola2018optimal}, the gradient with respect to $\alpha$ can be computed as
\begin{align}\label{eq:gradient}
    \nabla_\alpha J = \left[ \tfrac{\partial J}{\partial \alpha_1}, \tfrac{\partial J}{\partial \alpha_2}, \dots , \tfrac{\partial J}{\partial \alpha_n}\right]^\top,
\end{align}
where each component $\tfrac{\partial J}{\partial \alpha_i}$ is given as
\begin{align*}
\begin{split}
    \scriptstyle \tfrac{\partial J}{\partial \alpha_i} = 2 \cdot \text{trace}\left( \tfrac{\partial A_\mathrm{cl}}{\partial \alpha_i}PQ\right)+\text{trace}\left( \tfrac{\partial (B_\mathrm{cl}B_\mathrm{cl}^\top)}{\partial \alpha_i} Q\right)
    + \text{trace} \left( P\tfrac{\partial (C_\mathrm{cl}^\top C_\mathrm{cl})}{\partial \alpha_i}\right).\hspace{-1mm}
    \end{split}
\end{align*}

\section*{Acknowledgements}\fontdimen2\font=0.6ex
The authors wish to thank Simon Wenig from Mosaic Grid Solutions GmbH for his fruitful comments and discussions.

\renewcommand{\baselinestretch}{0.95}
\bibliographystyle{IEEEtran}
\fontdimen2\font=0.6ex
\bibliography{IEEEabrv,mybibliography}

\renewcommand{\baselinestretch}{1}
\begin{IEEEbiography}[{\includegraphics[width=1in,height=1.25in,clip,keepaspectratio]{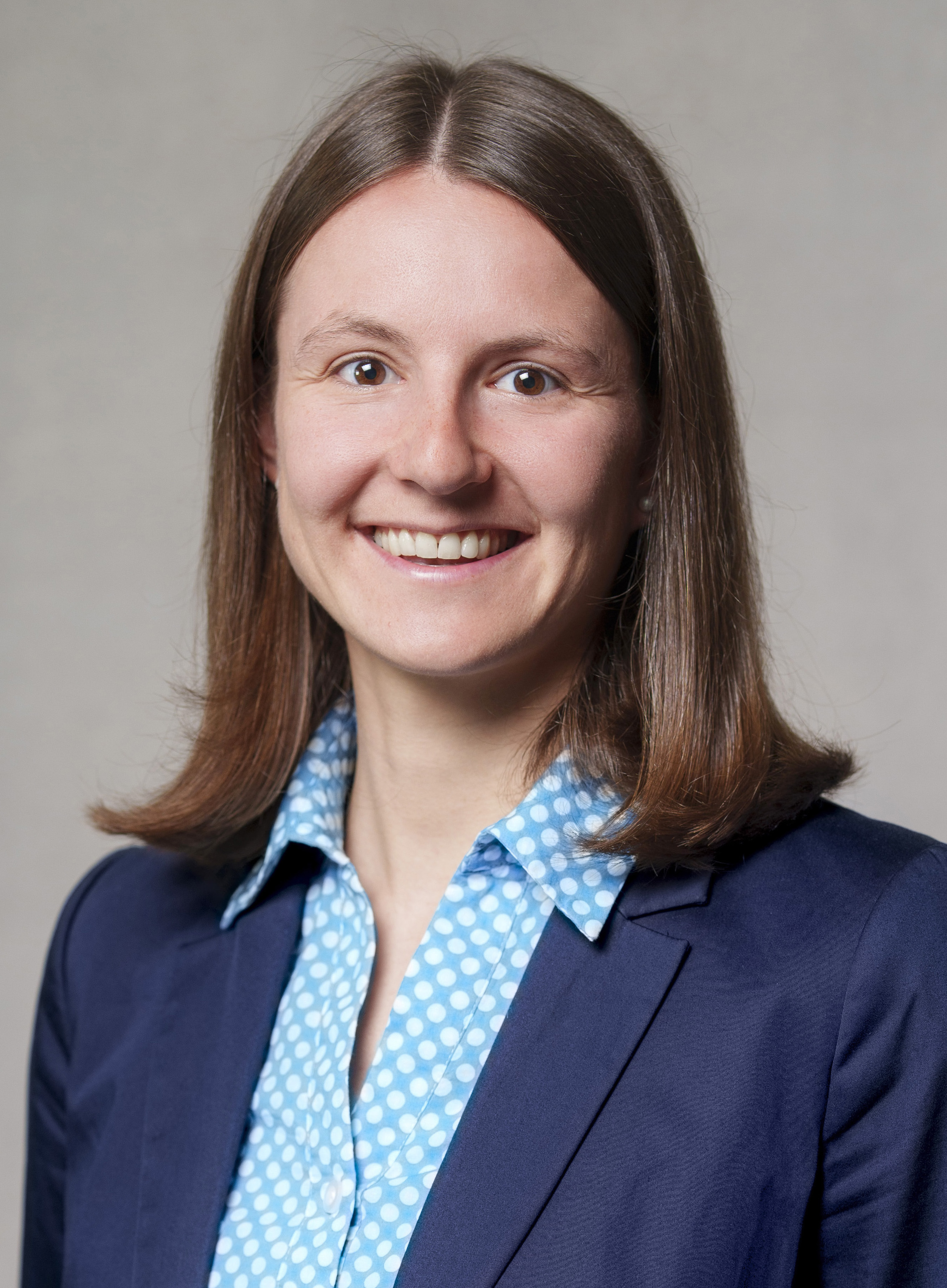}}]{Verena Häberle} is a Ph.D. student with the Automatic Control Laboratory at ETH Zurich, Switzerland, since June 2020. She received the B.Sc. and M.Sc. degree in electrical engineering and information technology from ETH Zurich, in 2018 and 2020, respectively. For her outstanding academic achievements during her Master's thesis at the Automatic Control Laboratory, ETH Zurich, under Professor Florian Dörfler, she was honored with the ETH Medal and the SGA Award from the Swiss Society of Automatic Control. Her research focuses on dynamic ancillary services provision, the control design of dynamic virtual power plants, as well as data-driven converter control  in future power systems.
\end{IEEEbiography}

\begin{IEEEbiography}[{\includegraphics[width=1in,height=1.25in,clip,keepaspectratio]{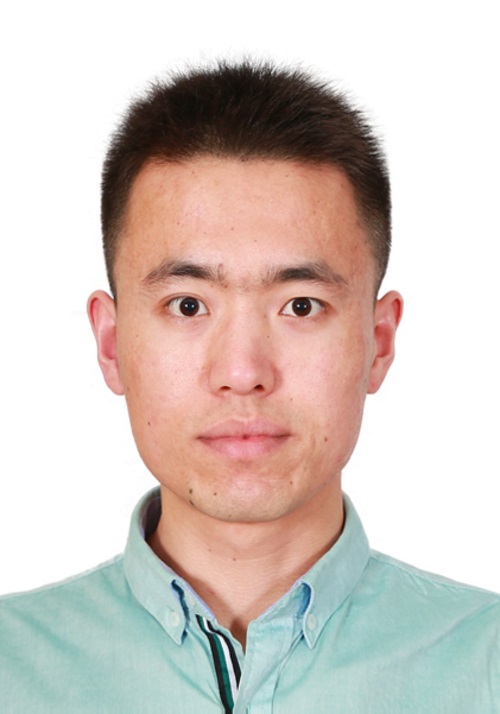}}]{Xiuqiang He} received his B.S. degree and Ph.D. degree in control science and engineering from Tsinghua University, China, in 2016 and 2021, respectively. Since 2021, he has been a Postdoctoral Researcher with the Automatic Control Laboratory, ETH Zürich, Switzerland. His current research interests include power system dynamics, stability, and control, involving multidisciplinary expertise in automatic control, power systems, power electronics, and renewable energy sources. Dr. He was the recipient of the Beijing Outstanding Graduates Award and the Outstanding Doctoral Dissertation Award from Tsinghua University.
\end{IEEEbiography}

\begin{IEEEbiography}[{\includegraphics[width=1in,height=1.25in,clip,keepaspectratio]{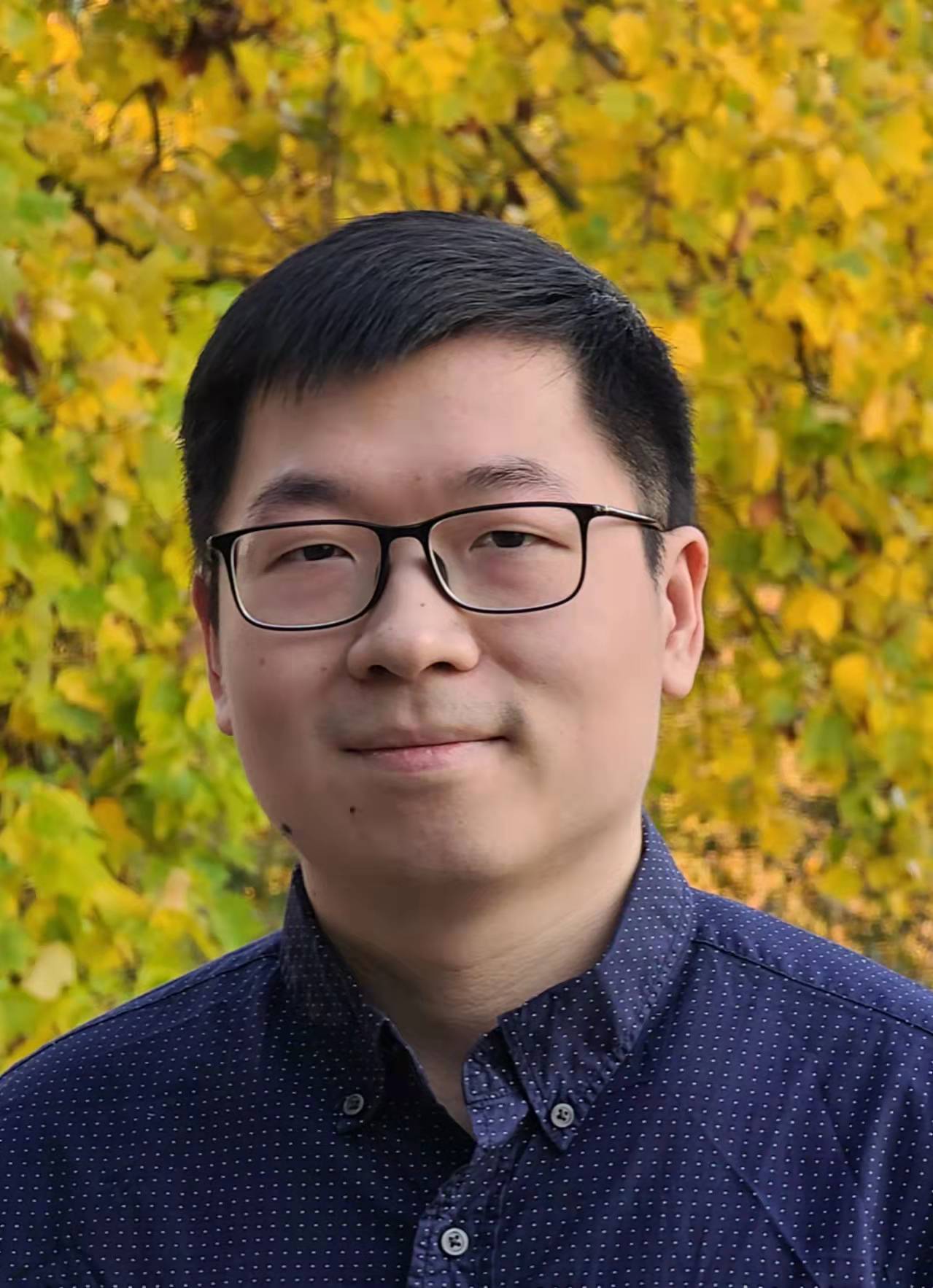}}]{Linbin Huang} received the B.Eng. and Ph.D. degrees from Zhejiang University, Hangzhou, China, in 2015 and 2020, respectively. Currently, he is a senior scientist with the Automatic Control Laboratory at ETH Zürich, 8092 Zürich, Switzerland. His research interests include power system stability, optimal control of power electronics, and data-driven control.
\end{IEEEbiography}

\begin{IEEEbiography}[{\includegraphics[width=1in,height=1.25in,clip,keepaspectratio]{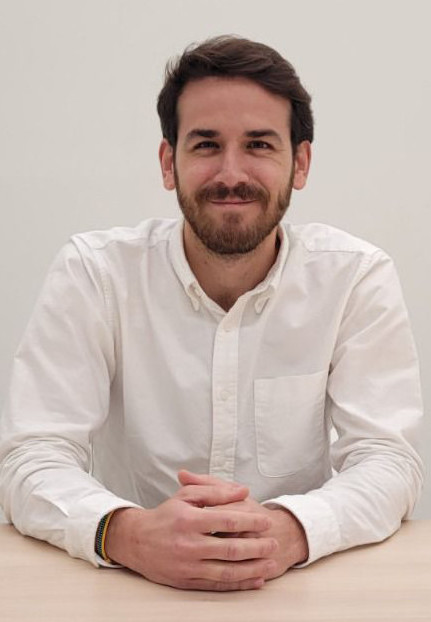}}]{Eduardo Prieto-Araujo} (S’12-M’16-SM’21) received the degree in industrial engineering from the School of Industrial Engineering of Barcelona (ETSEIB), Technical University of Catalonia (UPC), Barcelona, Spain, in 2011 and the Ph.D. degree in electrical engineering from the UPC in 2016. He joined CITCEA-UPC research group in 2010 and currently he is a Serra Hunter Associate Professor with the Electrical Engineering Department, UPC. During 2021, he was a visiting professor at the Automatic Control Laboratory, ETH Zurich. In 2022, he co-founded the start-up eRoots Analytics focused on the analysis of modern power systems. His main interests are renewable generation systems, control of power converters for HVDC applications, interaction analysis between converters and power electronics dominated power systems.
\end{IEEEbiography}

\begin{IEEEbiography}[{\includegraphics[width=1in,height=1.25in,clip,keepaspectratio]{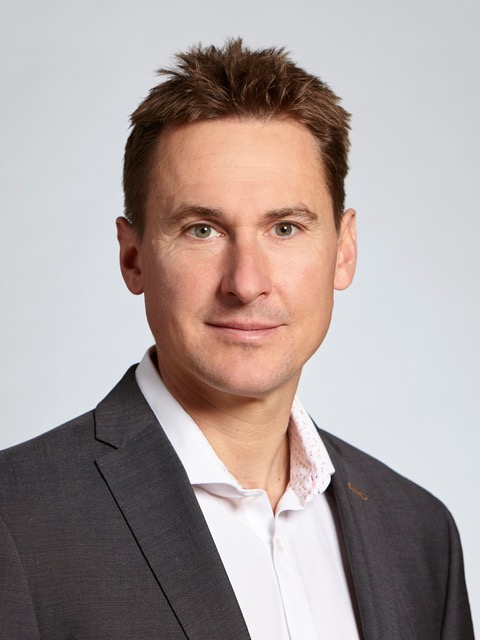}}]{Florian Dörfler} is a Full Professor at the Automatic Control Laboratory at ETH Zürich. He received his Ph.D. degree in Mechanical Engineering from the University of California at Santa Barbara in 2013, and a Diplom degree in Engineering Cybernetics from the University of Stuttgart in 2008. From 2013 to 2014 he was an Assistant Professor at the University of California Los Angeles. He has been serving as the Associate Head of the ETH Zürich Department of Information Technology and Electrical Engineering from 2021 until 2022. His primary research interests are centered around control, optimization, and system theory with applications in network systems, in particular electric power grids. He is a recipient of the distinguished young research awards by IFAC (Manfred Thoma Medal 2020) and EUCA (European Control Award 2020). His students were winners or finalists for Best Student Paper awards at the European Control Conference (2013, 2019), the American Control Conference (2016), the Conference on Decision and Control (2020), the PES General Meeting (2020), the PES PowerTech Conference (2017), the International Conference on Intelligent Transportation Systems (2021), and the IEEE CSS Swiss Chapter Young Author Best Journal Paper Award (2022). He is furthermore a recipient of the 2010 ACC Student Best Paper Award, the 2011 O. Hugo Schuck Best Paper Award, the 2012-2014 Automatica Best Paper Award, the 2016 IEEE Circuits and Systems Guillemin-Cauer Best Paper Award, the 2022 IEEE Transactions on Power Electronics Prize Paper Award, and the 2015 UCSB ME Best PhD award.
\end{IEEEbiography}

\end{document}